# Università degli studi "La Sapienza" di Roma

Facoltà di scienze Matematiche, Fisiche e Naturali

Corso di laurea specialistica in Astronomia e Astrofisica

Anno 2015/2016

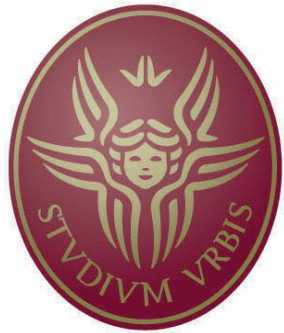

**Spectrophotometric correction of Vesta observations performed by the VIR Imaging spectrometer onboard Dawn mission**

Tesi di laurea specialistica di Pietro Scarica

| | |
|---|---|
| Relatore | Correlatori |
| Prof. Paolo De Bernardis | Maria Cristina De Sanctis |
| | Mauro Ciarniello |

# Contents

# INTRODUZIONE

Il remote sensing dei pianeti e dei corpi minori del sistema solare è un potente strumento per l'analisi delle loro proprietà fisiche e chimiche. I dati, acquisiti sotto condizioni di illuminazione e geometria di osservazione diverse (*angolo di incidenza, angolo di emissione e angolo di fase*), mostrano di essere dipendenti da tali condizioni. Per rimuovere la dipendenza della luce riflessa dalla superficie (la *riflettanza*), dalle condizioni di illuminazione e dalla geometria di osservazione, e per caratterizzare la variabilità intrinseca della superficie, i dati necessitano di una correzione fotometrica, che riporta le diverse osservazioni del corpo alle medesime condizioni di illuminazione, permettendo così un confronto tra le diverse acquisizioni. Tale processo può essere eseguito per mezzo di modelli empirici (*Minnaert, 1941; Akimov, 1988*) o basati sulla fisica (*Shkuratov et al., 1999; Hapke, 1981, 1984, 1986; Lambert, 1759; Seeliger, 1887*).

I dati fotometricamente corretti sono pronti per essere ulteriormente analizzati, per esempio producendo mappe della superficie, localizzando zone di interesse per futuri studi ed esplorazioni, confrontando i dati con misure di laboratorio. Inoltre i parametri dedotti per mezzo di una correzione fotometrica forniscono informazioni sui processi meccanici e fisici della superficie.

Lo scopo di questo lavoro di tesi è quello di correggere fotometricamente i dati spettroscopici dell'asteroide Vesta, osservato dallo strumento VIR a bordo della missione Nasa Dawn. Vesta, localizzato nella Fascia Principale degli Asteroidi, è il corpo genitore delle meteoriti della famiglia delle eucriti, diogeniti e howarditi (HED). Inoltre è il solo asteroide differenziato sopravvissuto quasi intatto, e quindi è un esempio unico tra tutti i corpi della Fascia degli Asteroidi.

Abbiamo usato il data-set fornito dallo Spettrometro a Immagine Visibile e Infrarosso (VIR: Visible and InfraRed Spectrometer) (*De Sanctis et al., 2010*), disegnato per la missione Nasa Dawn avente come target l'esplorazione di Vesta e Cerere.

La missione, lanciata il 27 Settembre 2007, è stata sviluppata per mezzo di una collaborazione internazionale di differenti parti. Il Jet Propulsion Laboratory della NASA ha fornito il coordinamento e la pianificazione generale della missione, i systems engineering del progetto, il sistema di volo e il sistema di propulsione a ioni. La Orbital Sciences Corporation si è occupata del design meccanico dello spacecraft. Il dipartimento per lo studio del sistema solare del Max Planck Institute e l'Agenzia Spaziale Tedesca (DLR) hanno fornito la Framing Camera (FC). Il Los Alamos National Laboratory ha provveduto al Gamma Ray and Neutron Detector (GraND). L'istituto Italiano Nazionale di Astrofisica (INAF) e l'Agenzia Spaziale Italiana (ASI) hanno fornito lo Spettrometro Visibile e Infrarosso (VIR).

Nel **Capitolo 1** è presentato l'asteroide Vesta nel suo aspetto fisico e nelle sue caratteristiche generali, nella composizione superficiale e nella sua struttura interna. Inoltre sono discussi gli obiettivi scientifici della missione Dawn e viene fornita una descrizione generale degli strumenti a bordo dello spacecraft.

Dopo una breve introduzione sulla spettroscopia (in particolare quella a immagine), nel **Capitolo 2** viene data una descrizione più accurata dello strumento VIR. L'esperimento VIR è uno spettrometro iperspettrale a immagine, in grado di accoppiare l'informazione spaziale (un'immagine) a quelle spettroscopiche (lungo un intervallo di lunghezze d'onda che va da 0.25-1.05 µm e da 1.0-5.0 µm, diviso in 864 bande). In questo secondo capitolo, tra gli obiettivi scientifici della missione sono discussi in particolare quelli riguardanti la composizione dell'asteroide.

Nel **Capitolo 3** vengono riportati i principali modelli fotometrici. In questa tesi, lo studio dei processi di scattering sulla superficie di Vesta è ottenuto attraverso l'applicazione della soluzione analitica dell'equazione del trasporto radiativo del modello di Hapke a cinque parametri (*Hapke, 2012*). In questo capitolo vengono discussi l'equazione del trasporto radiativo e alcuni modelli fotometrici, sia empirici sia basati su proprietà fisiche, con particolare attenzione al modello di Hapke adottato in questo lavoro di tesi. I diversi termini e i parametri che regolano tale modello sono presentati, con particolare focus sul loro significato fisico.

L'analisi dei dati è riportata nel **Capitolo 4**. Qui vengono presentati lo studio della copertura spaziale nelle osservazioni (in latitudine e longitudine), la copertura rispetto a particolari geometrie (angolo di fase, angolo di emissione, angolo di incidenza) e la selezione dei dati pre-processati. Successivamente viene fornito uno studio delle proprietà spettrofotometriche di Vesta, discutendo i set dei parametri fotometrici ottenuti per ciascuna delle lunghezza d'onda esaminate.

I parametri ottenuti nel capitolo precedente sono quindi utilizzati per correggere i set di dati. Nel **Capitolo 5** sono riportate le mappe di riflettanza fotometricamente corrette, sia nel visibile (0.751 µm) che nell'infrarosso (1.2 µm), e poste a confronto con lavori precedenti svolti (*Li et al., 2013, McCord et al., 2012, Longobardo et al., 2014*). Inoltre vengono presentate le mappe di differenza spettrale e in falsi colori (RGB), sono discussi i risultati ottenuti in questa tesi, ed è fornita una panoramica degli studi futuri che sarà possibile derivare a partire dalla correzione fotometrica operata.

# INTRODUCTION

Remote sensing of planetary and small solar system bodies is a powerful tool for the analysis of their physical and chemical properties. The data are usually acquired under different illumination and observation geometry (*incidence angle*, *emission angle* and *phase angle*) and show dependence from such conditions. In order to remove the dependence of the light reflected by the surface (the *reflectance*) from the illumination and observation geometry and to characterize the intrinsic variability of the observed surface, this data have to be interpreted through photometric correction, which scales different observations of the body to the same lighting conditions, permitting a comparison of different data. This process can be performed by means of empirical models (*Minnaert, 1941; Akimov, 1988*) or physically based ones (*Shkuratov et al., 1999; Hapke, 1981, 1984, 1986; Lambert, 1759; Seeliger, 1887*).

After photometric correction, the data are ready to be further analyzed, i.e. producing maps of the surface, locating areas of interest for future works and explorations, comparing the data with laboratory measurements. Moreover the parameters retrieved by photometric correction process give us information on the mechanical and physical nature of the surface.

The aim of this master degree thesis work is to produce a photometric correction of the spectroscopic data of asteroid Vesta, observed by the VIR instrument onboard Dawn NASA mission. Vesta, located in the Asteroid Main Belt, is the parent body of the eucrites, diogenites and howardites (HED) family of meteorites. It is the only nearly intact survived differentiated asteroid, and thus it is a unique example among all the bodies in the Main

Belt.

We used the data- set provided by the Visible and InfraRed Spectrometer (VIR) (*De Sanctis et al., 2010*), which is the Imaging Spectrometer designed for the Dawn NASA mission (mission targets: Vesta and Ceres).

The mission, launched in September 27, 2007, was developed as an international collaboration of different parts. The NASA's Jet Propulsion Laboratory gave overall planning and management, project systems engineering, the flight system, the scientific payload development and the ion propulsion system. The Orbital Sciences Corporation provided the mechanical design for the spacecraft. The Max Planck Institute for Solar System Studies research and the German Aerospace Center (DLR) supplied the Framing Camera (FC). The Los Alamos National Laboratory provided the Gamma Ray and Neutron Detector (GraND). The Italian National Institute for Astrophysics (INAF) and the Italian Space Agency (ASI) supplied the Visible and Infrared Spectrometer (VIR).

In **Chapter 1**, the asteroid Vesta is presented, in terms of its physical appearance, general characteristics, surface composition and internal structure. Along with this, the Dawn mission scientific objectives are discussed and an overview of the instruments onboard the spacecraft is given.

After a brief introduction to spectroscopy (in particular the imaging spectroscopy), a more proper description of the VIR instrument is presented in **Chapter 2.** The VIR experiment is a hyperspectral imaging spectrometer, which is able to couple the spatial information (an image) to the spectral one (along 864 bands in the 0.25-1.05 μm and 1.0-5.0 μm ranges). In this second chapter, the main scientific goals of the mission, in particular those concerning the asteroid composition, are discussed.

In **Chapter 3** the main photometric models are reported. The study of the scattering processes on Vesta's surface is obtained through the application of the analytical solution of radiative transfer equation given by a five parameters Hapke's model (*Hapke, 2012*). The radiative transfer equation, along with a summary of different – empirical and physically based – photometric models and in particular the adopted Hapke's model, are presented. The different terms and the parameters of the Hapke's model equation are discussed, in particular with respect to their physical meaning.

The data analysis is reported in **Chapter 4**. The study of the observation spatial coverage (in *latitudes* and *longitudes*) and the coverage with respect to some particular geometries (*phase angle*, *emission angle* and *incidence angle*), as well as the pre-processing data selections, are presented. The study of Vesta spectrophotometric properties is described with the discussion of the best-fit sets of the photometric parameters obtained for each of the wavelengths of interest.

Finally, the photometric parameters are used to correct the data-set. In **Chapter 5**, photometrically corrected reflectance maps in both VIS (0.751 μm) and IR (1.2 μm) ranges are shown, along with a comparison with previous works (*Li et al., 2013, McCord et al., 2012, Longobardo et al., 2014*). Then, maps of spectral differences and enhanced color (RGB) maps are produced. The obtained results are discussed as well as some hints for further studies of the asteroid surface.

# Chapter 1

# The scientific interest in Vesta and the Dawn mission

4 Vesta was discovered on 29 March 1807 by Heinrich Olbers, who was searching for the remnants of a theoretical destroyed planet after the recent discovery of Ceres, Pallas and Juno. The hunting for the missing body was pursuit since 1798, when a *"Celestial Police"* began a methodical investigation for an expected body in the gap between Mars and Jupiter, trying to corroborate the Titius-Bode law, the numerical progression that puts planetary distances into proportional numbers.

First of all was discovered Ceres, when the astronomer Giuseppe Piazzi found a moving object, as stated by himself: *"...on the evening of the 1st of January of the current year, together with several other stars, I sought for the 87th of the Catalogue of the Zodiacal stars of Mr la Caille. I then found it was preceded by another, which, according to my custom, I observed likewise, as it did not impede the principal observation. The light was a little faint, and of the colour of Jupiter, but similar to many others which generally are reckoned of the eighth magnitude. Therefore I had no doubt of its being any other than a fixed star..."* This faint star, not even reported on his charts, was not a comet – as previously announced by himself – but more a planet-like body. Then, basing on the limited tracking available, the mathematician C.F. Gauss developed an orbit calculation, that allowed to rediscover – one year later – the mysterious celestial body.

When Pallas came to the light, it suggested the possibility that another fragment of this *"broken planet"* would be discovered in the region of intersection between the orbit of Ceres and Pallas. There came Juno, much smaller than the others, and Vesta. After several decades, other bodies – supposed to be planets – were observed in the same region; as a result of the increasing number of new planets, they were all reclassified as asteroids.

**1.1 A general overview on Vesta before Dawn**

Vesta is an oblate spheroid, a tri-axial ellipsoid with a slightly irregular shape and a very large feature – a crater – at the south pole. It is the second larger body in the Main Asteroid Belt, with a mean diameter of 530 kilometers (*Thomas et al. 1997*), and a mass of $2.6 \times 10^{20}$ Kg (*Russell et al. 2012*). Those numbers credit Vesta to be 25% more massive than Pallas, even if only slightly larger; its density (3.4 g/cm$^3$) is higher than most asteroids and moons in the Solar System, and it is not so far from Mars' density (3.9 g/cm$^3$).

Even if Vesta seems small compared to the planets and our Moon, certainly it has some planetary characteristics, like large mountains-valleys systems and graben structures visible on large sections of the surface.

Located next to the Kirkwood gap (**Fig. 1**) corresponding to the strong Jupiter resonance 3:1, which depleted that region of the Main Asteroid Belt, Vesta orbits around the Sun in 3.6 years, with an eccentricity e = 0.09, an inclination i = 7.1° and a semi-mayor axis roughly about 2.4 AU (*JPL Small-body Database Browser*). Such proper orbital elements and the physical characteristics (**Tab. 1**) give the picture of a "quasi-planet" body, which straddles the line between planets and SSSBs (Small Solar System Bodies).

| Physical Parameter Table | | | | |
|---|---|---|---|---|
| Parameter | Value | Units | Sigma | Reference |
| absolute magnitude | 3.20 | mag | n/a | IRAS-A-FPA-3-RDR-IMPS-V6.0 |
| magnitude slope | 0.32 | | n/a | PDS3 (MPC 17257) |
| diameter | 530 | km | n/a | Thomas et al. (1997) |
| GM | 17.8 | km^3/s^2 | n/a | JPL IOM 312.F01-006 |
| rotation period | 5.342 | h | n/a | LCDB (Rev. 2016-July); Warner et al., 2009 |
| geometric albedo | 0.4228 | | 0.053 | IRAS-A-FPA-3-RDR-IMPS-V6.0 |
| B-V | .782 | mag | .020 | EAR-A-5-DDR-UBV-MEAN-VALUES-V1.1 |
| U-B | .492 | mag | .030 | EAR-A-5-DDR-UBV-MEAN-VALUES-V1.1 |
| Tholen spectral type | V | | n/a | EAR-A-5-DDR-TAXONOMY-V4.0 |
| SMASSII spectral type | V | | n/a | EAR-A-5-DDR-TAXONOMY-V4.0 |

Orbital Elements at Epoch 2457600.5 (2016-Jul-31.0) TDB
Reference: JPL 33 (heliocentric ecliptic J2000)

| Element | Value | Uncertainty (1-sigma) | Units |
|---|---|---|---|
| e | .089066737740174 62 | 1.0292e-08 | |
| a | 2.361348199280398 | 1.2317e-09 | au |
| q | 2.151030618501857 | 2.3887e-08 | au |
| i | 7.14040615716409 | 1.4093e-06 | deg |
| node | 103.8425122329109 | 2.7662e-06 | deg |
| peri | 151.1116347171285 | 7.5575e-06 | deg |
| M | 183.8768224741924 | 5.428e-06 | deg |
| $t_p$ | 2458248.914153744373 (2018-May-10.41415374) | 1.9724e-05 | JED |
| period | 1325.374085496326 | 1.037e-06 | d |
| | 3.63 | 2.839e-09 | yr |
| n | .2716214266896483 | 2.1251e-10 | deg/d |
| Q | 2.571665780058939 | 1.3414e-09 | au |

**Table 1 – General characteristics of Vesta:** Physical parameter table of Vesta (left), Orbital elements table of Vesta (right). *(JPL small-body database browser)*

Vesta is associated with a huge family of asteroids (families are group of asteroids that share the same orbital elements and are fragments of a larger asteroid, called parent body) and with a special class of meteorites, the HED (howardites, eucrites and diogenites). These meteorites, genetically related to each other, are derived from a differentiated body. The original connection was drawn from the similarity in spectral reflectance between Vesta and eucrites (*McCord et al. 1970*), a type of basaltic meteorites, that with diogenites and howardites compose the HED suite of meteorites. Many ground based spectroscopic studies have supported a link between Vesta and HED, showing the connection of spectra with the surface of Vesta, that appear to be dominated by eucrites, diogenites, and howardites (*Binzel et al. 1997; Gaffey 1997*).

Theories about the samples-transferring from Vesta to the Earth have been recently developed. In particular, it has been inferred that the Vestoids family (bodies dynamically linked to Vesta and similar in spectra to the HEDs) can act as a vector: the orbits of those Vestoids, extend between that of Vesta and both the 3:1 Jovian and *v6* resonances, which can act as escape slingshots from the main belt, allowing trajectories into the inner solar system. Several V-type asteroids – spectroscopically similar to Vesta – of ~1.0–3.4 km diameter, have been observed in near-Earth asteroids (*Cruikshank et al. 1991*), confirming these Vestoids as the most likely source of the HEDs.

Howardite-eucrite-diogenite (HED) meteorites provide the best samples available for a differentiated asteroid. Vesta, being associated with these meteorites, is the only "survived" differentiated asteroid, and thus is very special among the other main belt objects, dating back at the "Dawn" of the solar system formation. No other bodies of the solar system is as old as Vesta: it is the first small planet and it has the record of the processes acting in the early phases of the solar system formation.

This is the main reason for having a mission to explore Vesta: the Dawn mission.

**1.2 The Dawn mission**

The Dawn mission was selected as the 9th in the NASA Discovery Program, a long-term plan focused on the knowledge of the Solar System. It was launched in September 27, 2007, and orbited for 14 months around Vesta, since the July 16, 2011, then entered Ceres' orbit on March 6, 2015. The Dawn mission provides observations that advanced the knowledge of Vesta and gives a view into the early evolution of terrestrial planets.

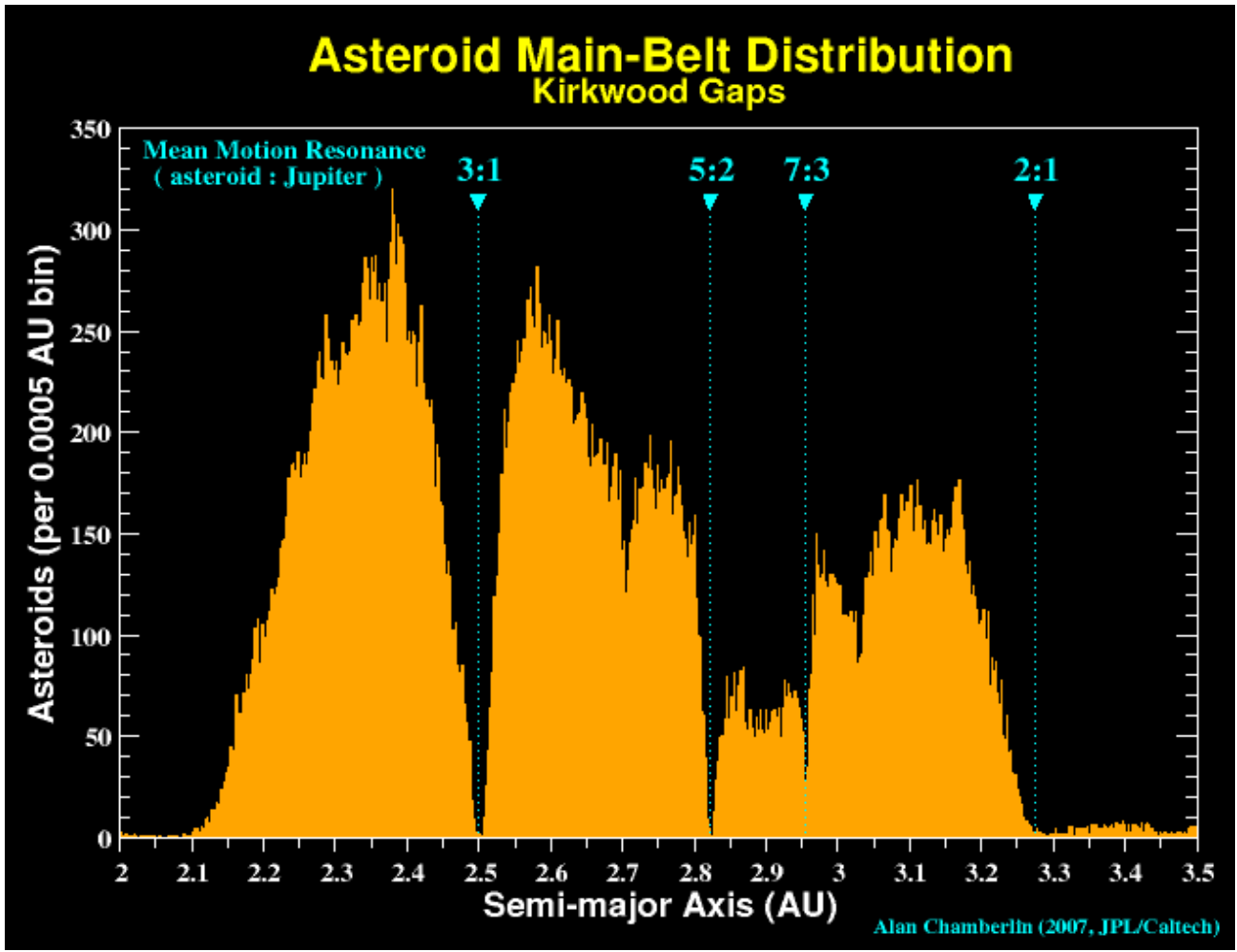

**Fig. 1 – Kirkwood gaps:** regions in resonance with Jupiter, representing a dip in the distribution of semi-major axis in the Main Asteroid Belt. Vesta is not far from the 3:1 resonance, populated only by the Alinda family, asteroids characterized by a large orbital eccentricity and close encounters with our planet. *(Chamberlin, 2007)*

The mission objective was to orbit two or more bodies in the Main Belt Asteroid and to characterize them through remote sensing. At the beginning, the project name was Main Belt Asteroid Rendezvous (MBAR), then became Dawn in the 2000 Discovery Step 1 proposal, to underline the view into the early Solar System that Dawn should have provided during its journey (*Thomas et al., 2011*).

Prior to this project, all our information about Vesta and Ceres relied on telescopic observations and inferred from the HED meteorites and theoretical models. Now detailed maps and spectra of the two main bodies of the Asteroid Main Belt are accessible, as well as information on the accretion that took place in the early Main Asteroid Belt and data on the differentiation processes. The mission is equipped with 3 instruments (a visual camera-FC, an imaging spectrometer-VIR and a gamma-ray and neutron detector-GRaND) and the radio system. Analysis of gravity (*Konopliv et al. 2011b*) and topography investigations (*Raymond et al. 2011*) give a clue about the internal structure and the nature of differentiation of planetary bodies; the gamma-ray and neutron detectors (GRaND) (*Prettyman et al. 2011*) supply elemental abundances, the Visual and InfraRed (VIR) (*De Sanctis et al. 2011*) imaging spectrometer provides mineralogical and chemical information, while the Framing Camera (FC) (*Sierks et al. 2011*) maps the surface morphology.

*1.2a Dawn: the spacecraft*

The mission was developed as an international collaboration of different parts. The NASA's Jet Propulsion Laboratory gave overall planning and management, project systems engineering, the flight system, the scientific payload development and the ion propulsion system, while the Orbital Sciences Corporation provided a mechanical design for the spacecraft (**Fig. 2**) based upon Orbital's STAR-2 series and furnished the overall assembly, including integration of the instruments, system-level tests, and launch operations (*Rayman et al., 2006*). Other counterparts supply the instruments onboard Dawn spacecraft: here it is a brief presentation of those instruments and a short description of their technical specifications.

The Max Planck Institute for Solar System Studies research and the German Aerospace Center (DLR) supplied the Framing Camera (FC). This camera covers the wavelenghts from the visible to the near-IR (430-980 nm), through an 8-position filter wheel which permits a selective imaging. It uses a 20 mm aperture, f/7.9 refractive optical system with a focal length of 150 mm. The field of view is 5.5° x 5.5°.

The Los Alamos National Laboratory furnished the Gamma Ray and Neutron Detector

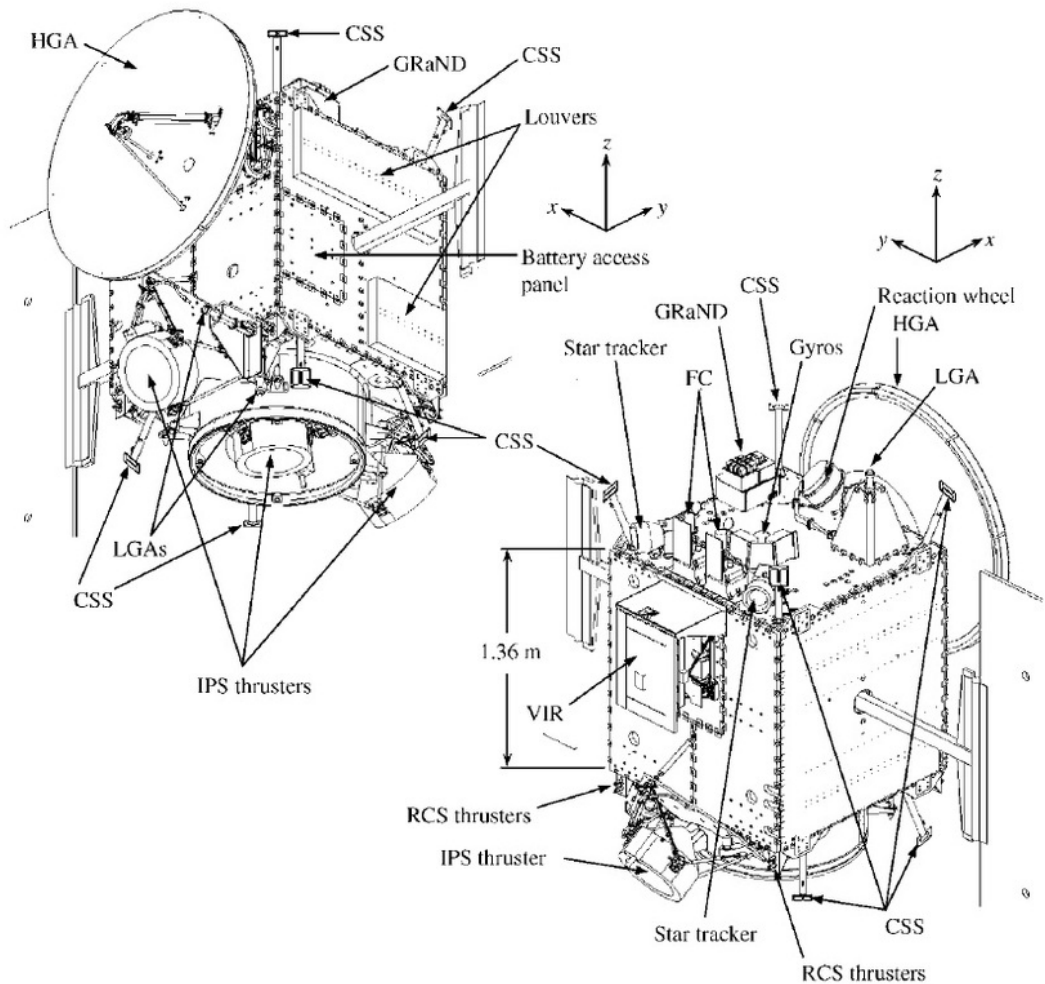

**Fig. 2 - The Dawn spacecraft:** It is made of aluminum and graphite composite, it has a dry mass of 747.1 kg and a mass of 1217.7 kg when fully fueled prior to launch. The spacecraft is a box-shaped design measuring 1.64m x 1.27m x 1.77m. The core of the spacecraft is a graphite composite cylinder, with xenon and titanium hydrazine tanks mounted inside. Access panels are constructed from an aluminum core with aluminum face sheets. *(De Sanctis et al., 2011)*

(GraND), which is used to illustrate the composition of the asteroids by studying energy and neutrons that emanate from them, in particular measuring the abundances in the top meter of the soil. It incorporates four principal channels and includes 21 sensors, which grant a wide field of view and are configured to permit separation of the signal from background noise. A photomultiplier tube measures the scintillation caused by the interaction of rays with a bismuth germanate (BGO) crystal. In addition, charge carriers created by these interacting rays, are detected through an array of semiconducting CdZnTe crystals. Epithermal and fast neutrons are sensed by their interactions with borated plastic scintillators.

The Italian National Institute for Astrophysics (INAF) and the Italian Space Agency (ASI) provided the Visible and Infrared Imaging Spectrometer (VIR). A more accurate description of the VIR Imaging Spectrometer, along with its scientific objectives, will be supplied in the next chapter.

**1.3 Vesta: what we know after Dawn results**

Here we will focus on the scientific knowledge of Vesta, the first target of Dawn mission. At first glance, the map of Vesta shows a strong hemispheres dichotomy in the asteroid's cratering record, reflected both in the composition and in the age: indeed, the southern hemisphere is much younger than the northern one, as results from crater age dates indicate.

The most prominent of the surface features are the two large impact structures in the southern hemisphere: the younger and larger Rheasilvia (around 500 Kilometres), and the older and more degraded Veneneia structure (around 400 Kilometres), partially overlaid by Rheasilvia itself.

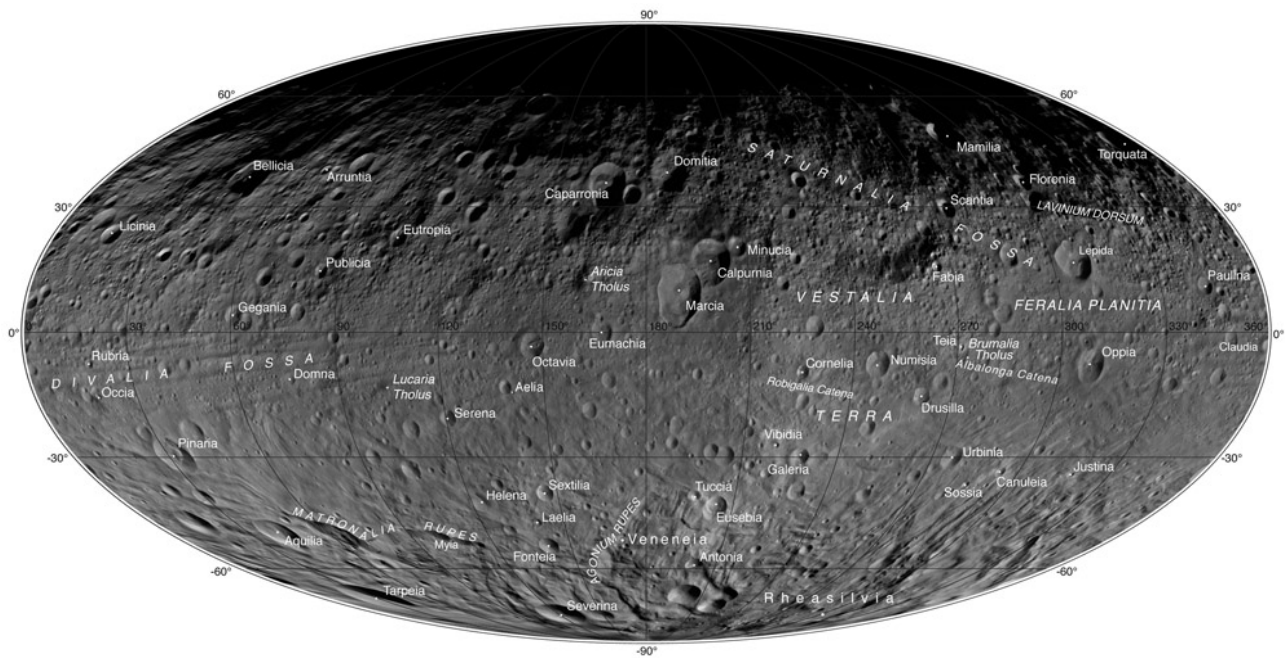

**Fig. 3 - Vesta's map as seen by DAWN -** the oldest, most heavily cratered surface at the north, Rheasilvia and Veneneia features in the south. *(http://www.clarksvilleonline.com/wp-content/uploads/2013/09/NASAs-Dawn-mission-photos-used-to-make-atlas-of-Asteroid-Vesta.jpg)*

Looking beyond the overall dichotomy, it is possible to divide the surface into three macroscopic types of terrain (**Fig. 3**): ancient and heavily-cratered regions (in the northern hemisphere), terrains full of ridges and troughs (on the equator and in the north) and younger terrains (in the Southern hemisphere) associated with the Rheasilvia and Veneneia craters (*Yingst et al., 2014*).

The overall-old and heavily-cratered regions include old and degraded craters, but also sharper and younger ones. Among the first, the most renown is the Feralia Planitia, an equatorial 270 Km-wide feature, though reshaped, even compressed in latitudes by the Rheasilvia impact. Among the second type – the sharper craters mentioned above – are counted the so-called "snowman craters", a group of three adjacent craters in Vesta's northern hemisphere named Calpurnia, Minurcia and Marcia, the last one being of particular importance for its pitted terrains and dark material and for the clear signature of OH, probably related to the delivery of hydrated material to Vesta, thanks to low-velocity carbonaceous impactors (*De Sanctis et al., 2012; McCord et al., 2012*).

The large sections of Vesta carved by concentric (on the equator) and linear (northern) troughs, are large-scale graben resulting from the two most important impacts in the history of Vesta. Bot the younger *"globe-girdling" (Yingst et al., 2014)* equatorial troughs and ridges and the linear system in the north, may be a sign of the differentiation of Vesta. Among the first type, the most prominent is Divalia Fossa, a 10-22 km wide and 465 km long chasm *(Yingst et al., 2014)*, straddled by Rubria and Occia craters and probably related to the Rheasilvia impact, while the largest of the second group, oriented NW-SE, is the Saturnalia Fossa, a 39.2 km wide and 458 km long chasm related to the Veneneia crater and probably partially-reactivated in the impact that originated the Rheasilvia formation *(Yingst et al., 2014)*.

Those two impacts produced a diogenite-rich southern hemisphere cleared of smaller craters (*De Sanctis et al. 2012; Schenk et al. 2012*), and in particular the younger Rheasilvia carved the overall shape of Vesta, generating a pronounced central mound by the rebound consequently to the impact.

*1.3a The interior*

Among all the asteroids in the Main Belt, Vesta may be a unique example of a differentiated body, "*a brick into the reconstruction of the processes that characterize terrestrial planet accretion*" *(Zuber et al. 2011)*.

Direct observations of some particular features – as the Rheasilvia mound terrain – may be a probe into Vesta's subsurface, while the large collection of meteorites samples named HED, most of which are thought to be related to Vesta itself, have provided – even before Dawn – a strong evidence for a planetary-scale differentiation, that could have produced a basaltic crust, a mantle and a core (**Fig. 4**). Observations of spectra (*McCord et al. 1970; Binzel and Xu 1993*) combined with the knowledge of mass (*Konopliv et al. 2011*) and size *(Thomas et al. 1997)* provides another window into Vesta's interior. Given the mass from perturbations by other asteroids and from perturbation of Mars' orbit by Vesta itself *(Michalak 2000; Baer and Chesley 2008; Kuzmanoski et al. 2010; Pitjeva 2005; Konopliv et al. 2006, 2011; Fienga et al. 2009)* and derived the best shape model from the Hubble Space Telescope (HST) *(Thomas et al. 1997)*, it was then possible to estimate the asteroid's bulk density.

In a layered model of Vesta, it shall be possible to distinguish between an upper and a lower crust, a mantle and a core. The theory of the interior and its developing into distinct layers is now confirmed with the new knowledge, thanks to Dawn mission. Indeed, it is possible to give a minimum-maximum range for the crust of Vesta, spanning from the 16 km-minimum tickness (*Ermakov et al., 2014*), obtained from the mean crustal thickness as a function of mantle and crustal densities, to the maximum tickness of 42 km (*Ruzicka et al.,*

*1997*), 41 km (*Mandler and Elkins-Tanton, 2013*), and 21 km (*Toplis et al., 2013*), estimated from geochemical models, the first two for both eucritic and diogenitic layer and the last one which refers only to the eucritic portion.

*1.3b Composition*

Pairing spectra with HED lithologies, it is shown that the dominant component on the surface is howardite-rich in eucrites. Pure diogenites and basaltic eucrites are less represented, maybe because they would appear in regions much smaller than the spatial resolution of VIR (*De Sanctis et al., 2013*). In particular diogenites are mainly observed in small areas into the deep of Rheasilvia basin (*Ammannito et al., 2013; McSween et al., 2013*).

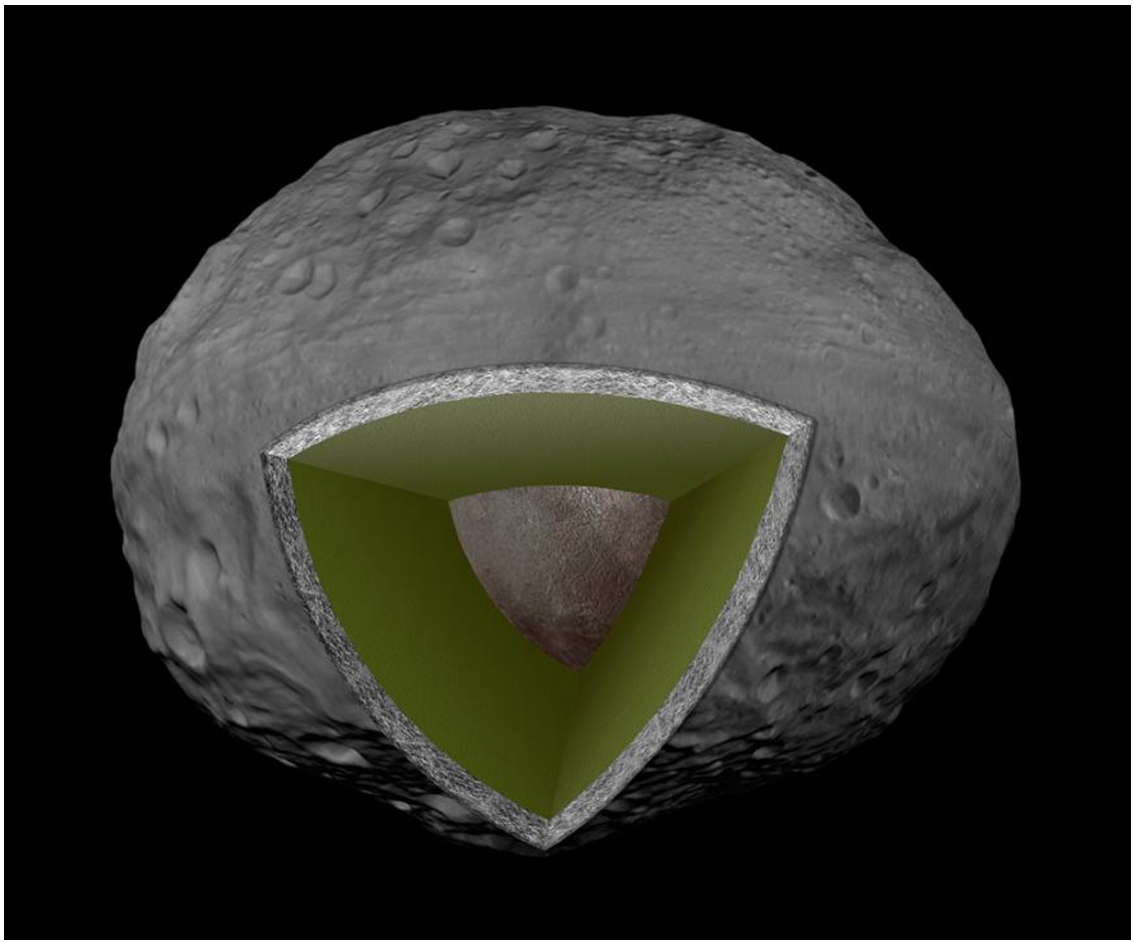

**Fig. 4 - Internal structure of Vesta:** rough representation of the layers structure of Vesta's interior. Iron sank into deep, forming a core (brown); a mantle covered the core of Vesta (green) and then formed a crust (grey), reshaped by impacts and lava-flowing from the mantle. (http://www.nasa.gov/mission_pages/dawn/multimedia/pia15510.html)

Under the howardites-rich in eucrite layer, there should be basaltic eucrites and – at a greater depth – plutons of cumulate eucrites and diogenites *(Zuber et al. 2011)*. The diogenites, in particular, should constitute the deepest component of the layered crust model *(Takeda, 1997)*.

This general overview of the surface of Vesta and its interior also gives the possibility to infer the evolution of the asteroid and imposes time constraints on some focal moments of the solar system in its early phases. Indeed, theories have been postulated about the formation of the different lithologies of Vesta.

The HED oxygen isotopic mixture, in addition to the composition of eucrites point towards a large-scale melting, thus suggesting a primordial magma ocean prior to cooling and differentiation. This large-scale melting has been mostly attributed to the presence of short-lived radioactive $^{26}$Al, probably injected by a supernova near the time of solar system formation, and the importance of this source has been supported by thermal models *(Ghosh and McSween 1998; McSween et al. 2003)*. After $^{26}$Al enriched the liquid phase and accumulated in the sub-surface, a magma ocean with a thickness of 1 to a few tens of km formed and a basaltic crust, whose thickness depended on the viscosity of silicate melt (*Neumann et al. 2013*). In this scenario, the formation of basaltic achondrites is due to residual liquids remaining at the end of the cooling phase of the inferred magma ocean *(Ikeda and Takeda 1985; Warren 1985; Longhi and Pan 1988)*.

Other theories requires partial melting of chondrites *(Consolmagno and Drake 1977; Stolper*

*1977)* and multiple heating events, which are challenging to confirm.

It has been studied the link between the evolution of the internal structure and the thermal heating of Vesta, considering different hypotheses, which constrains the time of formation of the asteroid. For a complete silicates melting, the asteroid should have formed before than 1.0 Ma after the injection of $^{26}$Al in the solar nebula, while, for a partial melt about 25%-50%, the formation of Vesta should have occurred up to 1.4 Ma after $^{26}$Al injection in the newborn Solar System *(Formisano et al. 2012)*.

# Chapter 2

# VIR and the spectroscopy

The Visible and InfraRed Spectrometer (VIR) is the Imaging Spectrometer designed for the Dawn mission, which provides the spectral images analyzed in details in **chapters 4** and **5**. In this section, a brief introduction to spectroscopy is given, along with a description of VIR design, its scientific objectives and a presentation of the instrument output.

**2.1 The spectroscopy**

Spectroscopy is the study of the interaction between matter and light as a function of wavelength (see **Fig. 5** for an example). The information carried by photons can be the product of three processes, scattering, emission and absorption, that occur during the interaction with matter. A spectrometer can measure these information; in this context it is useful to introduce the *reflectance* of the surface of a material, which gives its effectiveness in reflecting radiant energy. The reflectance will be widely used in the following chapters and a more proper definition will be given in **chapter 3**.

Here we follow the approach of *Clark, 1999*, which describes the capability of a spectrometer through four parameters.

*Spectral range.* The spectral ranges (wavelengths-range covered by the instruments) are commonly divided in: (a) ultraviolet (UV): from 0.001 to 0.4 µm, (b) visible: from 0.4 to 0.7 µm, (c) near-infrared (NIR): from 0.7 to 3.0 µm, (d) mid-infrared (MIR): from 3.0 to 30 µm,

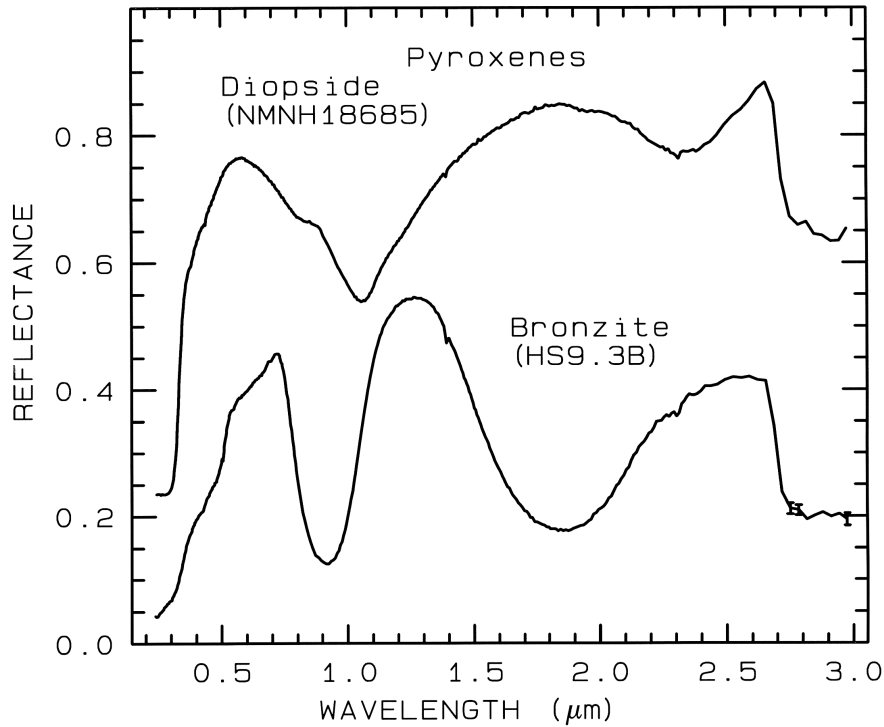

**Fig. 5 – Examples of spectra -** Reflectance spectra of two pyroxenes showing the change in Fe2+- absorption band position and shape with composition (Clark et al., 1993b). The reflectance spectrum curve is the plot of the reflectance as a function of wavelength.

and (e) far infrared (FIR): from 30 µm to 1 mm.

*Spectral bandwidth.* It is the width of an individual spectral channel. The narrower the spectral bandwidth, the narrower the absorption feature that the spectrometer is able to measure, if enough adjacent spectral samples are obtained. The shape of the bandpass profile is important: ideally, each spectrometer channel should reject all light except that from within a given narrow wavelength range, but in reality light may enter from out of the bandpass (e.g. inadequate blocking filters). The most common bandpass in spectrometers is a Gaussian profile, typically defined through the so called Full Width at Half Maximum (FWHM) (see **Fig. 6**).

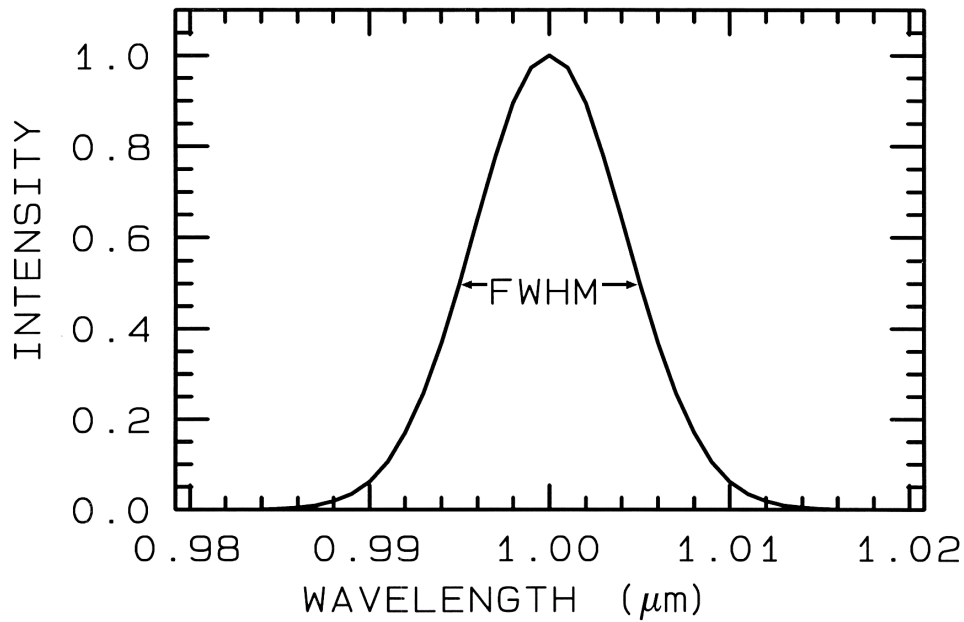

**Fig. 6 - A Gaussian profile** – A profile with a Full Width at Half Maximum (FWHM) of 10 nm is shown. *(Clark, 1999)*

*Spectral sampling.* It is the distance in wavelength between the spectral bandpass profiles for each channel in the spectrometer as a function of wavelength. It is related to the spectral resolution of a spectrometer.

*Signal-to-noise ratio.* The S/N quantifies how much is intelligible the signal respect to the background noise. The S/N required to solve a particular problem depends on the strength of the spectral features under study. It is dependent on the detector sensitivity, the spectral bandwidth, and intensity of the light reflected or emitted from the surface being measured.

2.1a Imaging Spectroscopy

Imaging Spectroscopy is the recording of a set of two-dimensional images of a certain object at different wavelengths within a given spectral range. It can be seen as the two-dimensional extension of the punctual spectroscopy. It can obtain a spectrum in each position of a large array of spatial positions, so that any spectral wavelength can be used to make a recognizable image. Through the analysis of the spectral features it is possible to map where those chemical bonds occur, and thus map materials.

Spectroscopy of natural surfaces is sensitive to specific chemical bonds in materials, whether solid – both crystalline and amorphous – liquid or gas. The main advantage is that it can be used both in the laboratory or from remote (e.g. other planets or asteroids), though it is very sensitive to small changes in the chemistry and structure of the material and sometimes it gives quite complex spectral signatures.

**2.2 The scientific goals of VIR**

One of the main goals of Dawn is to determine the mineral composition of the surface and to place it in geologic context. Visual and infrared spectroscopy allows determination of mineral composition for observed bodies and the imaging capability allows the determination of their spatial distribution. Coupling the knowledge of the chemical composition with the spatial distribution of different compounds, it is possible to infer the surface processes: mineralogical types are identified and compositional maps are drawn according to the spatial distribution of different materials. This allows to investigate how the observed surface evolved, and also the physical processes – like space weathering (*Blewett et al., 2015*)– affecting it.

The primary objectives of VIR mapping spectrometer are:

1. Identify different materials and mixtures and determine their spatial distribution;

2. Identify silicates, hydrates, ices and other minerals

3. Determine the overall continuum slopes of the spectra;

4. Mapping the presence and extent of space weather.

5. Map, with a spatial resolution of a few tens of meters, the asteroids and determine the spatial distribution of the various mineralogical types and their mixtures using both the spectral features and the overall brightness;

6. Determine the physical microstructure and nature of the surface particles by measuring the spectrophotometric phase curve.

In particular at Vesta, VIR has had these specific objectives:

7. Confirm whether or not a link exist between Vesta and the HED meteorites;

8. Obtain the first in-depth view of a planetary interior through the spectral imaging of Vesta's wide and deep impact basin;

9. Confirm the possible presence of hydrated minerals.

10. Reveal the nature of Vesta's ancient magma ocean or volcanic emplacement history;

11. Establish the abundances and mineralogy present on Vesta surface sufficiently to establish the source of meteorites recovered on Earth;

To fulfill this task, VIR analyzes asteroids' surface through both high spatial resolution imaging and high spectral resolution spectroscopy.

The VIR mapping spectrometer accomplished several scientific of the above listed objectives. However, some of the objectives were not completely fulfilled due to the lacking of a precise photometric correction of the dataset. In particular, the n.6 (*"Determine the physical microstructure and nature of the surface particles by measuring the spectrophotometric phase curve"*) was only partially achieved at Vesta.

For this reason, we implemented on the Vesta VIR dataset the photometric correction described in **Chapter 4** and **5** of this thesis work.

## 2.3 The VIR instrument

The study of the reflectance of a planet or an asteroid, allows to infer physical and chemical properties of the surface (*es. Rosetta-67P/Churyumov-Gerasimenko, Capaccioni et al., 2015*). Information of radiance is obtained by means of spectrometers, schematically composed by an optical system which collimates and disperses the spectrum and a detecting device.

An optical instrument using a single waveband is called "panchromatic" (*Willoughby et al., 1996*), while the one using multiple filters is called "multispectral imager"; instruments using tens or hundreds of spectral bands are called "hyperspectral imager". Hence, while multispectral imaging operates over several discrete spectral bands, an hyperspectral imaging system can perform imagery over hundreds of spectral bands, typically with bandwidths of few nanometers.

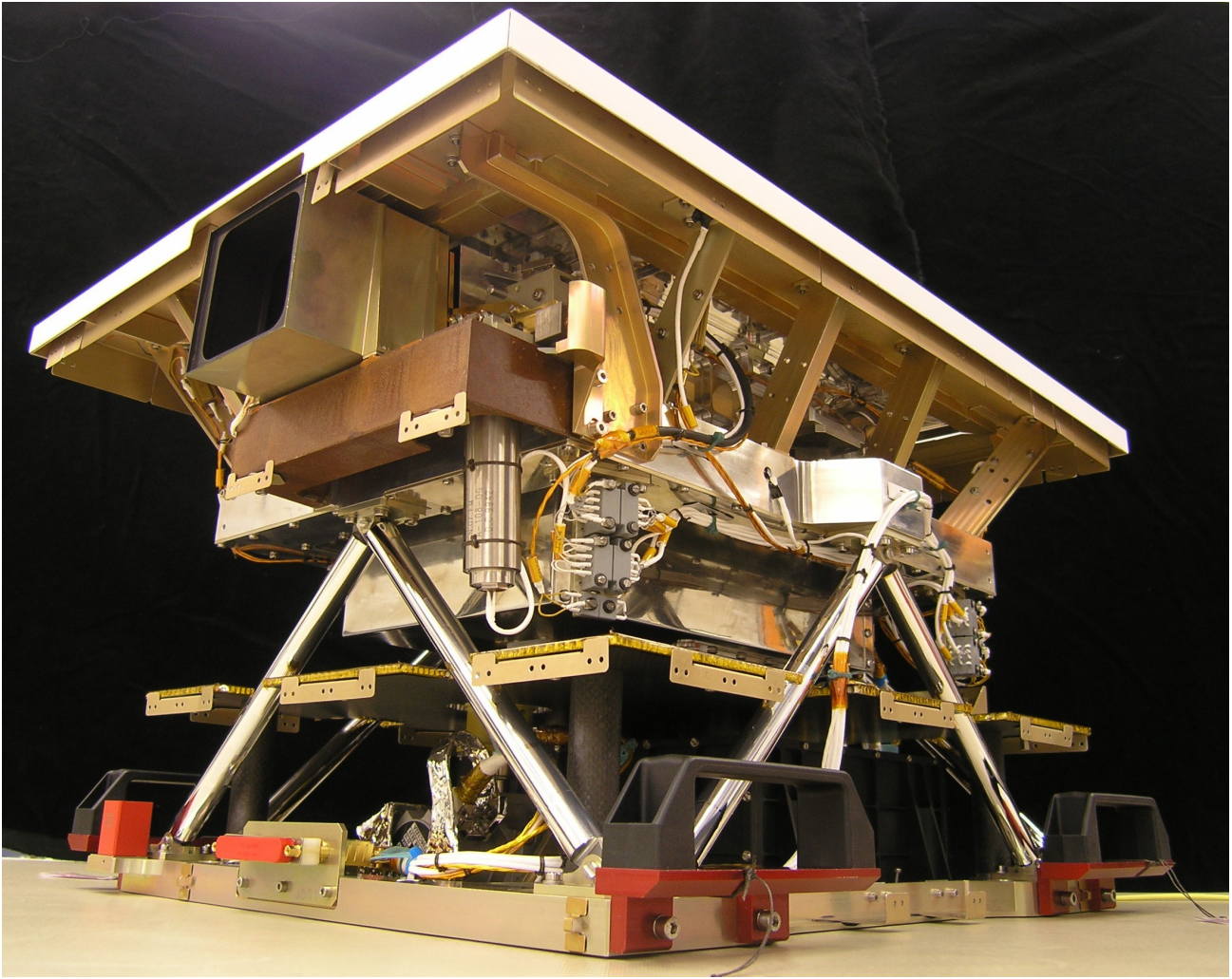

**Fig. 7 - VIR -** The Dawn mapping spectrometer VIR, a Visual and Infrared Imaging Spectrometer in a single optical head *(http://www.iaps.inaf.it/solarsystem/wp-content/gallery/dawn/vir2.jpg)*.

The VIR experiment (**Fig. 7**) is an hyperspectral Imaging Spectrometer (*De Sanctis et al., 2011*) adopting the design of similar instruments flying on Rosetta (*Coradini et al., 1998*) and Venus Express (*Piccioni et al., 2002*), with an optical layout based on the Cassini VIMS-V spectrometer (*Brown et al., 2004*). It is constituted by a Shafer telescope (*Shafer, 1982*) mated to an Offner slit spectrometer, with high spectral ($\Delta\lambda_{VIS}$ = 1.8nm/band, $\Delta\lambda_{IR}$ = 9.8nm/band) and spatial (IFOV = 250 µrad/pixel, FOV = 64 x 64 mrad) resolution.

| Table 2 - VIR components |||
|---|---|---|
| VIR OPTICAL HEAD | PROXIMITY ELECTRONICS | MAIN ELECTRONICS |
| Telescope (Shafer) optical components | Box structure | Digital processing units (DPUs) |
| Spectrometer (Offner) optical components | Mother board and connectors | S/C interface control units |
| Internal calibration system | CCD boards | Power supply for all the sub-units |
| Slit and shutter mechanism | IR board | Interface units |
| Visible detector assembly | Scan mirror and cover board | Coolers electronics |
| Infrared detector assembly | | |
| External and internal baffles | | |
| Cover unit | | |
| Radiators | | |

Spectral images are formed on two detectors to cover the 0.25-1.05 µm (VIS channel) and 1-5 µm ranges (IR channel). A diffraction grating disperses the incident beam, while the cooling system allows optimal performances; a calibration unit and two electronics modules calibrate the experiment and provide its general management.

**Tab. 2** is a list of VIR components, while **Tab. 3** summarizes in details VIR main characteristics.

Both telescope and spectrometer use spherical mirrors: no collimators, beam splitters and other intermediate objects are used and the optical system dimensions are reduced. Meanwhile, the presence of symmetric elements compensates coma and distortion and allows to balance aberrations of the telescope with the ones due to the spectrometer. The optical path of a light ray through the instrument optics, from the entrance slit to the exit, is described in **Fig. 8** (*De Sanctis et al., 2010*)**.** The input beam enters the system – and more precisely the telescope – through a slightly elliptical entrance pupil placed along the *Y* axis: the ray is parallel to the $Y'$ axis. It is reflected by the concave primary mirror M1, which is placed on a scan mechanism that allows it to move along the slit direction, with a maximum number of 256 spatial steps allowed, generating a FOV of 3.6°.

| | Table 3 - VIR main characteristics *(De Sanctis et al., 2010)* | | |
|---|---|---|---|
| | | VIR VIS | VIR IR |
| Spectral performances | Spectral range | 0.25-1.05 μm | 1.0-5.0 μm |
| | Spectral sampling | 1.8 nm/band | 9.8 nm/band |
| | Spectral resolution $\lambda/\Delta\lambda$ | 100-380 | 70-360 |
| Spatial performances | IFOV | 250 x 250 μrad | |
| | FOV | 64 x 64 mrad | |
| Telescope | Configuration | Shafer with Offner relay | |
| | Pupil area | 5.77 cm$^2$ | 17.71 cm$^2$ |
| | Pupil diameter | 27.1 mm | 47.5 mm |
| | Focal length | 152 mm | |
| | f/# | 5.6 | 3.2 |
| | AΩ | 3.6 x 10$^{-11}$ m$^2$ sr | 7.5 x 10$^{-11}$ m$^2$ sr |
| Spectrometer | Configuration | Offner | |
| | Slit dimensions | 38 μm x 9.53 mm | |
| | Grating groove densities | 267.9 mm$^{-1}$ | 53.8 mm$^{-1}$ |
| | Dispersion on focal plane | 52.6 nm/mm | 263.1 nm/mm |
| Focal planes | Type | CCD Thomson TH7896 | HgCdTe Raytheon |
| | Active size | 1024 x 508 pixels | 435 x 270 pixels |
| | Frame size | 432 x 256 | |
| | Elemental pixel size | 19 x 19 μm | 38 x 38 μm |
| | Acquired pixel size | 38 x 38 μm | |
| | Full well capacity | 10$^6$ e$^-$ | 2 x 10$^6$ e$^-$ |
| | Readout noise | 2 ADU | |
| | Dark current | <1 e$^-$/s | < 10 fA (@70K) |
| | A/D converter | 16 bit | |
| | Repetition time | 5-20-60 s | |
| Operative temperature | Focal plane | 150-190 K | 65-90 K |
| | Telescope and spectrometer | 100-150 K | |
| Mass | Optical Head and ME | 20 Kg | |
| Power | Average power | 52 W | |

After running into M1, the beam is then reflected by the small M2 flat folding mirror, the only flat surface of the system: this mirror, surrounded by several baffles, blocks parasitic light and reduces the internal straylight. The beam moves to the convex secondary mirror M3 and then is brought to the Offner slit spectrometer through the M4 and M5 Offner relay primary and secondary mirrors. After a new reflection operated by the spherical mirror M4, the beam is finally focused on the slit entrance of the spectrometer: here, the ray is reflected a first time by the Offner mirror M6 and is dispersed by the diffraction convex grating, which works as a Lyot stop – or a cold stop – useful in reducing the thermal background. Finally, after being diffracted, the beam meets for the second time the M6 mirror and reaches the two focal planes.

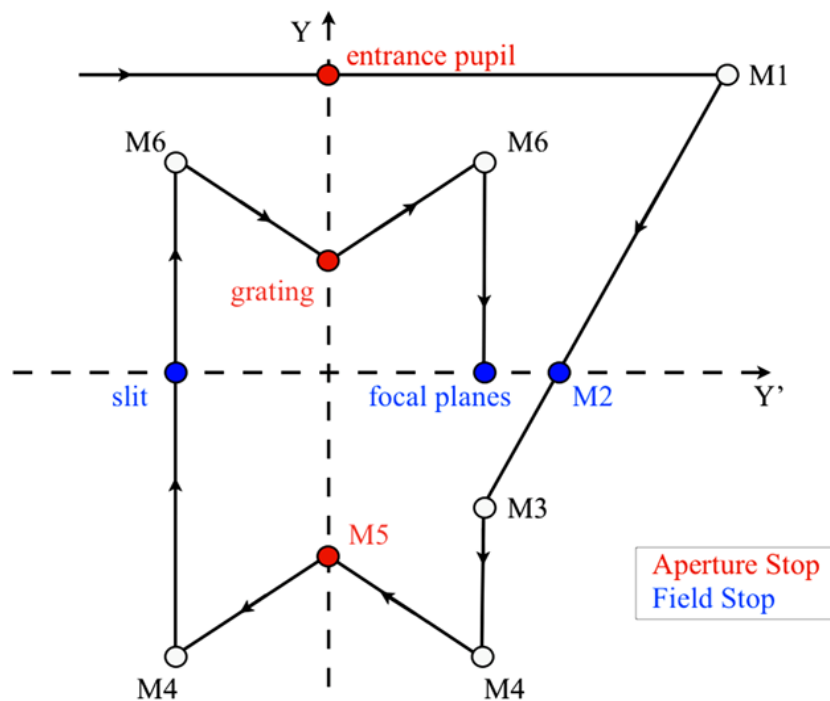

**Fig. 8 - VIR schematic -** a schematic representation of VIR, a Shafer Telescope coupled with an Offner Spectrometer. Red points correspond to the aperture stops of the system, the blue ones to the field stops. (*De Sanctis et al., 2010*)

*2.3a The Offner spectrometer*

The Offner spectrometer is the one in use for the VIR system. It offers some advantages over other spectrometers, as low chromatic aberrations, a compact size with reduced optical distortion and a high speed of acquisition. It is composed by two concave mirrors and one convex grating, acting as a set of equally spaced slits: each slit can be considered as a source of secondary waves.

In the classical approach the relation between the groove spacing and the angles of incident and diffracted light beams is given by the Bragg relation:

$$\sin(\theta) + \sin(\theta') = m\lambda p \tag{1}$$

being $\theta$ and $\theta'$ the angle of incident and diffracted beams, p the number of grooves per unit of length, m the order and $\lambda$ the wavelength.

*2.3b The diffraction grating*

The diffraction grating, shown in **Fig. 9,** is realized on a NG5 glass element, with grooves optimized to efficiently disperse the incident light across a wide spectral range (0.25–5 μm). This grating is divided in two concentric regions, known as Rowland's circles, corresponding to different groove densities: the central zone – with a higher groove density – corresponds to 33% of the surface and is devoted to disperse light in the VIS range (0.25–1.0 μm), while the remaining 67% - with a lower groove density – is used for the IR range (1–5 μm). The larger area devoted to IR, compensates for the lower solar irradiance in this spectral region (*De Sanctis et al., 2010*).

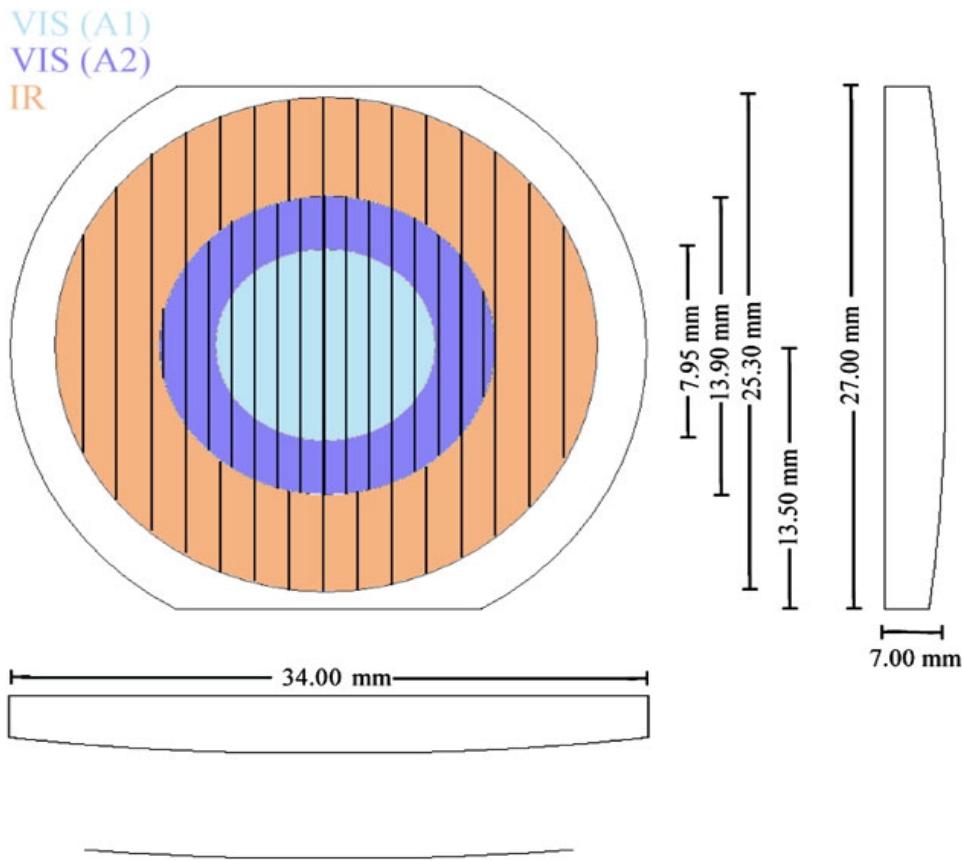

**FIG. 9 – Diffraction grating -** In cyan and blue the zones devoted to diffract the VISIBLE, in red the one used for the IR region. *(De Sanctis et al., 2010)*

*2.3c The focal planes*

The focal planes are a Thomson-CSF type TH 7896 CCD for the VIS channel and a HgCdTe bidimensional array for the IR channel to cover the whole operational range. The first, is a 1024×1024 pixels CCD, with the active region of the sensor being 1024×508 pixels wide. The combination of the elemental pixel size of 19 × 19 μm, with a fixed binning of 2×2, allows to double the acquired pitch to 38×38 μm and to match the VIS with the IR channel, giving the same FOV on the two channels. The latter, is a 270 × 435 pixels Mercury Cadmium Telluride array with a substrate layer having a cut-on at 0.895 μm and a cut-off at 5.03 μm, and is constituted by sensitive crystals grown with a Molecular Epitaxy

Process (MEP) to optimize the electrical uniformity of the detector. A charge trans-impedance amplifier (CTIA) grants high sensitivity at low photonic fluxes. The IR detector operates in good conditions at temperatures of the order of 75 K, establishing the necessity for an active cooling system (*De Sanctis et al., 2010*).

*2.3d The cooling system*

Thermal control on VIR operates through both a passive approach and an active one. An external radiator, located at the top of the optical head, grants a passive cooling of the instrument walls down to 130–140 K. In order to cool the system at lower temperatures – 70-80K – a cryocooler has been placed below the optical bench, to dissipate the heat by active cooling through the interface with the spacecraft (*De Sanctis et al., 2010*).

*2.3e The calibration unit*

The internal calibration unit is based on two paraboloid lamps, one for the VIS, the other for the IR. These two lamps, positioned in the telescope, very close to M1 and slightly angled with respect to the optical axis, provide a homogeneous spot of light across the whole FOV. The paraboloids concentrator are made of titanium, with a filament in tungsten 62 mm long and 30 μm in diameter. A metallic cover is able to close the optical entrance of the system when it needed and is used to perform the internal calibration. The internal calibration gives information on the spectral calibration of the system (*De Sanctis et al., 2010*).

*2.3f Electronics modules*

Two electronics modules manage the experiment (*De Sanctis et al., 2010*):

(1) the Proximity Electronics Module (PEM) reads the signals coming from the two different focal planes, generates the data, commands the scan unit mechanism and the metallic cover's actuator;

(2) the Main Electronics (ME) provides the power supply, manages the distribution unit and includes the Data Handling Support Unit, a Mass Memory (MM) – for data storage – and several cards devoted to the management of particular subsystems.

**2.4 The VIR data cubes**

During all the observational phases of the DAWN mission, VIR worked on the sunlight reflected from the two asteroids of interest and provided data stored in data cubes (**Fig. 10**). A VIR data cube is a 3-D array composed of a set of two-dimensional images of the target corresponding to each wavelength and then stacked together. Thus, the imaging spectrometer provides both spectral and spatial information, with each pixel (a spatial location corresponding to a specific sample and line) associated to a continuous sampled spectrum (*M.C.De Sanctis et al., 2010; M.T.Capria, 2014*).

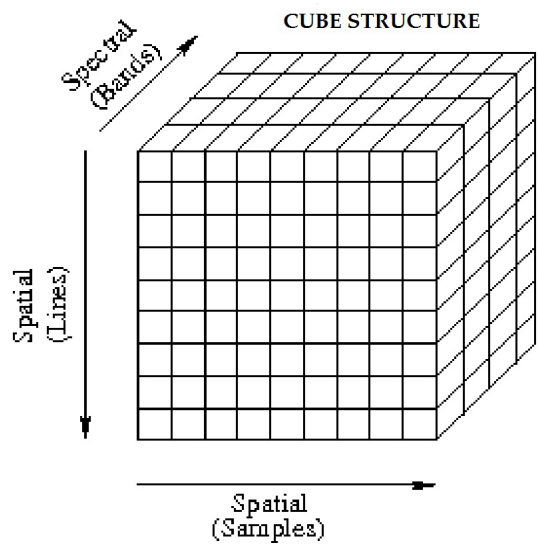

**Fig. 10 – Generic data-cube structure -** two spatial dimensions (samples x lines) and a spectral one (bands). (Adapted from *M.C.De Sanctis et al., 2010)*

Some systems acquire simultaneously the two spatial dimensions, while temporally sampling the spectral one (Whiskbroom scanning), others get only one spatial dimension and the spectral one, while temporally sampling the second spatial dimension (Pushbroom scanning).

The second technique – adopted in VIR – needs a field-limiting entrance slit and a dispersive element to allow the 2D detector to sample the spectral dimension and one spatial simultaneously (*Aikio, 2001*). The second spatial dimension, is typically generated by scanning or moving the instrument's field of view, thus building up a cube with line-per-line steps (**Fig. 11**).

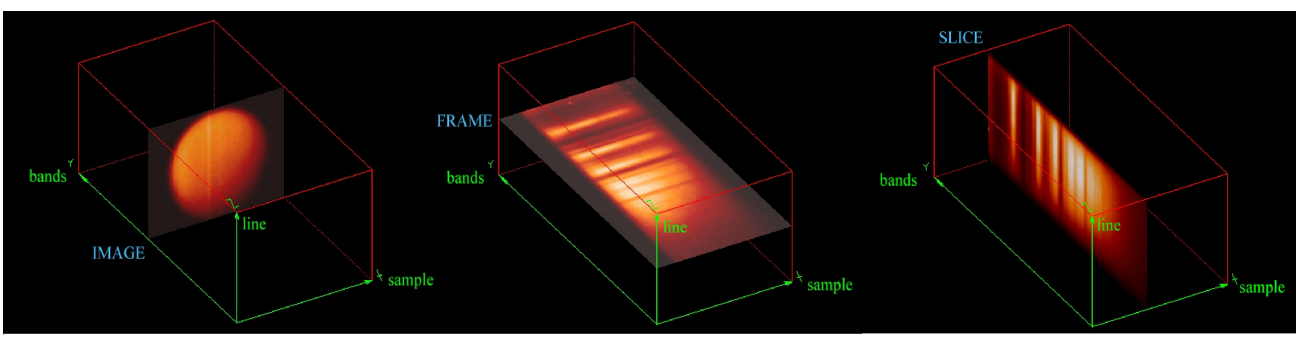

**Fig. 11 – Bi-dimensional data structure -** an "image" is defined by selecting a band, a "slice" is defined by selecting a sample, a "frame" is defined by selecting a line (fixed scan mirror). *(Adapted from Filacchione, 2006)*

These data-cubes are organized in archives composed by three different types of file:

- a .QUB file, containing the three-dimensional array which constitutes the core of the data;

- a .HDR file, which includes the instrument telemetry and other information as the physical storage order and the byte order. This information is needed to access the core;

- a .LBL file, containing general information, as start and stop time of acquisition and the celestial geometry.

# Chapter 3

# Photometric modeling and photometric correction

The amount of light scattered by the surface of a planetary body is a function of its physical properties and of the illumination and observing geometry. Its investigation is the task of planetary photometry. The comparison of the spectrophotometric output of an atmosphereless body with theoretical models – empirical, semi-empirical or physically based – that describe the light-matter interaction, allows one to characterize the surface composition and its properties like large scale roughness, regolith porosity and grain size. Along with this, the definition of a photometric model for a given target permits to perform a photometric correction of the measured signal decoupling the intrinsic albedo variability of the surface from the one induced by the observation geometry.

In this chapter we briefly discuss the main goals of photometric modeling and photometric correction; then we introduce the problem of the radiative transfer in particulate media and present a summary of different photometric models describing the spectrophotometric output of planetary surfaces, focusing on the Hapke's model, which has been adopted in this work.

**3.1 The importance of photometric modeling and photometric correction**

With the term "photometric correction" of remote sensing data we indicate a method to report observations of a given target to a common reference geometry. The removal of the effect of the observation geometry on the measured reflectance allows one to characterize the intrinsic variability of the surface albedo giving clues about the physical properties and mineralogical compositions of the surface. However, in order to perform a photometric correction, a "photometric model" able to reproduce the observed dependence of the reflectance on observation geometry is needed. Following *Li et al., (2015)*, the derivation of a photometric model and the application of a photometric correction are aimed to three main objectives:

- derivation of the optical and mechanical properties of a surface (particle size, porosity, roughness, complex refractive index, etc) by measuring the reflectance under various light scattering geometries.

- comparisons of observations taken under different geometries. This topic can be divided in three main points: (1) comparisons between different asteroid or planetary surfaces, observed under different geometries. For example, phenomena like the phase reddening (*Gehrels et al., 1964*), which causes the spectrum of a surface to have redder slopes and stronger spectral absorptions at higher phase angles (the angle between the incident light and the observing direction as seen from the target) can be corrected; (2) comparisons between different areas of the body observed under different illumination conditions: indeed, albedo and color maps of the whole surface at certain illumination geometries (*Li et al., 2013*) require photometric modeling; (3) comparisons with laboratory measurements.

- prediction of the reflectance at arbitrary illumination and viewing geometries for observations planning. This practical application is important for essentially all spacecraft missions with imaging or spectroscopic instruments.

According to the objectives presented above, we worked for the derivation of a photometric model of Vesta and to the photometric correction of VIR observations of the asteroid (**Chapter 4**), which is the main goal of this master degree thesis work.

### 3.2 Radiative transfer in a particulate medium

The propagation of an electromagnetic wave through a complex medium can be expressed by the equation of radiative transfer, which links together the change in radiant power with the processes that occur along its path. Below, the derivation of the radiative transfer equation (*Hapke, 1993*) is given.

Define a radiance field $I(s,\Omega)$, which is the intensity of an electromagnetic wave propagating into a direction $\Omega$ at a point s, that had at least one interaction with the particulate medium. In this notation the dependence from the wavelength it is not explicitly reported, while it is adopted the assumption that the medium inhomogeneities emit independently.

Take an infinitesimal cylinder of area *dA* and height *ds* (volume *dsdA*) (see **Fig. 12**), the radiant power at *s* passing through the base in a solid angle d$\Omega$ is

$$I(s,\Omega)dAd\Omega \;, \tag{2}$$

while that emerging from the top is the same entering the cylinder plus a differential increment:

$$I(s+ds,\Omega)dAd\Omega = \left[I(s,\Omega) + \frac{\delta I(s,\Omega)}{\delta s}ds\right]dAd\Omega \; ; \qquad (3)$$

then, subtracting the former from the latter, it is obtained

$$I(s+ds,\Omega)dAd\Omega - I(s,\Omega)dAd\Omega = \frac{(\delta I(s,\Omega))}{(\delta s)}dsdAd\Omega \; . \qquad (4)$$

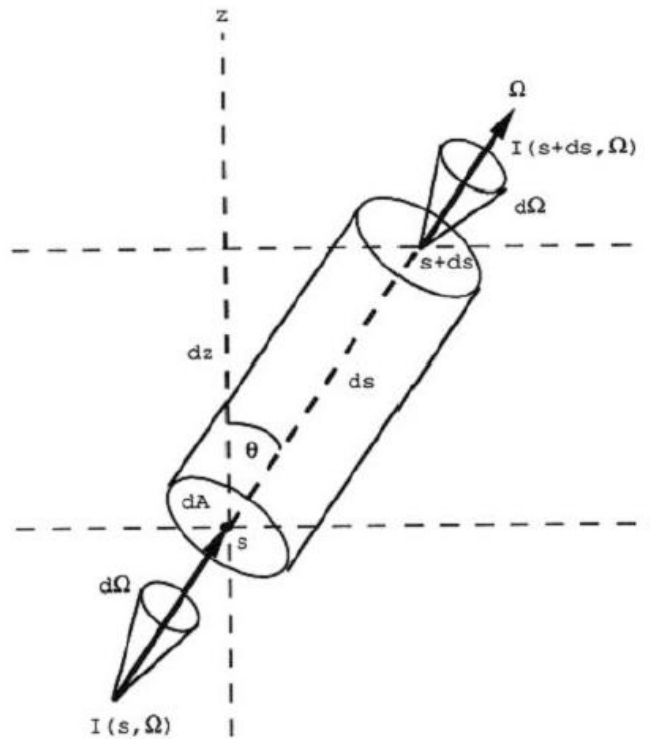

**Fig. 12 - Infinitesimal cylinder -** Schematic diagram of a radiation field passing through an infinitesimal volume. *(Hapke, 1993)*

This change in radiant power along the path is due to various processes, such as *absorption*, *scattering* and *emission*. The first two, can be analyzed in a unique process named *extinction*, characterized by the *medium extinction coefficient E(s, Ω)*, such as

$$\Delta P_E = -E(s,\Omega) I(s,\Omega) ds dA d\Omega, \qquad (5)$$

where the minus sign shows that the process decreases the power of the beam.

The term related to scattering can be expressed as:

$$\Delta P_S = \frac{1}{4\pi} G(s,\Omega',\Omega) I(s,\Omega') ds dA d\Omega' d\Omega; \qquad (6)$$

it can decrease or increase the power associated to the beam, accounting for both light scattered away from the propagation direction $\Omega$ and light scattered towards the line of propagation from different directions $\Omega'$; $G(s,\Omega',\Omega)$ is the *volume angular scattering coefficient*.

An integration over the whole solid angle $4\pi$, gives the *volume scattering coefficient*:

$$S(s,\Omega) = \frac{1}{4\pi} \int_{4\pi} G(s,\Omega',\Omega) I(s,\Omega') d\Omega', \qquad (7)$$

which gives a direct indication for the entire power scattered in $\Omega$.

The term related to emission is

$$\Delta P_F = F(s,\Omega) ds dA d\Omega, \qquad (8)$$

where *F(s, Ω)* is the *volume emission coefficient*, which is the power emitted per unit volume by the element at position *s* into unit solid angle Ω.

Then, equating the difference in radiant power along the cylinder to the sum $\Delta P_E + \Delta P_S + \Delta P_F$ of the three contributions, and dividing by *ds dA dΩ*, the equation of radiative transfer is obtained:

$$\frac{\delta I(s,\Omega)}{(\delta s)} = -E(s,\Omega)I(s,\Omega) + \frac{1}{4\pi}\int_{4\pi} G(s,\Omega',\Omega)I(s,\Omega')d\Omega' + F(s,\Omega) . \tag{9}$$

In the case of a horizontally stratified medium, the equation can be expressed as a function of the depth z, substituting $ds = \frac{dz}{\cos\theta}$. For our purpose, the equation is then divided by E(z), obtaining:

$$\frac{\cos\theta}{E(z)}\frac{\delta I(z,\Omega)}{(\delta z)} = -I(z,\Omega) + \frac{S(z)}{E(z)}\frac{1}{4\pi}\int_{4\pi}\frac{G(z,\Omega',\Omega)}{S(z)}I(z,\Omega')d\Omega' + \frac{F(z,\Omega)}{E(z)} , \tag{10}$$

that can be written also as:

$$-\cos\theta\frac{\delta I(\tau,\Omega)}{(\delta\tau)} = -I(\tau,\Omega) + \frac{w(\tau)}{4\pi}\int_{4\pi} p(\tau,\Omega',\Omega)I(\tau,\Omega')d\Omega' + J\frac{w(\tau)}{4\pi}p(\tau,\Omega_0,\Omega)e^{-\tau/\cos i} + f_T(\tau,\Omega)$$
(11)

representing the final form of the radiative transfer equation.

J is the incidence irradiance (or flux), and τ is the *optical depth*, a dimensionless number

expressed as:

$$\tau = \int_s^\infty E(s')ds' \cos\theta = \int_z^\infty E(z')dz' , \qquad (12)$$

or, in differential form:

$$d\tau = -E(s)ds \cos\theta = -E(z)dz . \qquad (13)$$

The larger the optical depth, the larger the amount of light absorbed by the medium. For example, it can be shown that light emitted upward from an altitude $z$ is reduced by a factor $e^{-\tau}$ while propagating towards the top of the medium.

$w(z)$ is the *single-scattering albedo*:

$$w(z) = \frac{S(z)}{E(z)} , \qquad (14)$$

which is the ratio of the scattering efficiency to the total extinction efficiency for an average particle composing the medium: it span from 0 for fully-absorbing particles to 1 for fully-scattering ones.

$p(z, \Omega', \Omega)$ is the *particle phase function*:

$$p(z, \Omega', \Omega) = \frac{G(z, \Omega', \Omega)}{S(z)} , \qquad (15)$$

which describes the directional scattering behavior of average grains; through this term it is possible to obtain the angular distribution of light intensity scattered by a particle at a given wavelength.

The last two terms refer to the *source function:*

$$f(z,\Omega) = \frac{F(z,\Omega)}{E(z)} = Je^{-\tau/\cos i} w(\tau) \rho(\tau,\Omega_0,\Omega) + f_T(\tau,\Omega) \; . \tag{16}$$

## 3.3 The bidirectional reflectance

The fundamental quantity describing light scattering from a surface is the reflectance, generally referred as the ratio of scattered radiance (or intensity) to incidence irradiance (or flux) (*Li et al., 2015*):

$$r\left[sr^{-1}\right] = \frac{I}{J} \frac{\left[Wm^{-2} sr^{-1}\right]}{\left[Wm^{-2}\right]} \; . \tag{17}$$

Reflectance quantities are defined by two prefixes to specify the collimation of incident light and the measurement conditions for scattered light; on modern usage (*Nicodemus, 1970; Nicodemus et al. 1977*) the word "reflectance" is preceded by two adjectives, the first referring to the source, and the second to the detector. The usual adjectives are, in decreasing collimation degree, "*directional*", "*conical*" or "*hemispherical*". If they are identical, the prefix *bi-* is used: thus the *bidirectional* reflectance is the same as the *directional-directional* reflectance, and describe perfectly collimated incident and emitted light. In the case of remote sensing observations from space missions, since the angle subtended by the Sun as seen by the observed target and the instrument acceptance angle are both small, the definition of bidirectional reflectance is adopted. Hereafter, for sake of brevity, we will refer to the bidirectional reflectance simply as $r$, instead of $r_{dd}$, where the d-d indicates the adjectives directional-directional.

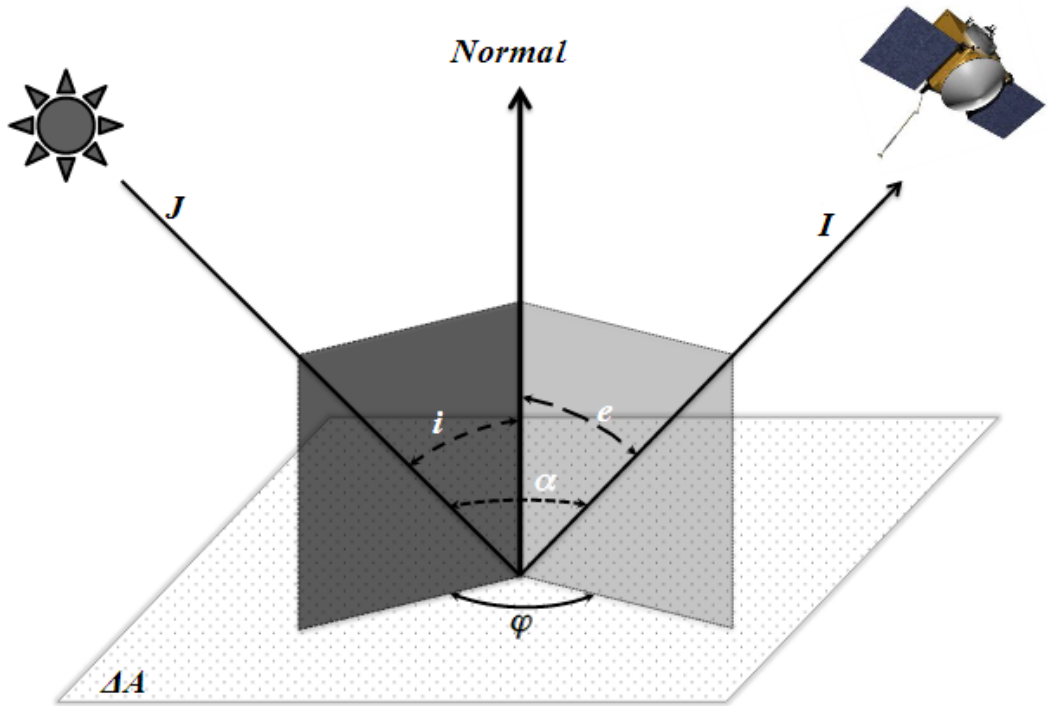

**Fig. 13 - Schematic representation of bidirectional reflectance geometry for a surface element ΔA** – the incidence (i), emission (e), phase ($\alpha$) and azimuth ($\phi$) angles are shown.

Given the surface element $\Delta A$ and the normal $N$ to this element, it is possible to identify the angles needed to specify the scattering geometry (**Fig. 13**). The first of those, $i$, is the *incidence angle* representing the angular separation between incident light and $N$; after the interaction with the medium, light emerges from the surface towards a direction that makes an *emission angle* $e$ with the normal; finally the *phase angle* $\alpha$, representing the angular separation between the directions to the source and the detector. The plane containing $N$ and the incident ray is the *plane of incidence*, the one containing the emerging ray and the normal to the surface is the *plane of emergence* while that containing both the incident and the emerging rays is the *scattering plane*. The *planes of incidence* and *emergence* are separated by the azimuth angle $\phi$.

In this context it is useful to refer to the cosines of these angles defined as $\mu_0 = \cos i$ and $\mu = \cos e$, and the bidirectional reflectance of the surface expressed as

$$r(\mu_0, \mu, \alpha, \lambda) = I(\mu_0, \mu, \alpha, \lambda) / J(\lambda), \tag{18}$$

where I is the radiance and J is the normal solar irradiance, which depends on the distance of the planet to the Sun and the wavelength. In the following sections, the dependence from the wavelength is not explicitly indicated and we will simply write $r(\mu_0, \mu, \alpha)$, instead of $r(\mu_0, \mu, \alpha, \lambda)$.

Along with the reflectance, two other quantities, typically related to laboratory measurements, are used two describe the photometric properties of a given sample: the *reflectance factor* (or *REFF*) and the *radiance factor* (or *RADF*). The *REFF* is the ratio of the reflected light from the sample to that from a perfectly diffuse reference surface (Lambertian surface, see **Sec. 3.3a**). Being the bidirectional reflectance of the Lambert surface $r = \frac{\mu_0}{\pi}$ expressed in unit of [$sr^{-1}$], the *REFF* is dimensionless. The *RADF* is the reflectance factor for normal illumination and in remote sensing measurements is usually referred as *I/F*, with:

$$I/F = \pi r. \tag{19}$$

*3.3a Physically based models of the bidirectional reflectance*

The main functional components of typical analytical models – of the bidirectional reflectance – have to describe:

(1) albedo and directional singly scattered light by an average regolith grain;

(2) how wave fronts are multiply scattered among and from an aggregate volume of average grains;

(3) the effects of macroscopic surface texture on reflected light;

(4) the *Opposition Effect*, a non-linear rise in the brightness observed at small phase angles on bodies covered by a particulate medium.

Different physically based models have been developed with the aim to describe this phenomena; below we give a brief description of a selection of them.

*Hapke model.* The widely used Hapke model (*Hapke, 1981; 1984; 1986*) has been discussed and improved in several studies (*Bowell et al., 1989; Helfenstein and Veverka, 1989; Helfenstein et al., 1997, Hapke 2002; 2008; 2012a*). Being this model the one in use for the spectrophotometric correction derived in this work, a more detailed description of its derivation is given in a separate section (**Sec. 3.4**).

*Shkuratov model.* The Shkuratov model (*Shkuratov et al., 1999*) describes a macroscopically fractal-like surface, and it includes a description of both *SHOE* (*Shadow Hiding Opposition*

*Effect*) and *CBOE* (*Coherent Backscattering Opposition Effect*), that characterize the Opposition Effect. The SHOE is the sudden increase in the observed reflectance at low phase angles due to the illumination of grains that would otherwise be in shadow; it is described by an exponential factor $f_{SHOE} = e^{-\alpha k}$. The CBOE is an additional effect which is observed when light scatters on grains of size comparable to the wavelength of the radiation; the peak in the reflectance will be produced by the constructive interference of direct and reverse paths at small phase angles and is described in this model by means of two parameters $\Delta_E$ and $d$.

The final equation of the Shkuratov model is given by the product of these two components with the normal reflectance coefficient $A_n$ – which represents the albedo at 0° phase angle – and the Akimov disk function (see **Sec. 3.3b**) $d(\Lambda, \beta, \alpha)$, that describes how the brightness of the surface at any given phase angle varies with photometric longitude $\Lambda$ and photometric latitude $\beta$:

$$r(\mu_0, \mu, \alpha) = A_n f_{SHOE}(\alpha) f_{CBOE}(\alpha) d(\Lambda, \beta, \alpha) \ . \tag{20}$$

This model depends on four-parameters: (1) $A_n$ is the normal albedo; (2) $k$, which decreases with increasing albedo to model the attenuation of shadow by multiple-scattering; (3) $\Delta_E$, the extinction mean free path of a photon; (4) $d$, that defines the separation distance of scatterers that contribute to coherent backscattering interactions.

*Lambert's law.* The diffuse-reflectance function, or Lambert's law (Lambert, 1759), refers to those surfaces which scatters radiation without any dependence on the emission e and the azimuth $\phi$ (**Fig. 13**). In this case the radiant intensity is directly proportional to the cosine of *i* and independent on the phase angle.

Bright and flat surfaces approximate this scattering behavior and a generic expression for their reflectance can be given by:

$$r_L(i,e,\psi) = K_L \cos i \,, \tag{21}$$

where $K_L$ is a constant.

The total power scattered by unit area of a Lambert surface into all directions of the upper hemisphere is:

$$P_L = \int_{2\pi} I(i,e,\psi) \mu d\Omega = \int_{e=0}^{\pi/2} \int_{\psi=0}^{2\pi} JK_L \cos i \cos e \sin e\, de\, d\psi = \pi JK_L \mu_0 \,. \tag{22}$$

Being the incident power per unit area $J\mu_0$, the fraction of the incident irradiance scattered by unit area of the surface back into the upper hemisphere is the Lambert albedo:

$$A_L = \frac{P_L}{J\mu_0} = K_L \pi \,, \tag{23}$$

so that

$$K_L = A_L/\pi \,. \tag{24}$$

Given this, the Lambert reflectance can be obtained by a substitution (*Schroder et al., 2013*):

$$r_L(i, e, \psi) = \frac{A_L}{\pi} \mu_0 . \tag{25}$$

When $A_L=1$, the surface is a perfectly diffusive surface.

*Lommel-Seeliger law.* The Lommel-Seeliger law (*Seeliger, 1887*) has the advantage of analytical simplicity, while representing an excellent first approximation to diffuse reflection. On the other side, this law has the disadvantage of its inability to display the Opposition Effect.

The Lommel-Seeliger law is based on a physical model of diffuse reflection in a semi-infinite particulate medium. It relies on the single scattering assumption, that means it does not consider the contribution of light undergoing multiple interactions with the particles in the medium (multiple scattering). This model assumes that light penetrates the surface, being attenuated exponentially through scattering and absorption as it propagates in the medium. Each element of volume encountered by the attenuated beam scatters part of its light isotropically: thus, only half is directed back towards the surface and this fraction will be further attenuated before emerging as diffuse reflected radiation.

The form of the disk function for the Lommel-Seeliger law is:

$$D(\mu_0, \mu) = \frac{2\mu_0}{\mu + \mu_0} . \tag{26}$$

It predicts a strong limb brightening at large phase angles, providing a good description of the light scattered by low-albedo bodies (i.e. Moon and Mercury), for which multiple scattering can be neglected.

It has been suggested a combination of the Lambert's law with the Lommel-Seeliger law (*Schroder et al., 2013*):

$$D(\mu_0, \mu, \alpha) = C_L \frac{2\mu_0}{\mu + \mu_0} + (1 - C_L)\mu_0 , \qquad (27)$$

where $C_L$ is a free parameter.

*3.3b Empirical expressions for the bidirectional reflectance*

Empirical models of the bidirectional reflectance are widely used for their mathematical simplicity, although no natural surfaces obey these law exactly.

In these formulations, the reflectance of the target is typically expressed as a function of two terms: the disk function $D(\mu_0, \mu)$, which describes how the reflectance varies over the planetary disk at constant phase angle as a function of $i$ and $e$, and the phase function $f(\alpha)$, accounting for the dependence on phase angle. The RADF can be expressed as

$$R = d(i,e) f(\alpha) , \qquad (28)$$

and it represents a good approximation when multiple scattering is negligible (*Li et al., 2015*). Defined $A_{eq}(\alpha)$ as the equigonal albedo,

$$A_{eq} = A_N f(\alpha) , \qquad (29)$$

where $A_N$ is the normal albedo and the phase function of a given body is normalized to unity at $\alpha=0°$, the globally averaged phase function can be expressed as a polynomial of degree $d$,

$$A_{eq}(\alpha) = \sum_{i=0}^{d} C_i \alpha^i,  \quad (30)$$

where $C_0 = A_N$. The coefficients $C_i$ provides the best fit for the average surface.

In the following subsections, a general overview of the most commonly used empirical models is presented.

*Minnaert's law.* Minnaert's law (*Minnaert, 1941*) is a generalization of Lambert's law, so that the power emitted per unit solid angle per unit area of the surface is proportional to $(\mu_0 \mu)^{c_M}$, instead of $(\mu_0 \mu)^1$. This leads to a disk function of the form

$$D(\mu_0, \mu) = \mu_0^{C_M} \mu^{C_M - 1},  \quad (31)$$

and to the bidirectional reflectance (*Schroder et al., 2013*)

$$r_M(i, e, \psi) = A_M \mu_0^{C_M} \mu^{C_M - 1}.  \quad (32)$$

$A_M$ is the albedo of this model, $C_M$ is a free parameter which depends on phase and azimuth angles. If $C_M = 1$, this law reduces to Lambert's law.

Minnaert's law is able to describe the variation of brightness of many surfaces over a limited range of angles, anyway it does not provide any link to the physical properties. Moreover, for observations at limb, characterized by $e = 90°$:

(1) if $C_M < 1$, the calculated brightness becomes infinite;

(2) if $C_M > 1$, the brightness would be zero: neither of these cases agrees with observations.

*Akimov's law.* This is a model (*Akimov, 1988*) with no parameters, useful for a rough surface, slightly undulated. It correctly removes the topography influence on reflectance at all wavelengths. The Akimov disk function is

$$D(\alpha,\beta,\gamma) = \cos\left(\frac{\alpha}{2}\right)\cos\left[\frac{\pi}{\pi-\alpha}\left(\gamma-\frac{\alpha}{2}\right)\right]\frac{(\cos\beta)^{\frac{\alpha}{\pi-\alpha}}}{\cos\gamma} \quad , \tag{33}$$

where $\beta$ is the photometric latitude and $\gamma$ is the longitude. The incidence and emergence angles are:

$$\mu_0 = \cos\beta\cos(\alpha-\gamma) \tag{34}$$

and

$$\mu = \cos\beta\cos\gamma \quad . \tag{35}$$

## 3.4 The Hapke model

The Hapke photometric model (*Hapke, 2012*) is a standard for the analysis of photometric data in planetary remote sensing studies. Its derivation is based on the solution of the radiative transfer equation and its primary applications regard imagery correction to standard lighting and viewing geometry for comparative purposes.

The general expression of the bidirectional reflectance of the Hapke model (*Hapke, 1993*) is presented below:

$$r(i,e,\alpha) = \frac{w(\lambda)}{4\pi} \frac{\mu_{0\text{eff}}}{\mu_{0\text{eff}} + \mu_{\text{eff}}} \left[ p(\alpha) B_{SH}(\alpha) + H(w,\mu_{0\text{eff}}) H(w,\mu_{\text{eff}}) - 1 \right] S(i,e,\alpha,\theta) . \qquad (36)$$

- $\mu_{0\text{eff}}$, $\mu_{\text{eff}}$ and the phase angle $\alpha$ characterize the observation geometry.

- p($\alpha$) describes the single scattering;

- $H(w,\mu_{0\text{eff}}) H(w,\mu_{\text{eff}}) - 1$ is the multiple scattering term;

- $B_{SH}(\alpha)$ accounts for the SHOE;

- S(i,e,$\alpha$,$\theta$) is the Shadowing function.

This equation expresses the bidirectional reflectance of a semi-infinite particulate medium as a function of the observing geometry and its physical properties. In the following sections we give a description of its derivation, as well as of its different terms.

*3.4a Scattering from a semi-infinite particulate medium*

Let's consider a semi-infinite particulate medium of isotropic scatterers and a thin layer added on top of this, with thickness $\Delta z$ and optical depth $\Delta\tau$. Assuming it is illuminated from the top with irradiance $J$, it is possible to recognize five different processes, proportional to $\Delta\tau$, characterizing the interaction of the light with the layer and the medium. Higher order interactions are not considered.

*Process (1).* Light passing through the layer is partially extinguished before and after the reflection on the reference surface (**Fig. 14a**). The emerging radiance depends on the incidence and emergence cosines and on the optical depth, as follows:

$$I = J e^{-\Delta\tau\left(\frac{1}{\mu_0}+\frac{1}{\mu}\right)} r(i,e) \approx J\left[1-\Delta\tau\left(\frac{1}{\mu_0}+\frac{1}{\mu}\right)\right] r(i,e) \; ; \tag{37}$$

and the variation obtained is

$$\Delta I_a = -J\Delta\tau\left(\frac{1}{\mu_0}+\frac{1}{\mu}\right) r(i,e) \; . \tag{38}$$

*Process (2).* Light does not reach the medium, but it is scattered back by the added layer (**Fig. 14b**). The radiance increment due to the scattering by all particles inside a cylinder with base $Q_S$ and height $\frac{\Delta z}{\mu}$ is

$$\Delta I_b = JN\sigma Q_s \frac{\Delta z}{\mu}\frac{1}{4\pi} = J\frac{w}{4\pi}\frac{\Delta\tau}{\mu} \; . \tag{39}$$

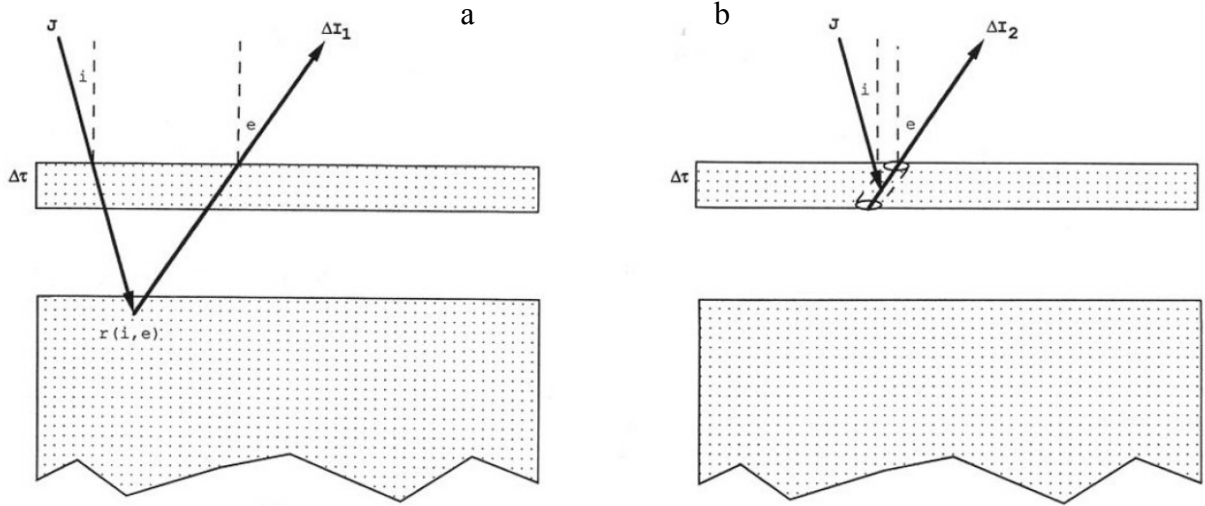

**Fig. 14 - Processes on the scattered radiation -** Processes (1) and (2) in left **(a)** and right **(b)** panel, respectively; description in the text. *(Adapted from Hapke, 1993; Ciarniello, 2012)*

*Process (3).* In this case (**Fig. 15a**) the radiation is scattered downward by the added layer; a fraction of the light is sent back to the observer from the medium, which has reflectance *r(i',e)*. For the light scattered by the layer, the radiance is

$$dI_{i'} = J \left(\frac{w}{4\pi}\right)\left(\frac{\Delta\tau}{\mu_0'}\right) d\Omega_{i'}, \qquad (40)$$

where $\mu_0'$ is the cosine of the angle that scattered light forms with the vertical and $d\Omega_{i'}$ is the solid angle increment.

Integrating,

$$\Delta I_C = J \frac{w}{2} \Delta\tau \int_{\mu_0'=0}^{1} r(\mu_0',\mu) d\frac{\mu_0'}{\mu_0'}. \qquad (41)$$

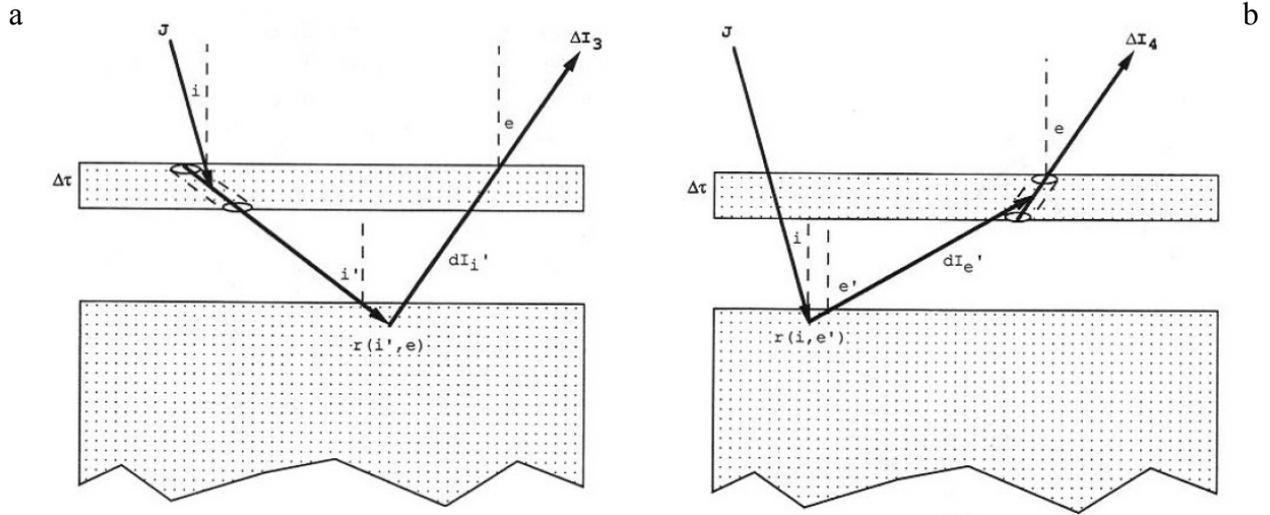

**Fig. 15 - Processes on the scattered radiation -** Processes (3) and (4) in left **(a)** and right **(b)** panel, respectively; description in the text. *(Adapted from Hapke, 1993; Ciarniello, 2012)*

*Process (4).* The same as the previous process, but the light undergoes scattering when leaving the medium (**Fig. 15b**):

$$\Delta I_d = J \frac{w}{2} \frac{\Delta \tau}{\mu} \int_{\mu'=0}^{1} r(\mu_0, \mu') d\mu' ,\qquad(42)$$

where $\mu' = \cos e'$.

*Process (5).* Light passing through the added layer is reflected from the surface and sent back again by the layer itself; then it leaves the medium (**Fig. 16**). In this case, the medium scatters a fraction of radiation $Jr(i,e'')d\Omega_{e''}$ in the direction $e''$, with azimuth angle $\psi_{e''}$. This light illuminates a cylindrical volume on the top layer, which sends back to the lower medium a fraction of radiation:

$$dI_{ie} = dI_{e''} \left(\frac{w}{4\pi}\right)\left(\frac{\Delta \tau}{\mu_0''}\right) d\Omega_{i''} ,\qquad(43)$$

where $\mu_0''=\cos i''$. An amount $dI_{ie}r(i'',e)$ is scattered again and leaves the medium with an emission angle *e*. Integrating over all the available directions $\Omega_{e''}$ and $\Omega_{i''}$ we have:

$$\Delta I_e = Jw\pi\Delta\tau \left[\int_{\mu_0''}^{1} r(\mu_0'',\mu)\frac{d\mu_0''}{\mu_0''}\right]\left[\int_{\mu''=0}^{1} r(\mu_0,\mu'')d\mu''\right]. \qquad (44)$$

Summing all the contributes from the five different processes and imposing

$$P(\mu_0,\mu) = \frac{4\pi}{w}\frac{\mu_0+\mu}{\mu_0}r(\mu_0,\mu), \qquad (45)$$

the equation becomes:

$$P(\mu_0,\mu) = \left[1+\frac{w}{2}\mu\int_0^1 \frac{P(\mu_0',\mu)}{\mu_0'+\mu}d\mu_0'\right]\left[1+\frac{w}{2}\mu_0\int_0^1 \frac{P(\mu_0,\mu')}{\mu_0+\mu'}d\mu'\right]. \qquad (46)$$

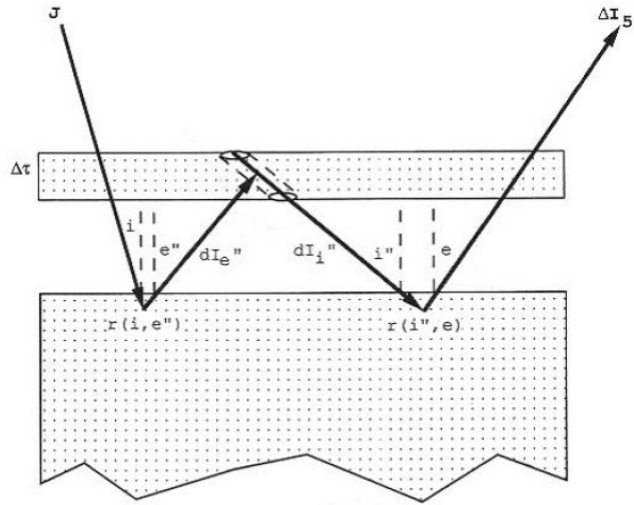

**Fig. 16 - Processes on the scattered radiation -** Process (5); description in the text. (Adapted from *Hapke, 1993; Ciarniello, 2012*)

The function P is symmetric in $\mu$ and $\mu_0$ and the two terms in the equation above are respectively functions of $\mu$ and $\mu_0$. Then, it is possible to substitute both of them with a function H(x), named Chandrasekhar function – where x=$\mu$ for the first term and x=$\mu_0$ for the second. The equation can be written as the product $P(\mu_0,\mu)=H(\mu_0)H(\mu)$, where H(x) solves the integral equation

$$H(x)=1+\frac{w}{2}xH(x)\int_0^1 \frac{H(x')}{x+x'}dx' \ . \tag{47}$$

It can be shown that an approximate analytical expression for the *H(x)* function is

$$H(x)\approx\frac{1+2x}{1+2\gamma x} \ , \tag{48}$$

where $\gamma=\sqrt{1-w}$; this approximation differs from the exact solution by less than 4%, as it is shown in **Fig. 17**, and has been adopted in this work. Given the substitution above, it is possible to explicit the bidirectional reflectance of a semi-infinite medium made of isotropic scatterers:

$$r(i,e,g)=\frac{w}{4\pi}\frac{\mu_0}{\mu+\mu_0}H(\mu_0)H(\mu) \ . \tag{49}$$

When w → 0, H(x) → 1 for all values of x: the equation reduces to the Lommel-Seeliger law, indicating that single scattering is the dominant process.

When x → 0, H(x) → 1: at glancing angles of incidence and emergence single scattering dominates the reflectance.

The solution of the equation strongly depends on the particle phase function only in the case of single scattering, while for multiple scattering the directional effects due to an anisotropic phase function tend to be averaged and the final result is similar to the one of isotropic scatterers. Given this, the solution can be extended to the case of anisotropic scatterers by treating the multiple scattering process assuming that the particles scatter isotropically and computing explicitly the contribute of single scattering with an anisotropic phase function:

$$r(i,e,g) = \frac{w}{4\pi} \frac{\mu_0}{\mu_0 + \mu} \left[ p(g) + H(\mu_0) H(\mu) - 1 \right]. \tag{50}$$

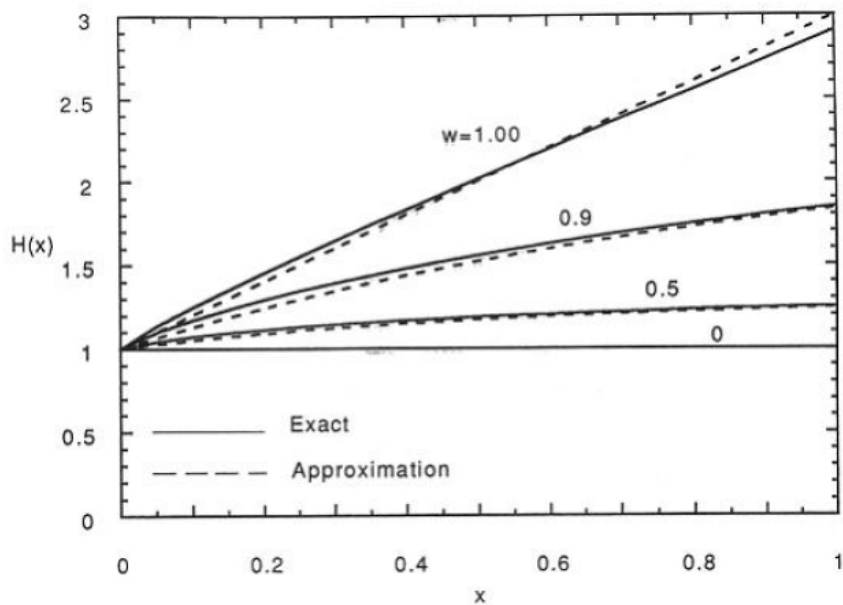

**Fig. 17 - H(x): comparison between the exact and approximated expressions -** The error due to the approximation (dashed line) is about 4% respect to the exact (solid line) one. (Adapted from *Hapke, 1993; Ciarniello 2012*)

*3.4b Single particle phase function*

The scattering properties of a single particle are described by the phase function $p(\alpha)$, which represent the angular pattern into which light is scattered by the particle itself. It depends on the physical characteristics of the particle and of the size parameter $X=2\pi r/\lambda$, where $r$ is the particle radius and $\lambda$ is the wavelength of the incident light. Although an analytical derivation of $p(\alpha)$ for isolated spheres is provided by the Mie theory (*Mie, 1908*), in the case of real regolith particles with $X \gg 1$ empirical formulations are commonly used. These expression depend on one or more parameters, which determine the phase function behavior.

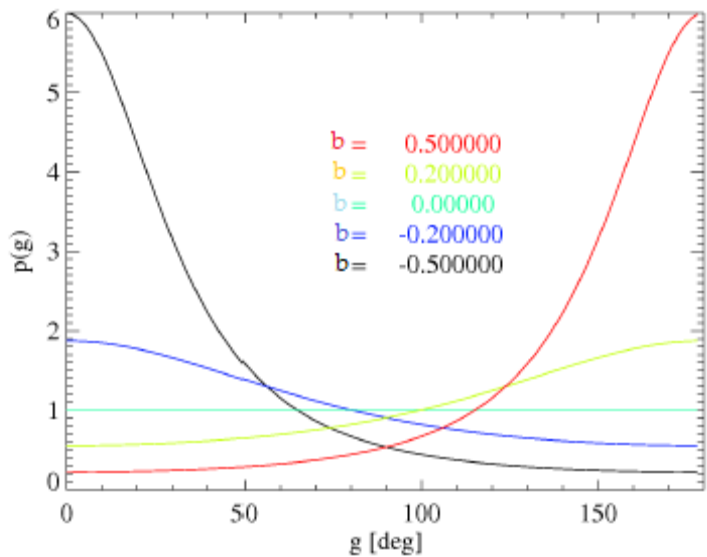

**Fig. 18 - Henyey-Greenstein phase function -** phase function for different values of b, from -0.5 (red) to 0.5 (black).

One of the most adopted formulation is the one derived by *Heyney and Greenstein (1941)*:

$$p(\alpha)=\frac{1-b^2}{1+b^2+2b\cos\alpha} , \qquad (51)$$

where *b* is the asymmetry parameter, which determines the behavior of *p(α)* and span from -1 to 1, ranging from backscattering, through isotropic (b=0), to forwardscattering (**Fig. 18**).

*3.4c The opposition effect*

The opposition effect (see **Section 3.3a**) is a steep luminosity increment at small phase angles, that is particularly pronounced in fine powders with a mean grain size less than about 20 μm (*Hapke, Van Horn, 1963*). The equation of radiative transfer itself is incapable of accounting for this effect. However, after the discovery by *Seeliger (1887, 1895)*, it has been treated by several authors, such as Hapke, Bobrov, Irvine, Lumme and Bowell (*Li et al., 2015*).

*SHOE.* In a medium made of particles larger than the wavelength the grains near the surface cast their shadows on deeper particles. When the surface is observed at large phase angles these shadows are visible, instead, at smaller phase angles, a larger fraction of the shadow is hidden by the casting particle itself and the light which has been able to reach a particle can travel along the same path in the opposite direction to reach the observer. This cannot happen for large *α* observations, thus producing a final brightening of the surface at small phase angles. It can be shown that this effect can be modeled by adding a term $B_S(α)$ to the single scattering term of the bidirectional reflectance:

$$r(i,e,\alpha) = \frac{w(\lambda)}{4\pi} \frac{\mu_{0\text{eff}}}{\mu_{0\text{eff}} + \mu_{\text{eff}}} \left[ p(\alpha) B_{SH}(\alpha) + H(w, \mu_{0\text{eff}}) H(w, \mu_{\text{eff}}) - 1 \right] ,  \quad (52)$$

where the analytical expression for $B_S(\alpha)$ is:

$$B_S(\alpha) = \sqrt{\frac{4\pi}{y}} e^{1/y} \left[ erf\left(\sqrt{\frac{4}{y}}\right) - erf\left(\sqrt{\frac{1}{y}}\right) \right] + e^{-3/y} - 1 \; , \tag{53}$$

with

$$y = 2\frac{\bar{\mu}}{\tau_1} = 2\frac{\tan(\alpha/2)}{N_E(\sigma \bar{Q}_E) \bar{a}_E} \; ; \tag{54}$$

where $erf(x)$ is the error function, $N_E$ is the distribution of particle sizes, $a_E$ is the average particles extinction radius and $\sigma Q_E$ is the cross-sectional area of an equivalent spherical particle. The expression for $B_S(\alpha)$ is difficult to compute because of the $erf(x)$ function; for this reason it is usually preferred the approximate expression:

$$B_S(\alpha) \approx (1+y)^{-1} = \left(1 + \frac{1}{h_S} \tan\frac{\alpha}{2}\right), \tag{55}$$

where

$$h_S = \frac{1}{2} N_E (\sigma \bar{Q}_E) \bar{a}_E \; . \tag{56}$$

The discrepancy from the analytical solution is less then 3%, as shown in **Fig. 19**.

In this derivation, only the single scattering process is considered, implicitly assuming that the rays are scattered by the particle exactly from the incidence point. However, in the case of transparent particles the light leaves the particle from a point which is not necessarily

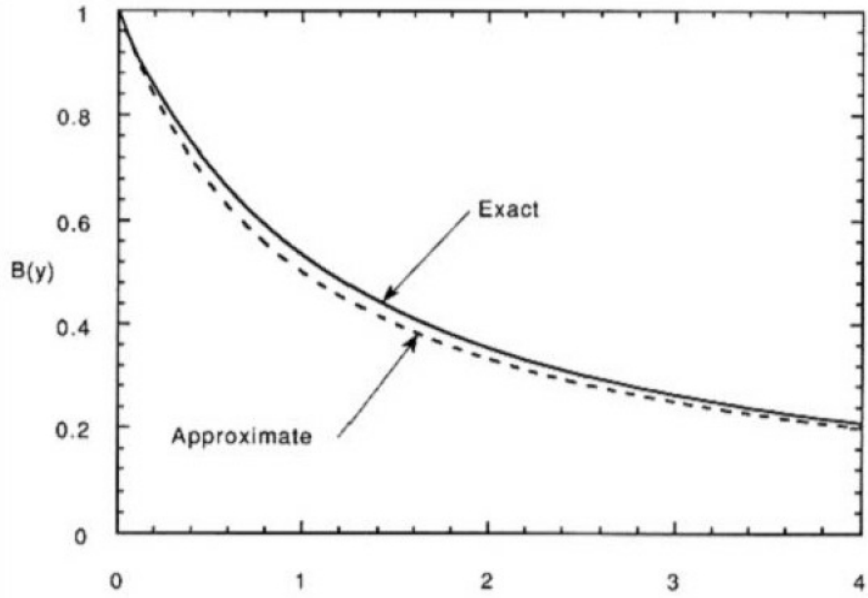

**Fig. 19 - B(y): comparison of the exact and the approximated expressions -** The error due to the approximation (dashed line) is about 3% respect to the exact (solid line) one. *(Adapted from Hapke, 1993; Ciarniello 2012)*

the incidence point and the probability that the emitted ray travels the same path of the incident ray, without interactions with other particles, decreases. This effect is described multiplying $B_S(\alpha)$ by a parameter accounting for the the amplitude of SHOE, named $B_{S0}$, which is an empirical term comprised between 0 and 1.

*CBOE.* This effect causes a prominent phase-curve spike that is observed within a few degrees of opposition (<2°), superimposed on SHOE and is described by:

$$B_{CB}(\alpha) = \frac{1 + \dfrac{1 - e^{-(1/h_{CB})\tan(\alpha/2)}}{(1/h_{CB})\tan(\alpha/2)}}{2\left[1 + (1/h_{CB})\tan(\alpha/2)\right]^2} , \qquad (57)$$

with $h_{CB} = \lambda / 4\pi L_T$ , where $\lambda$ is the wavelength and $L_T$ the photon mean free path length in the medium. However, the lack of data at lower phase angles did not allow us to constrain the coherent backscattering opposition effect.

*3.4d Large scale surface roughness*

The equation of the bidirectional reflectance derived above describes the spectrophotometric output of ideally flat surfaces. However, real planetary objects typically exhibit macroscopic surface roughness at scales larger than the wavelength. This can be modeled assuming the surface as made up of many facets, tilted with respect to the ideal flat surface, with a given slopes distribution. Macroscopic roughness modifies the resulting reflectance of the target, producing three major effects:

(1) scattering of light from one facet to another. This effect increases the reflectance, but it is small if either the albedo or the mean slope is small;

(2) unresolved shadows projected on one part of the surface by another. This effect decreases the reflectance;

(3) when the surface is illuminated at increasing zenith angles, the facets that are tilted away from the observer or from the lightsource are hidden or in shadow, so that the surfaces that are visible and illuminated are those that are tilted preferentially toward the detector or source.

In the Hapke' theory, these effects, with the exception of multiple scattering between facets (1), are modeled by the introduction of a shadowing function $S(i,e,\alpha,\theta)$, which is a multiplicative term that reduces the reflectance for increasing values of *i, e, α* and *θ*, where θ is the large scale surface roughness. This function has different formulations depending on whether *i* is larger or smaller than *e*.

*case i ≤ e.*

When $i \leq e$ the illumination shadow cast by a given object is always smaller than its visibility shadow; in this case the illumination shadow may be regarded as partially hidden in the visibility shadow and the shadowing function is:

$$S(i,e,\psi) \approx \frac{\mu_e}{\mu_e(0)} \frac{\mu_0}{\mu_{0e}(0)} \frac{\chi(\bar{\theta})}{1-f(\psi)+f(\psi)\chi(\bar{\theta})[\mu_0/\mu_{0e}(0)]} \quad (58)$$

where *f(ψ)* is the fraction of the illumination shadow hidden in the visibility shadow and is considered independent of *i* and *e*. χ(θ) is a function of *θ* through its tangent. $\mu_e$ and $\mu_{0e}$ in this context are the effective cosines, which are the cosines of the incidence and emission angle, modified for the effect of the surface roughness.

*case e ≤ i.* *f(ψ)* is now the fraction of the visibility shadow hidden in the illumination shadow. The formulation is the same given above, for $i \leq e$, with a different *f(ψ)*.

Introducing this final correction to the expression of **eq. 52**, we can write the final formulation for the bidirectional reflectance, as reported at the beginning of **Section 3.3**:

$$r(i,e,\alpha) = \frac{w(\lambda)}{4\pi} \frac{\mu_{0\text{eff}}}{\mu_{0\text{eff}}+\mu_{\text{eff}}} \left[ p(\alpha) B_{SH}(\alpha) + H(w,\mu_{0\text{eff}}) H(w,\mu_{\text{eff}}) - 1 \right] S(i,e,\alpha,\theta) \quad (59)$$

# Chapter 4

# Data analysis

In this chapter, we will show the data analysis performed. The strategy used to map the surface during the different mission phases is presented (**section 4.1**), while in **section 4.2** it is shown the statistical analysis needed to characterize the data coverage in longitudes, latitudes and phase angles and the selection of these data. We chose a set of wavelengths over the Infrared and Visible ranges (**section 4.3**) and performed the photometric correction over the data (**section 4.4**) using the Hapke's model described in **chapter 3;** in **section 4.5** we discussed the model and measurements errors.

**4.1 Mapping strategy**

The mission to Vesta is characterized by four main science data acquisition phases at different orbital distances from the target and consequently at different observations spatial resolutions (**Table 4**). These phases are the Vesta Science Survey (VSS), Vesta Science HAMO-1 (VSH), Vesta Science LAMO (VSL) and Vesta Science HAMO-2 (VH2), while data are also acquired as Dawn approaches the body. Each of these phases had different scientific goals.

During the Vesta Approach phase (VSA), Dawn spacecraft progressively reduced its distance from 160000 to 3000 km. Later, it maneuvered to enter in a polar orbit around the target. After a transfer trajectory, the spacecraft moved into the Vesta Survey orbit (VSS phase), from where the VIR instrument and the FC accomplished an extensive mapping, covering the surface at low spatial resolution (*De Sanctis et al., 2010*).

| Table 4 Mission phases at Vesta. | | | |
|---|---|---|---|
| **Mission Phase** | **Start at** | **Distance** | **Nominal Resolution** |
| VSA | 03-05-2011 | / | / |
| VSS | 11-08-2011 | 3000 km | 700 m/pixel |
| VTH | 03-09-2011 | / | / |
| VSH | 29-09-2011 | 950 km | 180 m/pixel |
| VSL | 12-12-2011 | 465 km | 90 m/pixel |
| VH2 | 15-06-2012 | 950 km | 170 m/pixel |
| VTC | 25-07-2012 | / | / |

A transfer trajectory (Vesta Transfer to HAMO, VTH) carried Dawn to the high altitude mapping orbit (HAMO), starting the VSH phase, and after 30 days to the low altitude mapping orbit (LAMO), entering in the VSL phase. Then the spacecraft moved back again to the high altitude mapping orbit (HAMO-2) and finally entered the Vesta to Ceres phase (VTC), that guided the mission to Ceres, its second target.

The main constraints on the acquisition strategy for different orbits are imposed by the limitations of the on board memory (Virtual Recorder of S/C and VIR memory), illumination conditions, orbit maintenance maneuver, download, and by the need to collect images for optical navigation (*De Sanctis et al., 2010*). VIR collected spectra of Vesta using different observation strategies in the different orbits.

During the Survey phase, the main goals were to obtain a global mapping of the surface with the VIR mapping spectrometer at low resolution and to produce global images with the Framing Camera. This has been achieved through a series of different consecutive mapping cycles (named C1, C2 ... C7). Most of the surface was covered during Survey, but the north polar regions were not fully mapped due to lighting conditions (presence of enduring shadows and high incidence angles).

During the two HAMO phases (VSH and VH2, from hereafter HAMO-1 and HAMO-2, respectively), VIR data were collected to sample the spectral variability at scales smaller than the ones achieved during the global Survey mapping and to provide high-resolution coverage for areas of interest. In every cycle of VSH, VIR acquired data in pushbroom configuration in the southern hemisphere. The variety of different spacecraft orientations required for the camera mapping during HAMO allowed the VIR instrument to effectively use the pushbroom mode over a wide range of southern latitudes (5° - 65° South).

At last, the LAMO phase of the mission has primarily been used to collect gamma ray and neutron spectra, as well as to determine the gravity field of the asteroid Vesta.

**4.2 Preliminary studies**

The full archive, containing the Dawn VIR cubes from all the orbital phases at Vesta, is publicly available at http://sbn.psi.edu/archive/dawn/vir/DWNVVIR_I1A/ .

All data files are stored in the 'DATA' branch of the directory tree. The data are organized into sub-directories by phase and activity. DATA sub-directories are named according to the convention:

$$\text{“yyyymmdd\_nnnnn”,}$$

where:

"yyyymmdd" = phase/activity start date,

"nnnnn" = phase/activity name,

es. "20110510_OPNAV_002/".

While data cube are named according to the convention:

"VIR_sss_ll_v_sctime_z"

es "VIR_IR_1B_1_366390345_refl_1.QUB".

First, we studied the whole available data-set from a statistical point of view, in order to characterize the spatial coverage and the coverage with respect to some particular geometries (incidence, emission and phase angles). In fact, the better is the coverage in – phase, emission and incidence – angles, the better would be the correction produced by means of a fitting model; moreover, a better coverage in latitude and longitude, would allow us to correct a larger region of the observed surface. As shown below, these criteria are met by observations acquired during the Survey, HAMO-1 and HAMO-2 phases, respectively. These data, consequently have been selected to characterize Vesta's spectrophotometric properties and derive the photometric correction, that is the main focus of this work. HAMO-1 and HAMO-2 – which have a similar spatial resolution – are put together. Hereafter we'll simply refer to HAMO, which will represent the sum of HAMO-1 and HAMO-2 data-sets.

The preliminary analysis here presented, consisted of three "steps":

*step (i)* - unpacking and reading of the data collection for the entire sequence, by means of a proper IDL procedure;

*step (ii)* - organization and storage of the extracted information in easy-access variables containing the spectral reflectance and the observations geometries (longitude, latitude, emission angle, incidence angle, phase angle, distance from the surface);

*step (iii)* - filtering out of corrupted data.

*SURVEY.*

The Survey data-set is composed by a total of 3152280 pixels. These observations span from -89.9° and 44.6° in latitude and over the whole range of longitudes. Emergence and incidence angles span in a large interval, as expected for this mission sequence (because we have a high probability to see also pixels at the limb); similarly, the phase angles cover a wide range (see **Table 5)**.

| Table 5 Survey observation geometries and surface coverage | | |
|---|---|---|
| | MIN | MAX |
| PHASE ANGLE | 12.14° | 78.05° |
| INCIDENCE ANGLE | 0.01° | 110.07° |
| EMISSION ANGLE | 0.01° | 87.17° |
| LATITUDE ANGLE | -89.93° | 44.56° |
| LONGITUDE ANGLE | 0.00° | 360.00° |

Below, we give a detailed description of each single analytical "step" described in the previous section:

*"steps" (i-ii)* - After a first "pre-processing" phase, useful to acquire knowledge of the data storage organization and the general form of a cube (see **Section 2.4** for details), data have been collected and the information contained in cubes has been read through a proper IDL procedure. This information is stored in a bi-dimensional array or "table", easy accessible from other external procedures. This array is of the form *array(k,n)*, where *k* is an index that identifies the spectral or spatial information and *n* indicates the observed pixel.

In **Fig. 20** it is shown the reflectance map from Survey observations data at a given "test-wavelength" (0.796 µm), as derived from the re-projection on a latitude-longitude grid of the pixels stored in the corresponding "table"; each pixel of the map is associated to the mean value of all the observed reflectances falling into it.

Darker – or brighter – areas, are due to lower – or higher reflectance, caused either by different observation geometries or possibly by intrinsic differences of the surface albedo. Areas where cubes – taken under different illumination conditions – overlap, can be recognized (in particular between -30°/0° latitude interval). Some artifacts appear in the southern hemisphere (30°/150° longitude; -30°/-60° latitude) and a strong gradient in reflectance just above the equator, in the North-South direction, which are clearly related to observation geometry issues and need to be minimized by means of a photometric correction.

It can be noted that three regions are not well covered:

(1) the whole North Pole and half of the northern hemisphere (approximately above 30°);

(2) a small portion of the South Pole, which shows a lack of data, partially related to stretching due to cylindrical coordinates;

(3) lack of observations in a region just below the equator (between 150° and 225° longitude).

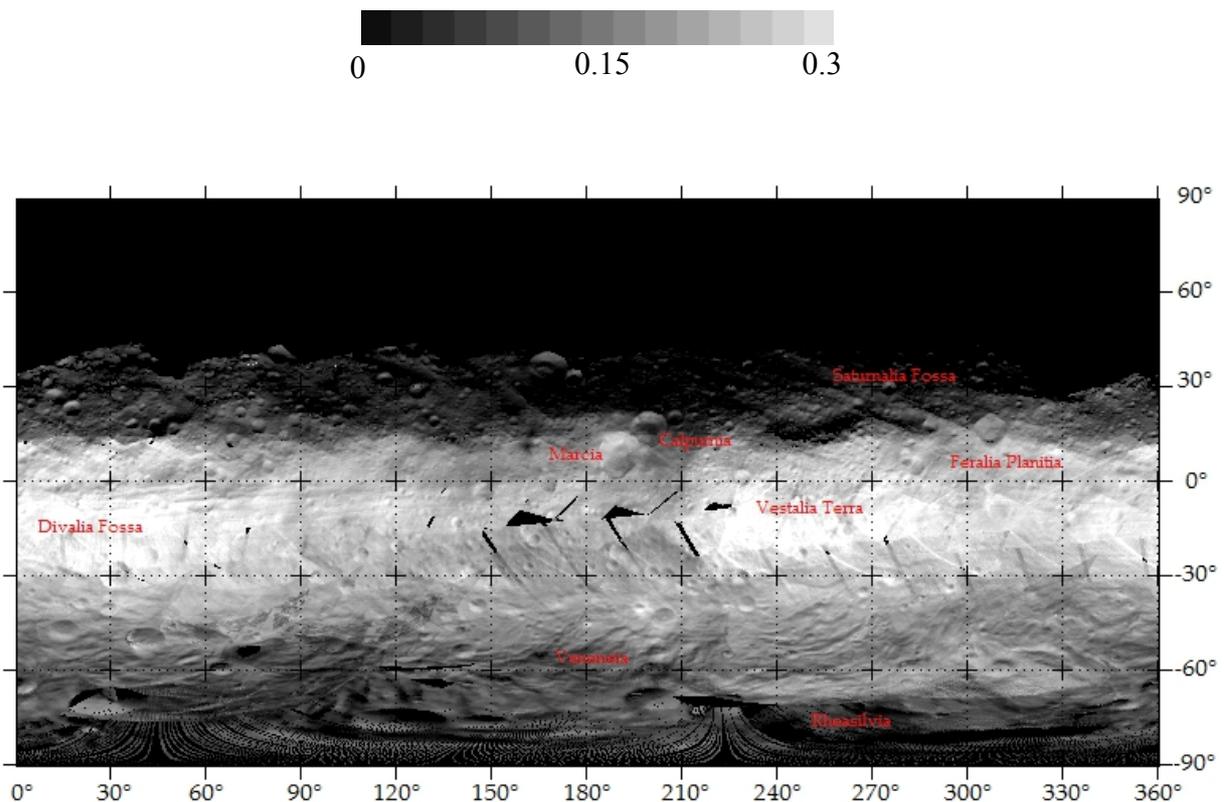

**Fig. 20 – Reflectance map from SURVEY data –** Reflectance map from Survey observations before photometric correction, at 0.796 μm. Values on the x axis are the longitudes of Vesta, while on the y axis latitudes are reported.

This is confirmed in **Fig. 21,** where the observations redundancy map is reported. The region just below the equator is indeed characterized by a worst statistic than the other regions covered by VIR observations. The coverage in longitudes shown is consistent with the mapping strategy, because each cycle of the Survey should have covered at least one time the whole range 0°-360°.

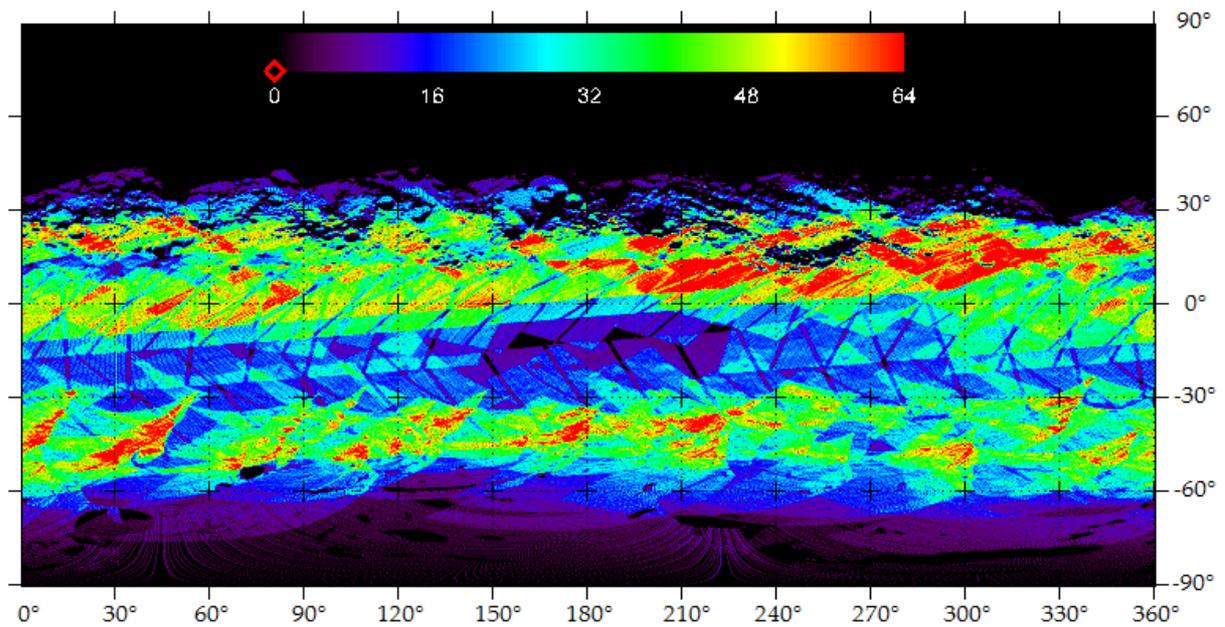

**Fig. 21 - Redundancy map from Survey data –** on the x axis there are the longitudes, on the y axis the latitudes. Blue areas are less covered, yellow and red areas are the most covered (the color bar indicates the number of occurrences inside a 0.5° x 0.5° bin).

*Step (iii)* Pixels acquired with a very low signal (reflectance< 0.001) have been filtered out to limit the occurrence of observations in total or partial shadow; similarly, pixels with extremely large values of the reflectance can be considered unphysical, and possibly related to instrumental effects. We found that these pixels can be sorted out assuming an upper limit of 1.

In order to limit the effect of unfavorable geometries, that can be affected by relatively large errors, a limit on the largest values of the incidence and emission angle has also been set. A typical choice in this kind of studies is to impose a cut on those pixels observed with *e* and *i* > 70° (*Ciarniello et al., 2016*). Assuming these selection criteria, approximately the 10% of the 3152280 initial pixels has been eliminated. After that, the data-set has been analyzed again and compared to the previous, unfiltered one:

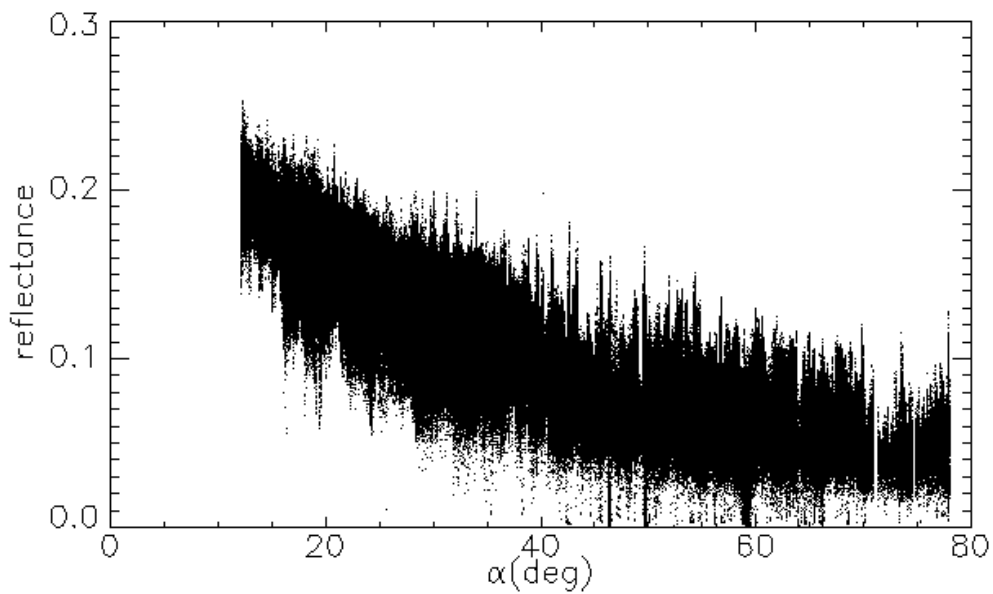

**Fig. 22 – Reflectance scatter plot from Survey data-** reflectance (y axis) vs phase angle (x axis) for each pixel of the Survey after data selection at 0.796 μm. There is a clear monotonic decreasing trend.

- a scatter plot of reflectance as a function of the phase angle at 0.796 µm (**Fig. 22**) provides a direct view of the filtered data and it can be noted that the two quantities are anti-correlated;

- in **Fig. 23** we report histograms of the number of observations as a function of the phase angles before and after the filtering of the data-set. In panel **a** and **b**, it is shown that filters are more selective at high phases, because of the cut of high emergence and incidence angles observations and shadowed pixels;

- a density plot, like the one presented in **Fig. 24** (details in the caption), is comprehensive of both information. Areas in light red correspond to a larger pixels density (in the phase angle – reflectance plane), areas in dark red correspond to a lower pixels density.

The wide range of phase angle (12°-78°) in the Survey sequence allows a good modeling of the phase curve. However, data at very low phase angles are missing, thus preventing a proper characterization of the Coherent Backscattering Opposition Effect, that, for this reason, is not included in our modeling.

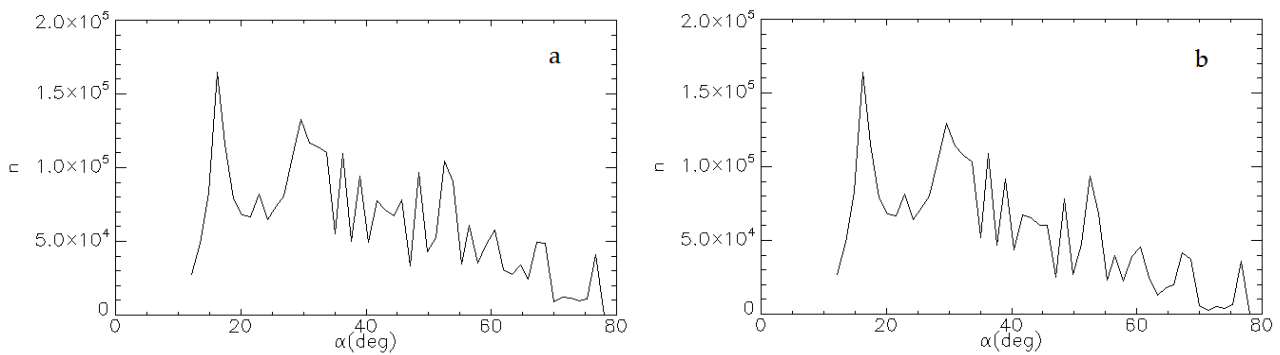

**Fig. 23 - Number of observations as a function of the phase angle, Survey -** Observations phase angle distributions before **(a)** and after **(b)** data selection. The ≈10% filtered out data do not strongly affect the general distribution. At larger phase angles (> 45°) the values are approximately a 20% lower in case **(b)**.

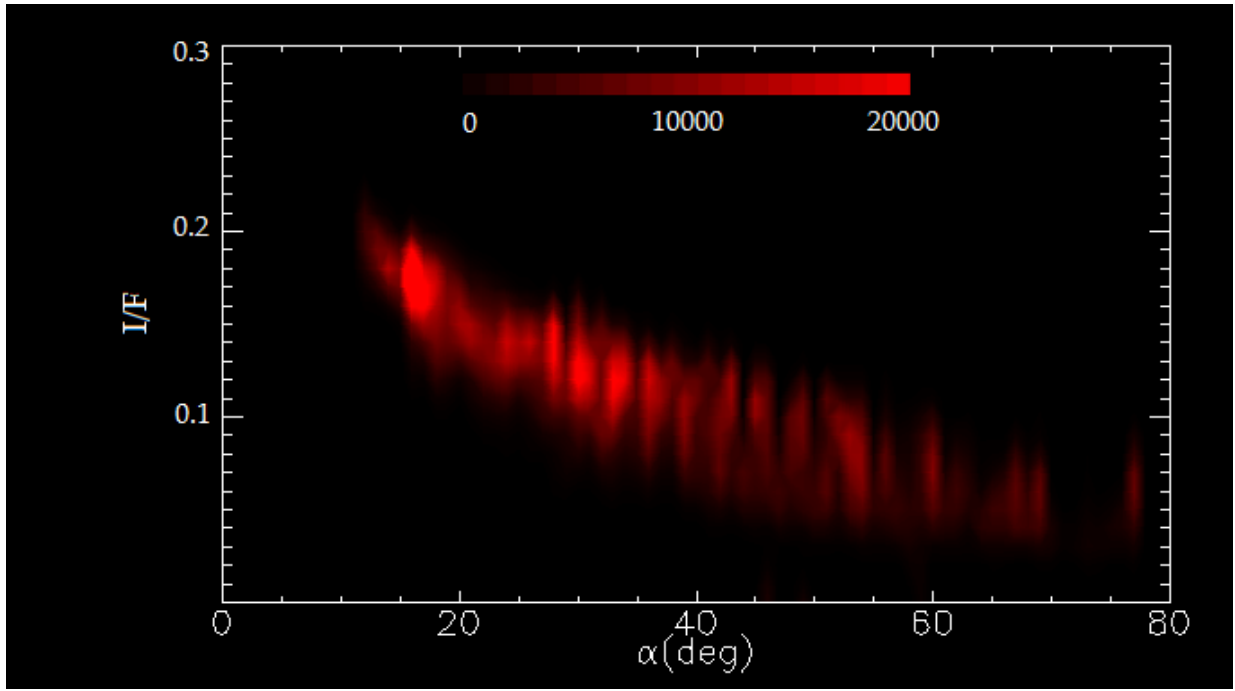

**Fig. 24 - Reflectance density plot, Survey data -** density plot of reflectance (y axis) vs phase angle (x axis) for the Survey phase a at 0.796 μm, after filtering of the data-set. It can be noted the general trend of reflectance and the coverage in phase angle.

*HAMO.*

The same approach adopted for the Survey observations has been used for the Hamo. The huge quantity of pixels (12372940), even greater than the Survey data-set, is useful to provide a better statistic, although, as shown below, the limited range of phase angles do not allow us to constrain a large part of the reflectance curve.

*Steps (i-ii)* Observations geometry properties are reported in **Table 6** for a direct comparison with the Survey data in **Table 5**. The main differences between the two data-sets, are represented by smaller phase angles range in the case of HAMO (≈40° with respect to the ≈65° of the Survey) and a better coverage in latitude (≈70° in the Northern Hemisphere respect to the ≈45° of the Survey).

| Table 6 HAMO observation geometries and surface coverage | | |
|---|---|---|
| | MIN | MAX |
| PHASE ANGLE | 27.87° | 66.74° |
| INCIDENCE ANGLE | 0.08° | 121.33° |
| EMISSION ANGLE | 0.01° | 89.56° |
| LATITUDE ANGLE | -89.93° | 70.75° |
| LONGITUDE ANGLE | 0.00° | 360.00° |

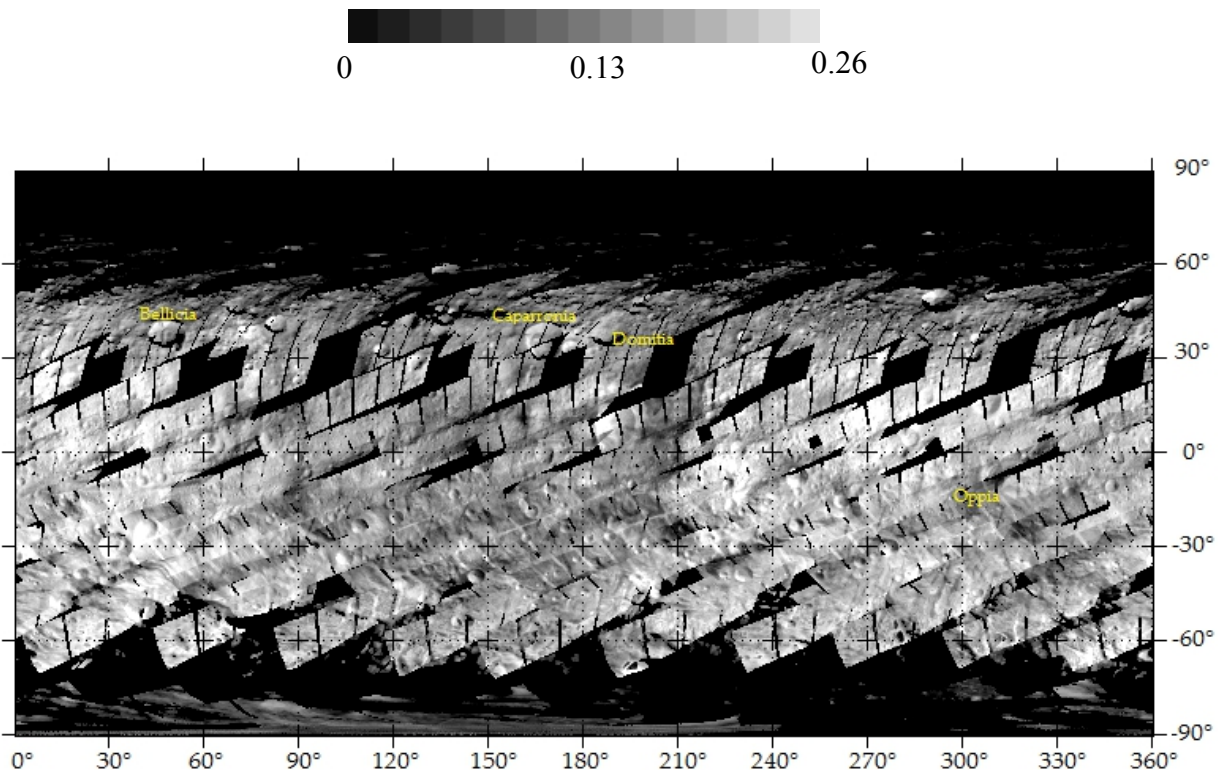

**Fig. 25 - Reflectance map from HAMO data –** reflectance map from Hamo observations before photometric correction, at 0.796 μm. Values on the x axis are the longitudes of Vesta, while on the y axis latitudes are reported. A larger number of small scales structures are visible with respect to the Survey phase, thanks to the higher resolution.

In **Fig. 25** we report the reflectance map from Hamo data at 0.796 μm, before photometric correction. With respect to the Survey case (**Fig. 20**), the overlapping areas between cubes seems to be more coherent, possibly because of the smaller range of phase angles spanned by the observations.

In **Fig. 26** the redundancy for the Hamo phase is shown; the scale in the color bar is different from the one used for the Survey case (**Fig. 21**). As in the Survey, there is a trend in pixel distribution with the latitude, while there is no obvious trend in pixel distribution with the longitude; the main difference is represented by the better coverage at the equatorial latitudes (-40°/15°) for the Hame phase.

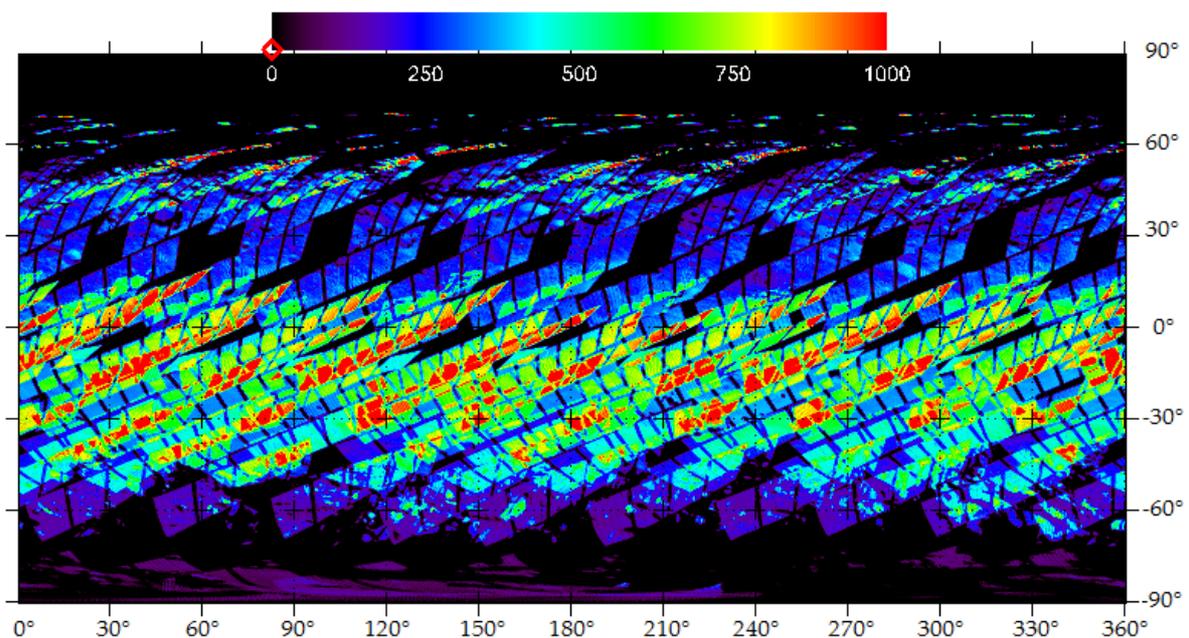

**Fig. 26 - Redundancy map from Hamo data** – longitudes and latitudes are indicated on the x and y axis respectively. Blue areas are less covered, yellow and red Blue areas are less covered, yellow and red areas are the most covered (the color bar indicates the number of occurrences inside a 0.5° x 0.5° bin).

*Step (iii)* For the HAMO case we adopt the same constraints on the reflectance, incidence and emission angles of the Survey phase:

$$0.001 < \text{reflectance} < 1.0$$

$$e < 70°, i < 70°$$

Assuming this selection criteria, approximately the 17% of the data has been eliminated. After these selections, the data-set has been analyzed again and compared to the previous, unfiltered one:

- in **Fig. 27** it is shown a scatter plot of reflectance as a function of the phase angle (details in the caption);

- in **Fig. 28** we report histograms of phase angle distributions before (panel **a**) and after (panel **b**) the selection (details in the caption);

- in **Fig. 29** a density plot of reflectance as a function of the phase angle is shown (details in the caption).

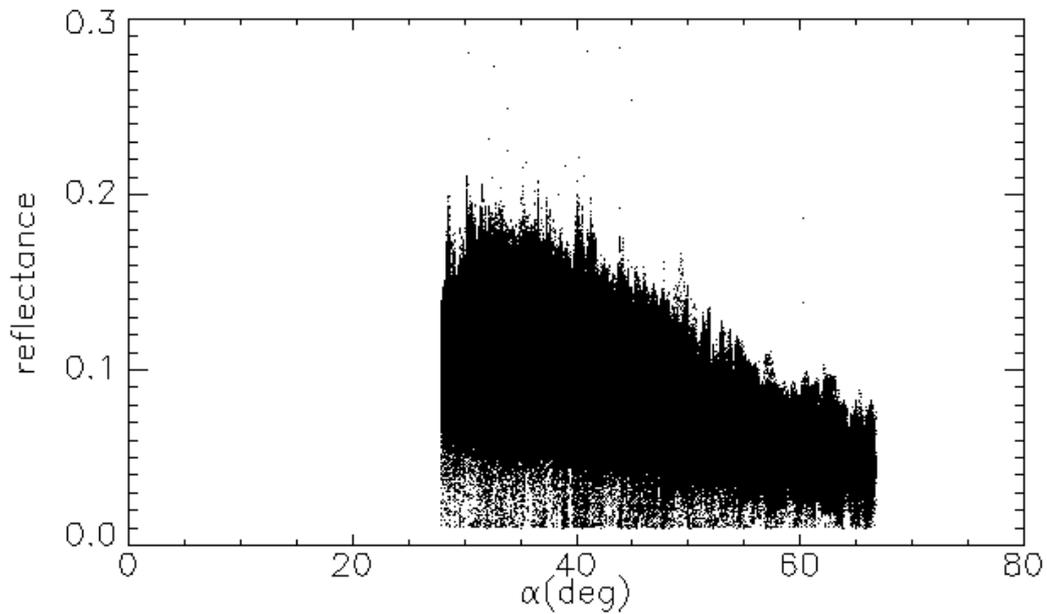

**Fig. 27 - Reflectance scatter plot from Hamo data** – reflectance of each pixel (y axis) vs phase angle (x axis) for the Hamo after data selection at 0.796 μm. The distribution is much more condensed than the Survey, spanning over a smaller phase angle range (27°-66°).

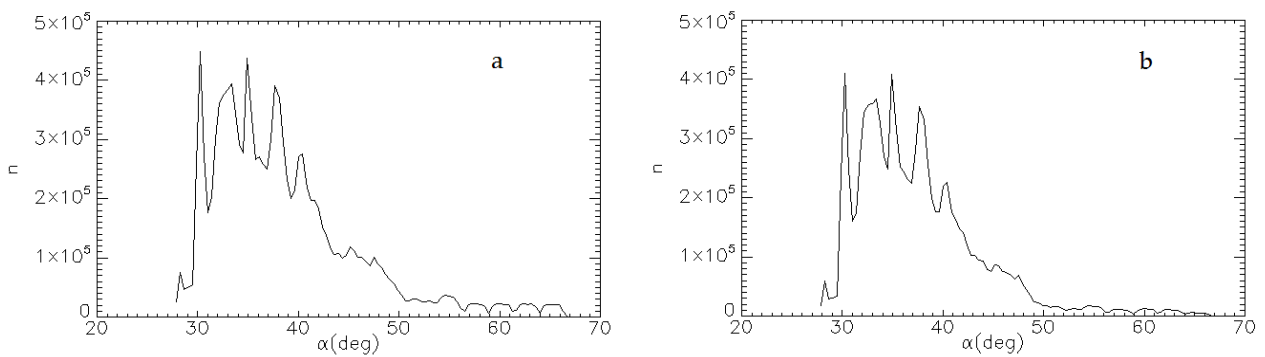

**Fig. 28 - Number of observations as a function of the phase angle, Hamo -** Observations phase angle distributions before **(a)** and after **(b)** data selection. The ≈17% filtered out data partially affect the general distribution at lower phase angles, but the cut is much more important over 50°, where it affects about a 50% of the data.

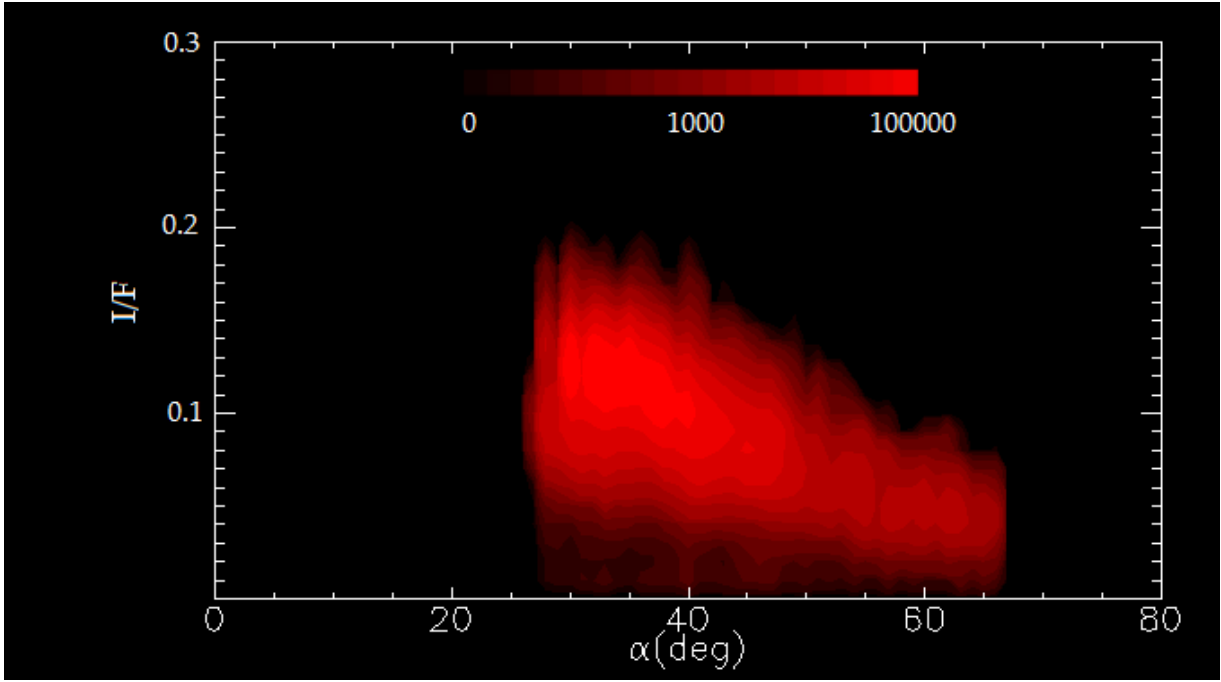

**Fig. 29 – Reflectance density plot Hamo -** density plot of reflectance (y axis) vs phase angle (x axis) for the Hamo phase at 0.796 µm, after filtering of the data-set. It can be noted the general trend of reflectance and the coverage in phase angle. Note that the color scale is non-linear.

The reduction of the number of pixels at large phase angles after the filtering (**Fig. 28**) is probably related, like in the Survey, to the constraints imposed on the incidence and emission angles and the reflectance. Both **Fig. 27** and **Fig. 29** show a monotonic trend of the reflectance with the phase angle.

## 4.3 Spectral wavelength selection

The total VIR spectral range covers two intervals (0.25 – 1.05 μm and 1 – 5 μm) spanning from the Visible to near Infrared (see **section 2.3**), with a total of 864 spectral channel. A study of Vesta spectrophotometric properties in each spectral channel would be impractical from a computational time point of view. For this reason, we performed our analysis on a selected set of the wavelengths, which yet provide a satisfactory description of Vesta spectral properties. In **Fig. 30** we report the overall spectrum of Vesta from VIR data (average reflectance as a function of wavelengths).

VIR spectrum of Vesta's surface shows absorption bands centered at 0.9 and 1.9 μm, interpreted as due to the presence of ironbearing low-calcium pyroxenes (*De Sanctis et al. 2012*). Located approximately at 0.7 μm, 1.4 μm, 2.5 μm, the order sorting filters on the VIR detectors can be recognized: their precise position is reported in **Table 7**. Above 3.5 µm, the thermal emission of the surface dominates the spectrum. Here, we will focus on the discussion of the spectral behaviors below 3.5 µm, which pertains to the reflected part of the radiance coming from Vesta's surface.

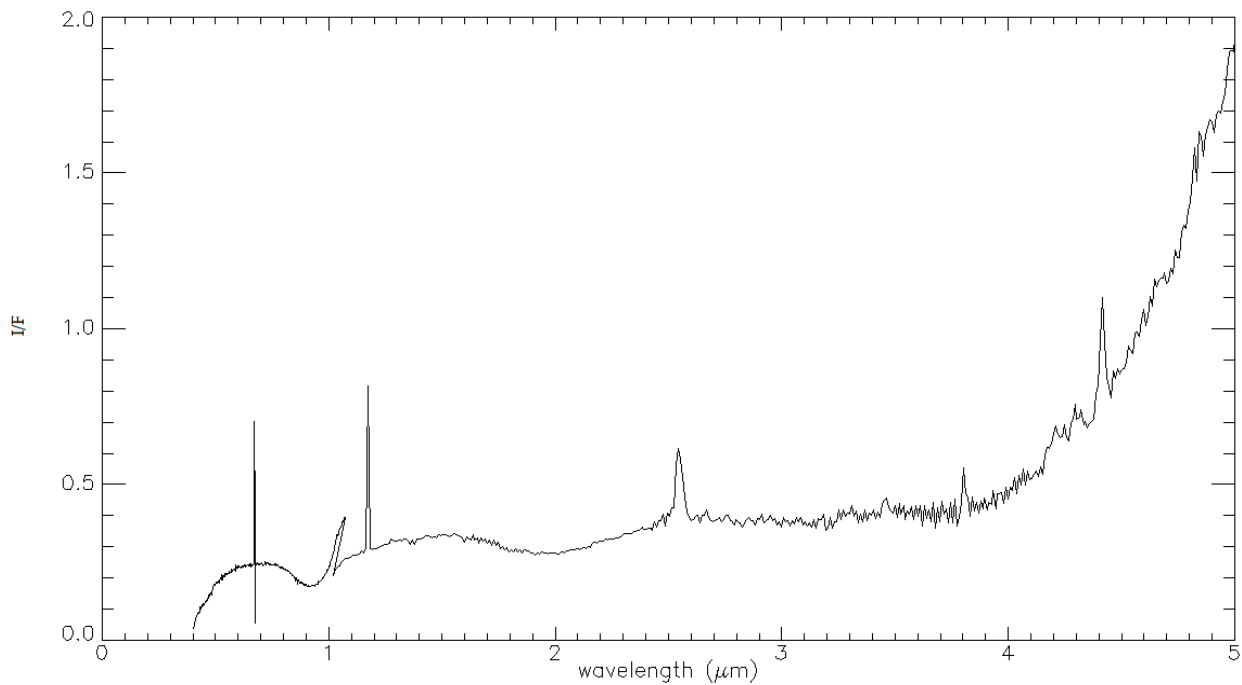

**Fig. 30 - Vesta Spectrum -** overall spectrum of Vesta as obtained by VIR. On the x axis there are the wavelengths, on the y the measured reflectance.

| **Table 7 – Filters positions –** positions of VIR filters in Visibile and Infrared regions |
|:---:|
| 0.662 – 0.687 [µm] - VIS |
| 1.418 – 1.560 [µm] - IR |
| 2.411 – 2.610 [µm] - IR |
| 3.736 – 3.883 [µm] - IR |
| 4.350 – 4.450 [µm] - IR |

The wavelenghts at which the analysis is performed, are selected with the aim to satisfy the following requirements:

- cover all of the Vesta's spectrum, from 0.2 to 3.5 µm, in order to study the behavior of our photometric parameters and intrinsic properties of the observed surface at different wavelengths;

- avoid the four filters listed in **Table 7**, in order to minimize instrumental artifacts;

- reproduce the general trend of the continuum and that of the main absorption features at 0.9 and 1.9 µm: for those bands we chose at least one point for both shoulders, one at the minimum, and one other for both the sides, between the shoulders and the minimum.

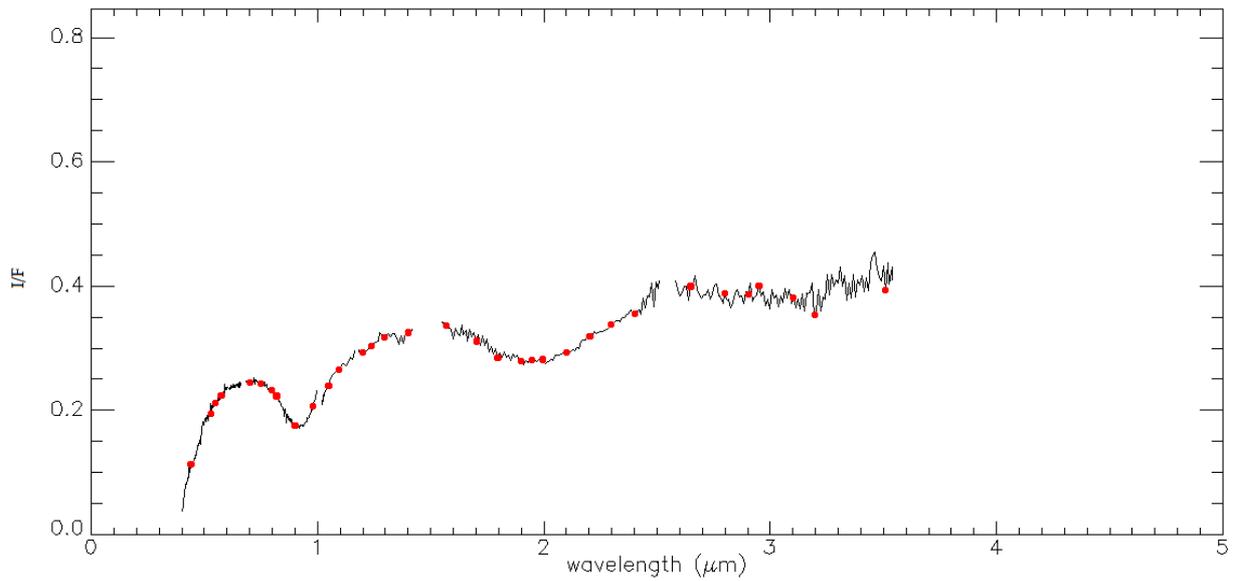

**Fig. 31 - Vesta Spectrum and selected wavelengths -** overall spectrum of Vesta as obtained by VIR.. Band filters are removed, as well as the thermal emission part; in red, the selected wavelengths. Longward of 2.5 µm, instrumental artifacts produce spurious spectral variability.

| Table 8 – selected wavelengths – | | |
|---|---|---|
| wavelengths in the first column span all over the visible range of VIR, from 0.441 µm to the first pyroxene absorption bad (BI); that in the second column goes from the first (BI) to the second (BII) absorption band; while in the third column they span from BII to the cut that we operated in the spectrum, corresponding to the thermal range (TR). | | |
| VIS - BI (µm) | BI -BII (µm) | BII – TR (µm) |
| 0.441 | 1.049 | 2.099 |
| 0.531 | 1.096 | 2.203 |
| 0.550 | 1.200 | 2.298 |
| 0.575 | 1.238 | 2.402 |
| 0.700 | 1.295 | 2.648 |
| 0.751 | 1.399 | 2.799 |
| 0.796 | 1.569 | 2.903 |
| 0.819 | 1.702 | 2.950 |
| 0.900 | 1.796 | 3.102 |
| 0.980 | 1.900 | 3.196 |
|  | 1.948 | 3.509 |
|  | 1.995 |  |

The wavelengths of interest are reported in **Table 8**, while in **Fig. 31**, they are superimposed (red dots) on the average Vesta's spectrum.

**4.4 The photometric parameters**

In this work we applied the Hapke's model (*Hapke, 2012*) (see **Section 3.4**) to perform the photometric correction of VIR selected data-set for all the chosen wavelengths. Here we briefly summarize the adopted formulation:

$$r(i,e,\alpha) = \frac{w(\lambda)}{4\pi} \frac{\mu_{0\text{eff}}}{\mu_{0\text{eff}} + \mu_{\text{eff}}} \left[ p(\alpha) B_{SH}(\alpha) + H(w, \mu_{0\text{eff}}) H(w, \mu_{\text{eff}}) - 1 \right] S(i,e,\alpha,\theta) \; ; \tag{60}$$

- w is the single-scattering albedo (SSA);

- $\mu_{0\text{eff}}$ and $\mu_{\text{eff}}$ are the effective cosines of the incidence and emission angles;

- $p(\alpha) = \dfrac{1 - b^2}{(1 + 2b\cos\alpha + b^2)^{3/2}}$ is the one-parameter formulation for the phase function, where b ranges in the interval [0;1];

- $B_{SH}(\alpha) = \dfrac{B_0}{1 + 1/h \tan(\alpha/2)}$ is the Shadow Hiding Opposition Effect term, depending on the opposition surge angular width h and amplitude $B_0$.

- H(w,x) is the Chandrasekhar's function for multiple-scattering contribution;

- $S(i,e,\alpha,\theta)$ is the Shadowing function depending on $\theta$, the average surface roughness slope.

For each wavelength, a fit procedure gives a best-set of parameters *(w, $B_0$, h, b, θ)* which reproduces the surface photometric behavior. We used the MPFIT package found in the Markwardt IDL library (*Markwardt, 2008*) to search for the best-fit parameter set that minimizes the $\chi^2$, defined as:

$$\chi^2 = \sum \left[ r(i,e,\alpha) - r_{measured} \right]^2 , \qquad (61)$$

where r(i,e,α) is the modeled I/F and $r_{measured}$ is the observed I/F; the sum is obtained over all data points. Below, the parameters obtained for both Survey and Hamo for the chosen wavelengths, are presented (**Table 9** and **Table 10**) and discussed.

As stated above (**Section 4.2**), the lack of data points at small phase angles (<2°) prevented us to model the CBOE. Consequently, the $B_{CB}$ term of the Hapke's model has been assumed equal to 1, since its effect at larger phase angles can be neglected; similarly, with the given phase angle coverage, also a proper characterization of SHOE is not possible. However, SHOE can partially affect the phase curve up to 20°-30°: for this reason in our analysis we assume the value of the opposition effect amplitude derived by *Helfenstein* and *Veverka, 1989*, $B_0$=1.03, letting the angular width *h*, free to vary.

| Table 9 – Hapke parameters, Survey – $B_0$=1.03 | | | | | |
|---|---|---|---|---|---|
| λ (μm) | w | $B_0$ | h | b | θ (°) |
| 0.441 | 0.319 | 1.030 | 0.065 | -0.236 | 23.510 |
| 0.531 | 0.489 | 1.030 | 0.074 | -0.243 | 25.373 |
| 0.550 | 0.517 | 1.030 | 0.073 | -0.249 | 24.216 |
| 0.575 | 0.538 | 1.030 | 0.075 | -0.248 | 24.893 |
| 0.700 | 0.573 | 1.030 | 0.074 | -0.250 | 24.722 |
| 0.751 | 0.568 | 1.030 | 0.076 | -0.251 | 24.985 |
| 0.796 | 0.538 | 1.030 | 0.079 | -0.259 | 24.484 |
| 0.819 | 0.513 | 1.030 | 0.082 | -0.264 | 24.672 |
| 0.900 | 0.409 | 1.030 | 0.085 | -0.274 | 25.204 |
| 0.980 | 0.453 | 1.030 | 0.085 | -0.285 | 23.717 |
| 1.049 | 0.617 | 1.030 | 0.033 | -0.266 | 21.443 |
| 1.096 | 0.668 | 1.030 | 0.019 | -0.265 | 19.341 |
| 1.200 | 0.711 | 1.030 | 0.020 | -0.264 | 19.917 |
| 1.238 | 0.725 | 1.030 | 0.020 | -0.266 | 20.211 |
| 1.295 | 0.743 | 1.030 | 0.019 | -0.269 | 19.228 |
| 1.399 | 0.755 | 1.030 | 0.013 | -0.275 | 17.436 |
| 1.569 | 0.760 | 1.030 | 0.014 | -0.281 | 18.040 |
| 1.702 | 0.718 | 1.030 | 0.021 | -0.280 | 17.682 |
| 1.796 | 0.680 | 1.030 | 0.022 | -0.279 | 19.373 |
| 1.900 | 0.663 | 1.030 | 0.029 | -0.289 | 21.219 |
| 1.948 | 0.669 | 1.030 | 0.024 | -0.281 | 18.379 |
| 1.995 | 0.667 | 1.030 | 0.030 | -0.281 | 20.767 |
| 2.099 | 0.691 | 1.030 | 0.022 | -0.282 | 16.051 |
| 2.203 | 0.723 | 1.030 | 0.024 | -0.287 | 17.244 |
| 2.298 | 0.748 | 1.030 | 0.024 | -0.290 | 17.758 |
| 2.402 | 0.775 | 1.030 | 0.020 | -0.286 | 14.199 |
| 2.648 | 0.818 | 1.030 | 0.017 | -0.293 | 10.830 |
| 2.799 | 0.812 | 1.030 | 0.019 | -0.286 | 13.904 |
| 2.903 | 0.806 | 1.030 | 0.026 | -0.288 | 17.343 |
| 2.950 | 0.801 | 1.030 | 0.079 | -0.277 | 13.681 |
| 3.102 | 0.788 | 1.030 | 0.067 | -0.271 | 12.907 |
| 3.196 | 0.757 | 1.030 | 0.083 | -0.268 | 14.526 |
| 3.509 | 0.787 | 1.030 | 0.085 | -0.290 | 17.707 |

| Table 10 – Hapke parameters, Hamo – $B_0$ =1.03 | | | | | |
|---|---|---|---|---|---|
| λ (μm) | w | $B_0$ | h | b | θ (°) |
| 0.441 | 0.291 | 1.030 | 0.095 | -0.263 | 21.467 |
| 0.531 | 0.455 | 1.030 | 0.081 | -0.279 | 20.841 |
| 0.550 | 0.486 | 1.030 | 0.076 | -0.282 | 20.336 |
| 0.575 | 0.504 | 1.030 | 0.078 | -0.283 | 20.302 |
| 0.700 | 0.540 | 1.030 | 0.074 | -0.285 | 20.375 |
| 0.751 | 0.539 | 1.030 | 0.074 | -0.281 | 20.510 |
| 0.796 | 0.525 | 1.030 | 0.063 | -0.279 | 20.252 |
| 0.819 | 0.507 | 1.030 | 0.057 | -0.278 | 20.248 |
| 0.900 | 0.418 | 1.030 | 0.041 | -0.283 | 21.107 |
| 0.980 | 0.463 | 1.030 | 0.051 | -0.273 | 20.246 |
| 1.049 | 0.594 | 1.030 | 0.042 | -0.292 | 18.530 |
| 1.096 | 0.643 | 1.030 | 0.042 | -0.286 | 16.579 |
| 1.200 | 0.676 | 1.030 | 0.041 | -0.296 | 16,748 |
| 1.238 | 0.690 | 1.030 | 0.037 | -0.303 | 16.679 |
| 1.295 | 0.713 | 1.030 | 0.035 | -0.304 | 15.514 |
| 1.399 | 0.732 | 1.030 | 0.035 | -0.293 | 13.177 |
| 1.569 | 0.746 | 1.030 | 0.027 | -0.295 | 12.078 |
| 1.702 | 0.720 | 1.030 | 0.0001 | -0.320 | 13.317 |
| 1.796 | 0.683 | 1.030 | 0.0001 | -0.313 | 14.825 |
| 1.900 | 0.665 | 1.030 | 0.0001 | -0.320 | 16.696 |
| 1.948 | 0.678 | 1.030 | 0.0001 | -0.311 | 14.685 |
| 1.995 | 0.672 | 1.030 | 0.0001 | -0.318 | 16.125 |
| 2.099 | 0.703 | 1.030 | 0.0001 | -0.309 | 12.926 |
| 2.203 | 0.727 | 1.030 | 0.0001 | -0.323 | 13.784 |
| 2.298 | 0.747 | 1.030 | 0.0001 | -0.332 | 14.297 |
| 2.402 | 0.776 | 1.030 | 0.0003 | -0.327 | 11.335 |
| 2.648 | 0.810 | 1.030 | 0.031 | -0.306 | 9.747 |
| 2.799 | 0.800 | 1.030 | 0.029 | -0.308 | 10.556 |
| 2.903 | 0.788 | 1.030 | 0.011 | -0.332 | 12.976 |
| 2.950 | 0.795 | 1.030 | 0.063 | -0.284 | 10.177 |
| 3.102 | 0.759 | 1.030 | 0.096 | -0.257 | 10.803 |
| 3.196 | 0.751 | 1.030 | 0.048 | -0.273 | 11.805 |
| 3.509 | 0.795 | 1.030 | 0.0001 | -0.343 | 14.841 |

Plots in **Fig.32** show the fitted parameters as a function of the wavelengths:

- w (the SSA) shows a trend with the wavelength, very similar to the spectrum in **Fig. 30**: also the two signatures at 0.9 and 1.9 µm are clearly recognizable;

- h presents discontinuities at 1.0 µm and 3.0 µm, due to the inability of the fit procedure to univocally characterize this parameter because of the lack of data at small phase angles;

- θ shows a decreasing trend with wavelength (above 1.0 µm) and a certain degree of anti-correlation with w;

- b has a general decreasing trend too; the two signatures at 0.9 and 1.9 µm (especially the first) seems to affect the parameter.

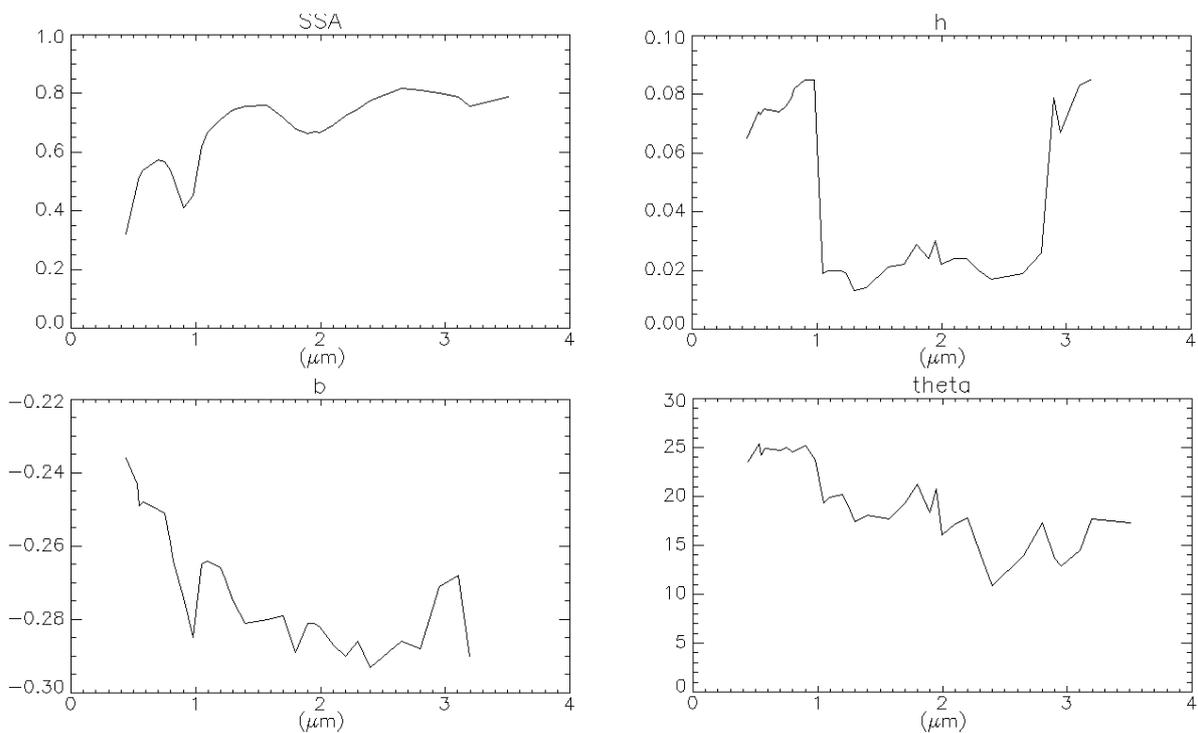

**Fig. 32 – Hapke's model parameters for Survey data –** SSA (top panel, left), h (top panel, right), b (bottom panel, left), θ (bottom panel, right); on the x axis the wavelengths. See the text for details.

In the Hamo case, the strong variation observed in the Survey for the h parameter around 1.0 μm is still visible (see **Fig. 33**). h and b show some similarities: they both have a general decreasing trend with the wavelength, and an abrupt increase at 3.0 μm, suggesting a possible degeneration between the two parameters. Both θ and the SSA maintain the general trend of the Survey case, showing no significant differences.

A direct comparison of the parameters obtained for the two different data-sets is shown in **Fig. 34**.

The derived w spectrum is very similar in the two cases. b and θ, respectively, exhibit the same spectral behavior in Survey and Hamo, with the Survey-parameters slightly larger than the Hamo-ones. h shows in both cases an abrupt variation at 1.0 μm and at 3.0 μm; it is also the parameter which shows more discrepancies between the two cases, because of the incapability to give a proper description of the Opposition Effect within the phase angle range of the two data-sets.

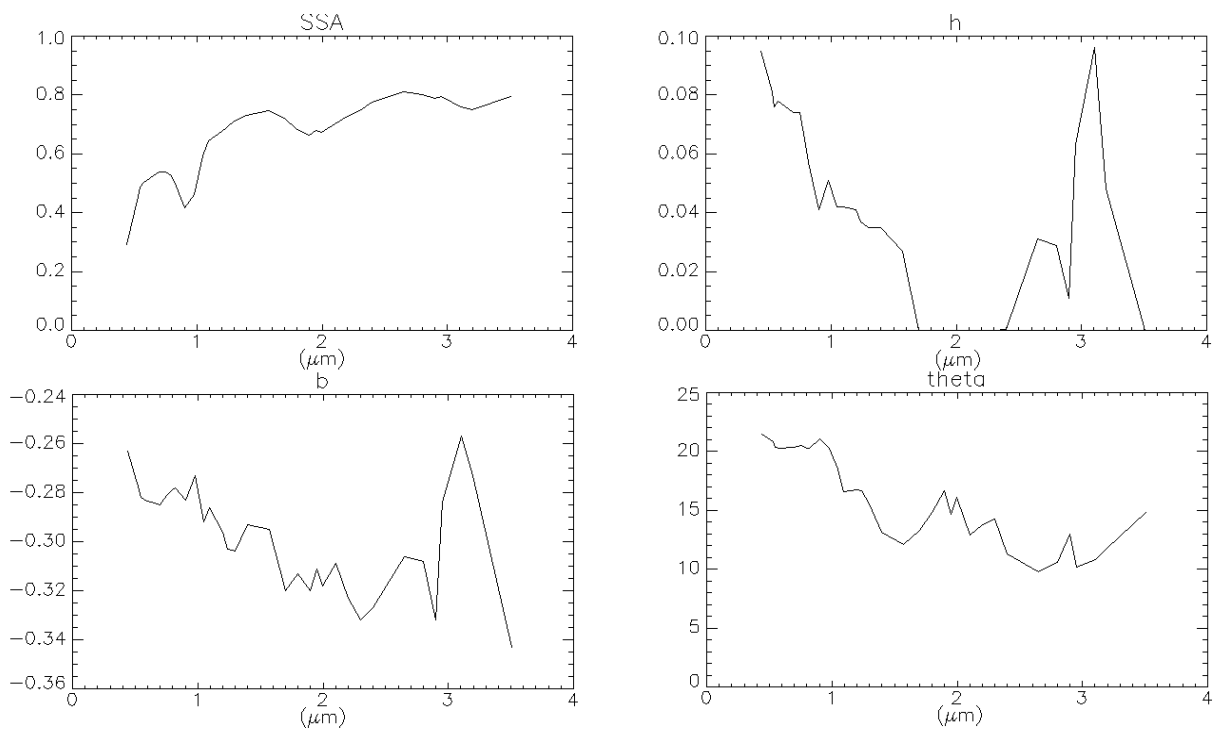

**Fig. 33 - Hapke's model parameters for Hamo data** – SSA (top panel, left), h (top panel, right), b (bottom panel, left), θ (bottom panel, right); on the x axis the wavelengths. See the text for details

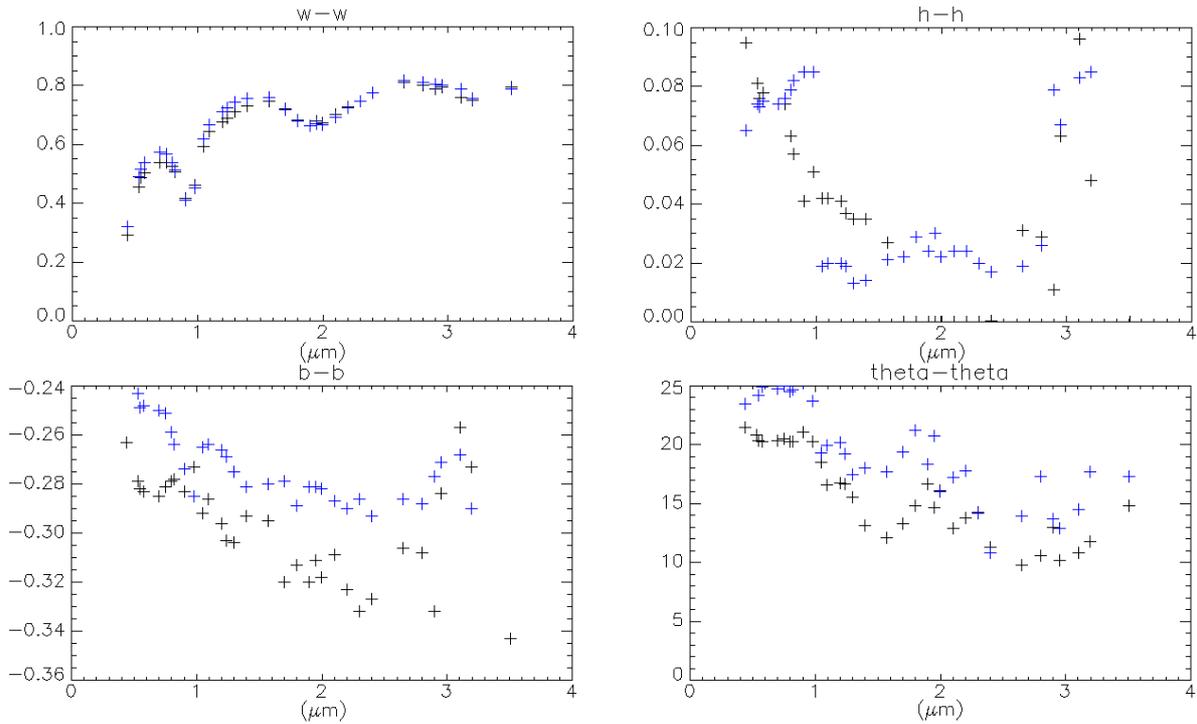

**Fig. 34 - Hamo parameters vs Survey parameters –** Parameters obtained from Hamo (black) are plotted with the parameters obtained from Survey (blue). On the x axis the wavelengths. See the text for details.

*4.4a Surface roughness*

As described in **Chapter 3,** the value of the θ parameter is related to the large scale roughness of the target, representing the average slope of the structures on its surface. Given this, θ is assumed to have no spectral dependence, being related to terrain morphology at scales much larger than the wavelength (*Hapke, 2012*). However, results described in **Section 4.4** indicate a residual dependence of θ with the wavelength, either due to degeneration with the others fitted parameter, or to the presence of inter-facets scattering. The latter is due to light scattered multiple times between the different elements ("facets") of the surface reducing the shadows produced by roughness. Assuming the former case, a typical approach in this kind of analysis is to force the θ parameter to be independent on wavelengths. This can be done by fixing the θ values to the average obtained for the Survey and the Hamo

$$\theta_{SURVEY}=20.2\pm3.4 ,$$

$$\theta_{HAMO}=17.2\pm2.9 ,$$

and repeating the same fitting procedure as before for the free parameters. Alternatively, this decreasing trend is compatible with the fact that w is larger in the IR range, allowing for a higher contribution of inter-facets scattering and with the observed anti-correlation between w and θ. Assuming that the spectral dependence of θ in our study is real (see **Fig. 32** and **33**) and related to inter-facets it is possible to remove the dependence of the large scale roughness from the albedo. It can be done estimating the true value of θ as

$$\theta=(1-r_0)\theta_P , \qquad (62)$$

where $\theta_P$ is the true roughness and θ is the one derived through our photometric correction; $r_0$ is the diffusive reflectance.

In **Fig. 35** and **36**, it is shown that the trend of the true values of θ (green points) is more flat with respect to the modeled ones (black points), being the maximal fluctuation of the new values of θ ≈30%, with respect to the ≈50% of the roughness directly obtained by the model. However, it is still possible to recognize a trend with the albedo. We can infer that either **eq. 62** is unable to remove the dependence from the albedo, or that this can't be fully explained through inter-facets scattering.

In *Li et al., (2013)*, the decreasing trend of **Fig. 32** and **33**, was not observed: it was supposed that either the multiple scattering on Vesta is not strong enough, or the roughness on the surface is caused by topographic shadows at scales much larger than the wavelengths of the FC data. We confirm that this trend is not recognizable in the VIS range, and it becomes clear only in the IR region, inferring for the presence of shadows at scales greater than 1.0 μm.

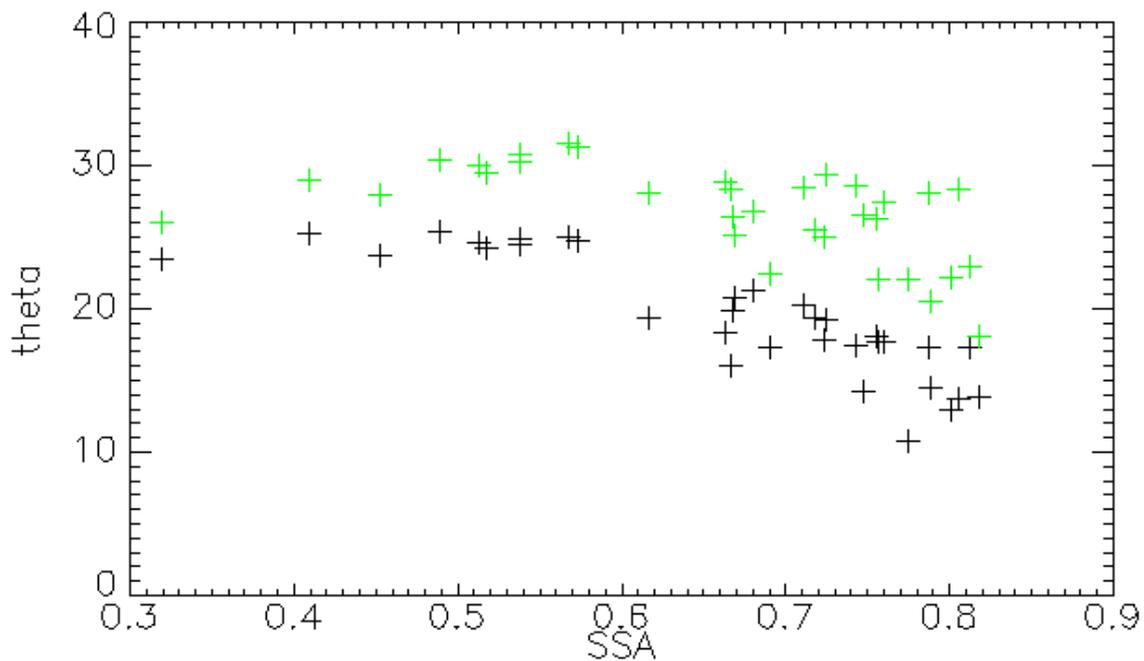

**Fig. 35 - comparison of the true roughness with respect to the measured one, Survey -** on the x axis there is the single scattering albedo, on the y axis the large scale roughness. Black points correspond to the measured values of roughness, while the green points correspond to the true roughness estimated for each of the wavelengths.

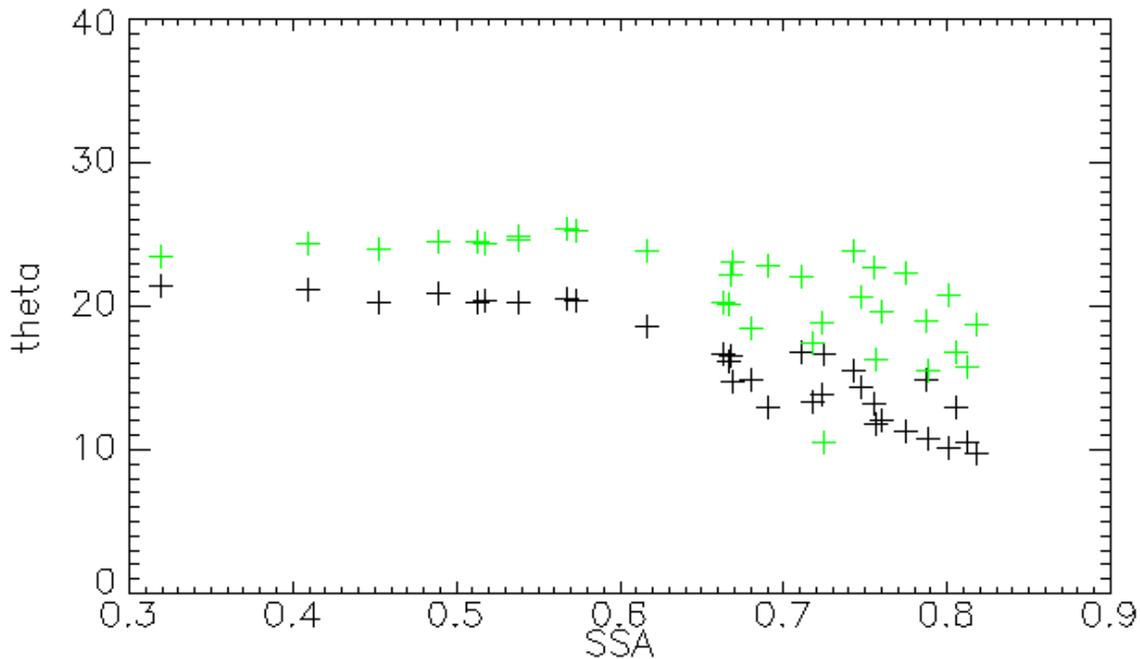

**Fig. 36 - comparison of the true roughness with respect to the measured one, HAMO -** on the x axis there is the single scattering albedo, on the y axis the large scale roughness. Black points correspond to the measured values of roughness, while the green points correspond to the true roughness estimated for each of the wavelengths.

*4.4b SSA*

As inferred by discussions of **Fig. 32** and **33**, w is representative of the average spectral properties of the surface of Vesta and it is linked to its composition. An increasing reflectance corresponds to an increasing SSA and viceversa. It ranges from ≈ 0.3 to ≈ 0.8, with a typical discrepancy $\frac{\Delta w}{w}$ of about 1% between θ-free and θ-fixed parameters (for both Survey and HAMO) and of 3.3% between Survey and HAMO.

*4.4c The phase function*

The adopted single particle phase function is given by the Henyey-Greenstein formulation (*Henyey and Greenstein, 1941*) depending on one parameter b. The general decreasing trend of b with wavelengths shown for our two sets of data (**Fig. 32** and **33**) indicates that longer wavelengths correspond to a more back-scattering phase function; from about 2.5 µm to 3 µm this general trend is reversed. As an examples, in **Fig. 37** the phase function curves for the Survey sequence at four different wavelengths are compared; different colors are used for each of these wavelengths.

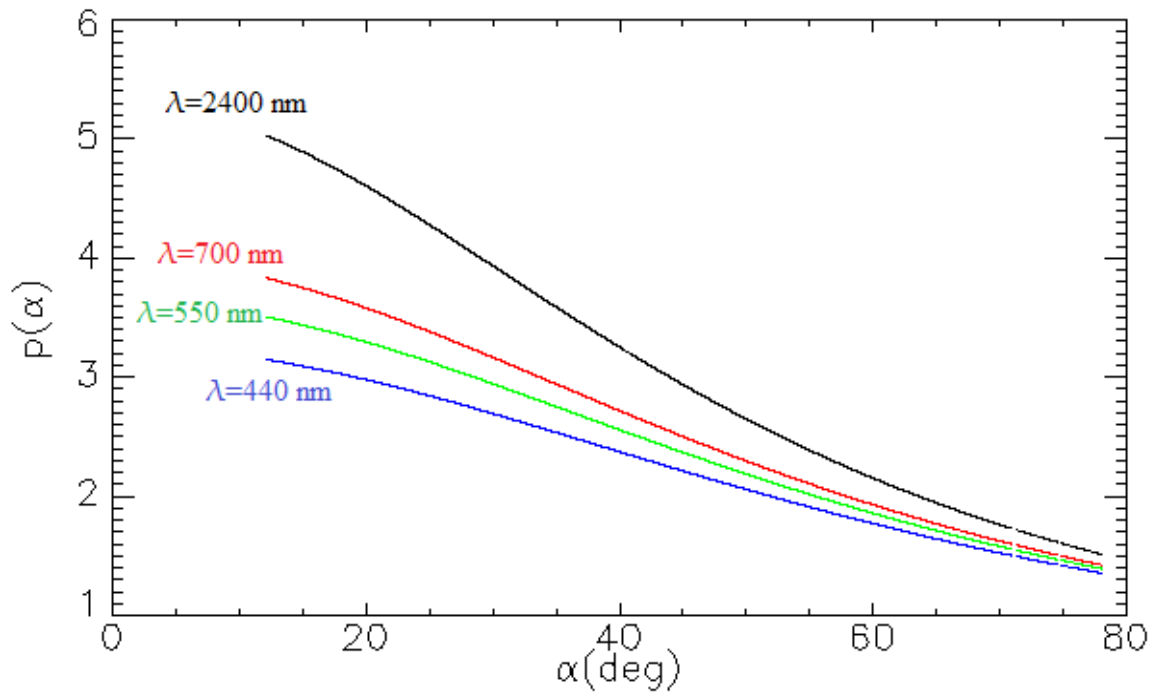

**Fig. 37 - p(α) vs phase angle, Survey -** The colored curves are the phase function at: 440 nm (blue), 550 nm (green), 800 nm (red) and 2400 nm (black) for the Survey sequence; respectively, b is -0.236, -0.249, -0.259 and -0.286.

*4.4d The Shadow Hiding Opposition Term*

The amplitude $B_0$ (fixed) and the width h, cannot be well constrained due to the lack of data at low phase angles. In the analysis shown above, for both Survey and HAMO, the h parameter appears to be poorly constrained and its dependence from the wavelength is mainly driven by fit degeneration with other parameters.

The h trend derived for the Survey case is similar to that obtained by *Li et al., (2013)* in the visible range (**Fig. 38**), with a constant profile below 750 nm – with the exception of 438 nm – and an increasing trend at longer wavelengths.

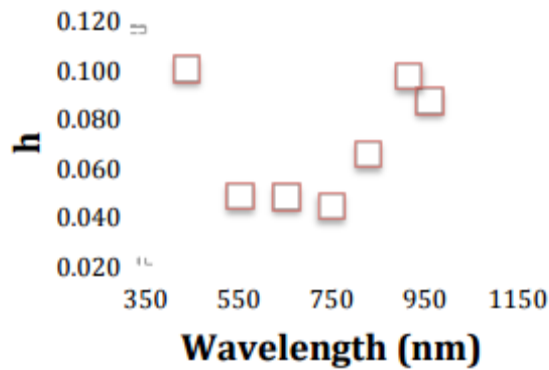

**Fig. 38** - h parameters obtained by *Li et al. 2013* (adapted from *Li et al., 2013*).

In their work, *Li et al., (2013)* decided to fix the h-parameter to 0.07, averaging the values obtained over different wavelengths. Here we adopt the same strategy, which is also compatible with the fact that the h parameter should not depend on the wavelength, being the SHOE angular width only related to porosity of the regolith. For a further comparison with *Li et al., (2013)*, we also changed the value of $B_0$ to 1.7, as derived from their analysis. In **Tables 11** and **12**, the obtained parameters are reported.

In general, fixing two parameters makes more variable the others, because all the fluctuations (previously described by four parameters) are now characterized only through these three parameters. In **Fig.39**, the parameters obtained for a fixed h ("h-fixed") and a free h ("h-free"), are over-plotted. The results of this approach can be summarized in the points below:

- w shows no significant discrepancies with the previous analysis and it is not affected by new values of h and $B_0$: this indicates that the degeneration between the Opposition Effect parameters and w is limited;

- the trend of θ with wavelength in **Fig. 39** is qualitatively similar to the one derived in the previous analysis; the new values are 5% higher in the Visible range, and 10% in the Infrared. The θ values in **Table 11** for the wavelengths 2.648 μm and 2.799 μm are artifacts of the fit procedure; the null values obtained, are imposed as a limit inside our procedure, to avoid negative and physically meaningless surface roughness slopes;

- the b parameter is the more affected by this choice of the SHOE parameters: its trend with wavelength appears to be coupled with w. It may be the hint for a direct physical connection between b and w; in this case, at increasing values of albedo, we would have a larger forward scattering mechanism, because the particles would be more transparent to light. It may be also due to a compensation effect between parameters, due to our new choice of h.

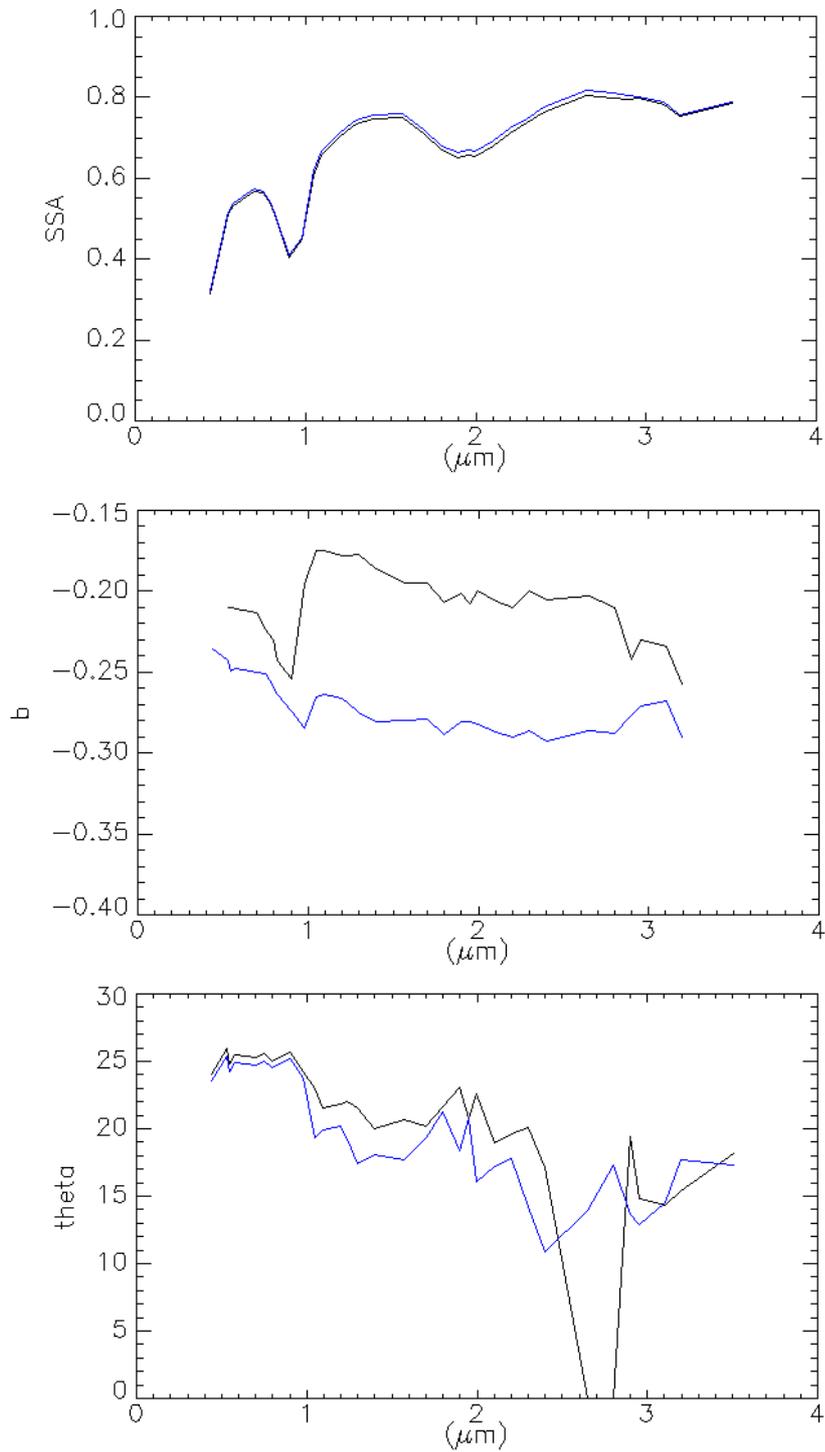

**Fig. 39 - Comparison of h-fixed parameters vs h-free parameters, Survey -** h-fixed parameters (black lines) are overplotted to the h-free ones (blue lines). From the top to the bottom: SSA, b and θ.

| Table 11 – Hapke parameters, Survey – $B_0$ =1.7, h =0.07 ||||||
|---|---|---|---|---|---|
| λ (μm) | w | $B_0$ | h | b | θ (°) |
| 0.441 | 0.313 | 1.700 | 0.070 | -0.192 | 24.080 |
| 0.531 | 0.485 | 1.700 | 0.070 | -0.204 | 25.926 |
| 0.550 | 0.512 | 1.700 | 0.070 | -0.210 | 24.793 |
| 0.575 | 0.533 | 1.700 | 0.070 | -0.210 | 25.435 |
| 0.700 | 0.568 | 1.700 | 0.070 | -0.211 | 25.304 |
| 0.751 | 0.563 | 1.700 | 0.070 | -0.213 | 25.575 |
| 0.796 | 0.534 | 1.700 | 0.070 | -0.224 | 25.012 |
| 0.819 | 0.509 | 1.700 | 0.070 | -0.230 | 25.209 |
| 0.900 | 0.405 | 1.700 | 0.070 | -0.243 | 25.651 |
| 0.980 | 0.449 | 1.700 | 0.070 | -0.254 | 24.248 |
| 1.049 | 0.605 | 1.700 | 0.070 | -0.196 | 22.941 |
| 1.096 | 0.658 | 1.700 | 0.070 | -0.175 | 21.541 |
| 1.200 | 0.702 | 1.700 | 0.070 | -0.175 | 21.867 |
| 1.238 | 0.716 | 1.700 | 0.070 | -0.178 | 22.052 |
| 1.295 | 0.735 | 1.700 | 0.070 | -0.178 | 21.570 |
| 1.399 | 0.746 | 1.700 | 0.070 | -0.177 | 19.975 |
| 1.569 | 0.751 | 1.700 | 0.070 | -0.186 | 20.679 |
| 1.702 | 0.707 | 1.700 | 0.070 | -0.195 | 20.176 |
| 1.796 | 0.669 | 1.700 | 0.070 | -0.195 | 21.656 |
| 1.900 | 0.651 | 1.700 | 0.070 | -0.207 | 23.111 |
| 1.948 | 0.656 | 1.700 | 0.070 | -0.201 | 20.778 |
| 1.995 | 0.655 | 1.700 | 0.070 | -0.208 | 22.569 |
| 2.099 | 0.679 | 1.700 | 0.070 | -0.200 | 18.911 |
| 2.203 | 0.712 | 1.700 | 0.070 | -0.206 | 19.650 |
| 2.298 | 0.737 | 1.700 | 0.070 | -0.210 | 20.051 |
| 2.402 | 0.764 | 1.700 | 0.070 | -0.200 | 17.140 |
| 2.648 | 0.805 | 1.700 | 0.070 | -0.205 | 0.000 |
| 2.799 | 0.799 | 1.700 | 0.070 | -0.203 | 0.000 |
| 2.903 | 0.796 | 1.700 | 0.070 | -0.210 | 19.423 |
| 2.950 | 0.797 | 1.700 | 0.070 | -0.242 | 14.802 |
| 3.102 | 0.783 | 1.700 | 0.070 | -0.230 | 14.332 |
| 3.196 | 0.754 | 1.700 | 0.070 | -0.234 | 15.404 |
| 3.509 | 0.784 | 1.700 | 0.070 | -0.257 | 18.213 |

| Table 12 – Hapke parameters, Hamo – $B_0$ =1.7, h =0.07 | | | | | |
|---|---|---|---|---|---|
| λ (μm) | w | $B_0$ | h | b | θ (°) |
| 0.441 | 0.287 | 1.700 | 0.070 | -0.244 | 21.598 |
| 0.531 | 0.446 | 1.700 | 0.070 | -0.255 | 21.099 |
| 0.550 | 0.476 | 1.700 | 0.070 | -0.255 | 20.667 |
| 0.575 | 0.494 | 1.700 | 0.070 | -0.257 | 20.651 |
| 0.700 | 0.529 | 1.700 | 0.070 | -0.257 | 20.772 |
| 0.751 | 0.528 | 1.700 | 0.070 | -0.253 | 20.893 |
| 0.796 | 0.511 | 1.700 | 0.070 | -0.245 | 20.682 |
| 0.819 | 0.492 | 1.700 | 0.070 | -0.241 | 20.699 |
| 0.900 | 0.399 | 1.700 | 0.070 | -0.238 | 21.602 |
| 0.980 | 0.448 | 1.700 | 0.070 | -0.233 | 20.705 |
| 1.049 | 0.575 | 1.700 | 0.070 | -0.245 | 19.364 |
| 1.096 | 0.625 | 1.700 | 0.070 | -0.239 | 17.644 |
| 1.200 | 0.658 | 1.700 | 0.070 | -0.247 | 17.853 |
| 1.238 | 0.671 | 1.700 | 0.070 | -0.253 | 17.942 |
| 1.295 | 0.694 | 1.700 | 0.070 | -0.251 | 16.940 |
| 1.399 | 0.715 | 1.700 | 0.070 | -0.239 | 14.849 |
| 1.569 | 0.729 | 1.700 | 0.070 | -0.233 | 14.041 |
| 1.702 | 0.695 | 1.700 | 0.070 | -0.230 | 15.490 |
| 1.796 | 0.658 | 1.700 | 0.070 | -0.224 | 16.629 |
| 1.900 | 0.637 | 1.700 | 0.070 | -0.233 | 18.366 |
| 1.948 | 0.652 | 1.700 | 0.070 | -0.221 | 16.424 |
| 1.995 | 0.644 | 1.700 | 0.070 | -0.230 | 17.845 |
| 2.099 | 0.679 | 1.700 | 0.070 | -0.218 | 15.003 |
| 2.203 | 0.702 | 1.700 | 0.070 | -0.234 | 15.970 |
| 2.298 | 0.721 | 1.700 | 0.070 | -0.243 | 16.507 |
| 2.402 | 0.753 | 1.700 | 0.070 | -0.237 | 14.154 |
| 2.648 | 0.795 | 1.700 | 0.070 | -0.246 | 12.355 |
| 2.799 | 0.784 | 1.700 | 0.070 | -0.246 | 12.913 |
| 2.903 | 0.767 | 1.700 | 0.070 | -0.254 | 15.354 |
| 2.950 | 0.786 | 1.700 | 0.070 | -0.247 | 11.662 |
| 3.102 | 0.756 | 1.700 | 0.070 | -0.236 | 11.235 |
| 3.196 | 0.739 | 1.700 | 0.070 | -0.227 | 13.185 |
| 3.509 | 0.772 | 1.700 | 0.070 | -0.254 | 17.330 |

## 4.5 The photometric corrected dataset

The whole data-set for both Survey and Hamo is photometrically corrected using separately the derived sets of parameters ("θ-free" and "h-free"). This is done by obtaining the photometrically corrected radiance factor, which is the measured I/F reported to standard geometry (i = 30°, e = 0°, g = 30°) by means of the relation:

$$I/F = \frac{I/F_{mis}(i,e,\alpha)}{I/F_{mod}(i,e,\alpha)} I/F_{mod}(30°, 0°, 30°) . \qquad (63)$$

The effect of the ratio $\frac{I/F_{mis}}{I/F_{mod}}$ (measured values over modeled ones) is to correct the reflectance distribution of **Fig. 24** and **Fig. 29**, removing the dependence on incidence, emission and phase angle, while I/F $_{mod}$ (30°, 0°, 30°) is the modeled radiance at typical laboratory observing condition.

In **Fig. 40** the result of the photometric correction over the Survey data set is shown. A direct comparison with **Fig. 24** allows us to recognize the same density distribution in the phase-angle vs reflectance plane with removed dependence on phase angle.

A linear fit over the distribution in **Fig. 40** (the blue line) is compared to the same obtained on I/F$_{meas}$:

- corrected data-set → (I/F=0.20-$\alpha$*8.88*E-5),
- uncorrected data-set → (I/F=0.26-$\alpha$*2.81*E-3).

It can be noted how the angular coefficient changes of about two orders of magnitudes.

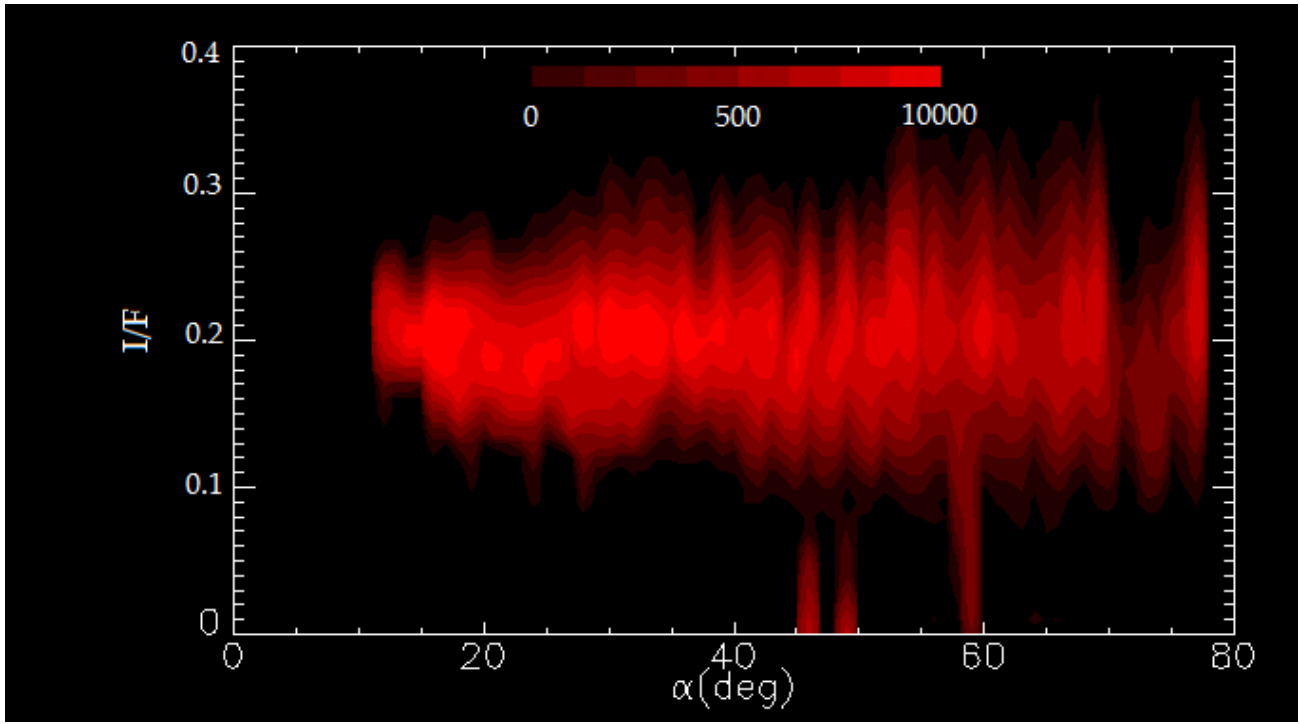

**Fig. 40 – correct I/F for Survey data –** density plot for I/F at 0.796 μm after the correction, for the entire Survey data-set. Brighter regions are more populated than the darker ones. Note that color scale is non-linear.

The histogram in **Fig. 41,** represents the distribution of the photometrically corrected reflectance for the Survey sequence. The presence of a shoulder in the histogram, allow us to infer the presence of two populations associated to different terrains: the brighter peaked at 0.21, the darker at 0.18. The brightest population is associated to a wider area of Vesta.

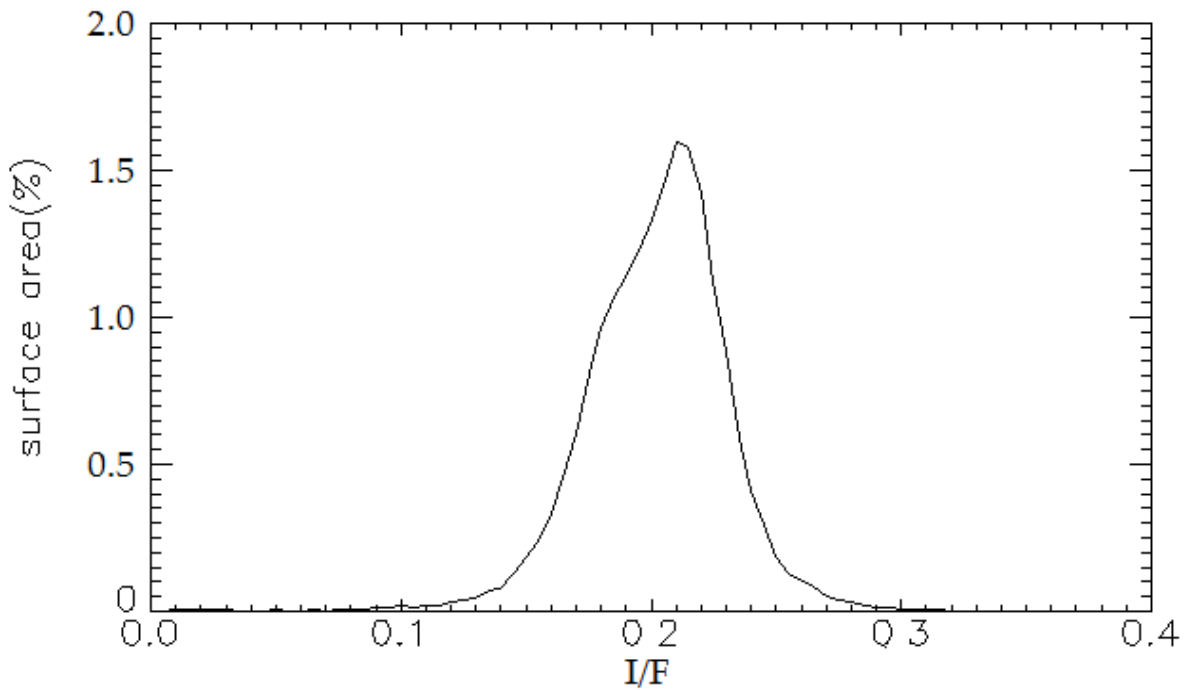

**Fig. 41 - Histogram of I/F after correction at 796 nm, Survey** – on the y axis it is reported the surface area (expressed in %).

In **Fig. 42** and **43** the distributions of the corrected reflectance as a function of incidence and emission angles are shown, respectively. In the first plot no occurrences are visible over ~ 60° degrees due to the adopted observation strategy, while the second one shows an increased scattering of points over the limit of selection of 70° degrees, due to a higher S/N ratio for these illumination conditions and to the presence of permanent shadows. In both cases the correction seems to act in the same manner, flattening the distribution and minimizing the dependence of the reflectance with respect to incidence and emission angles.

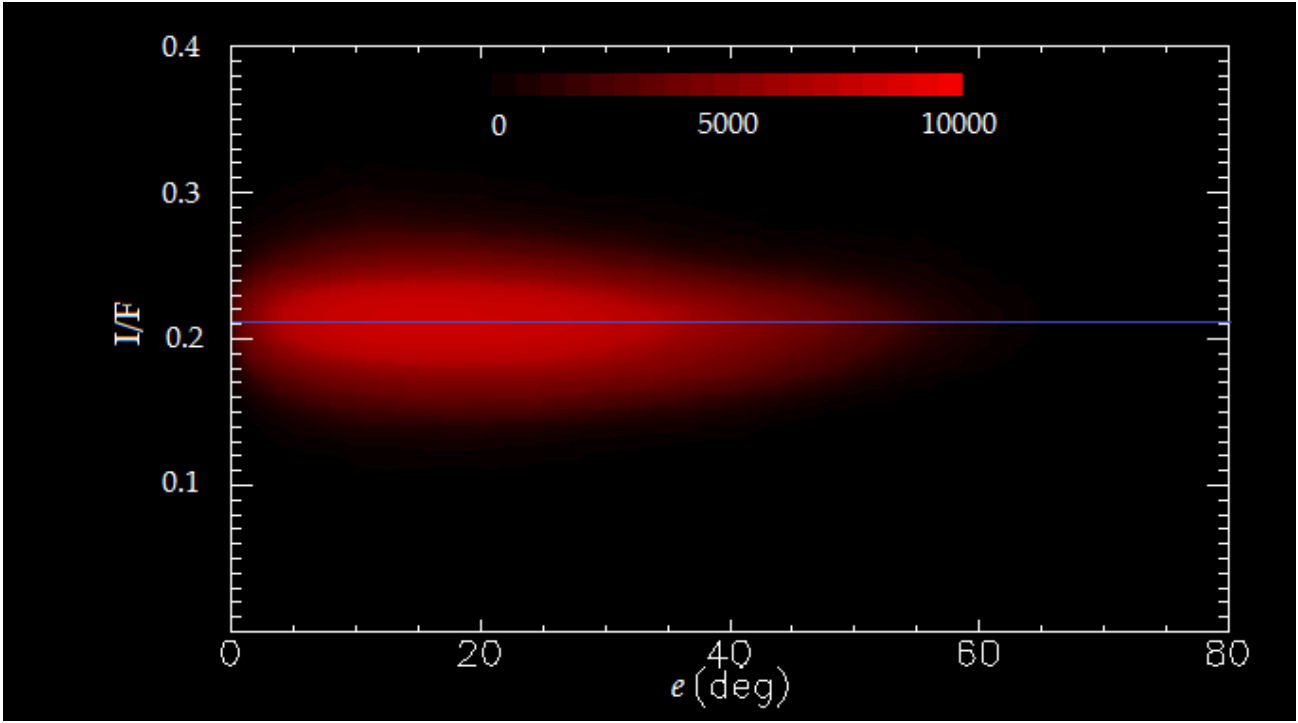

**Fig. 42 – Density plot of reflectance vs emission angle, Survey –** Density plot of the corrected reflectance (y axis) over the emission angle (x axis) at 0.796 μm. A linear fit of the distribution is shown (I/F=0.21 -e*1.37*E-4). It indicates that the distribution is flat with respect to the x axis.

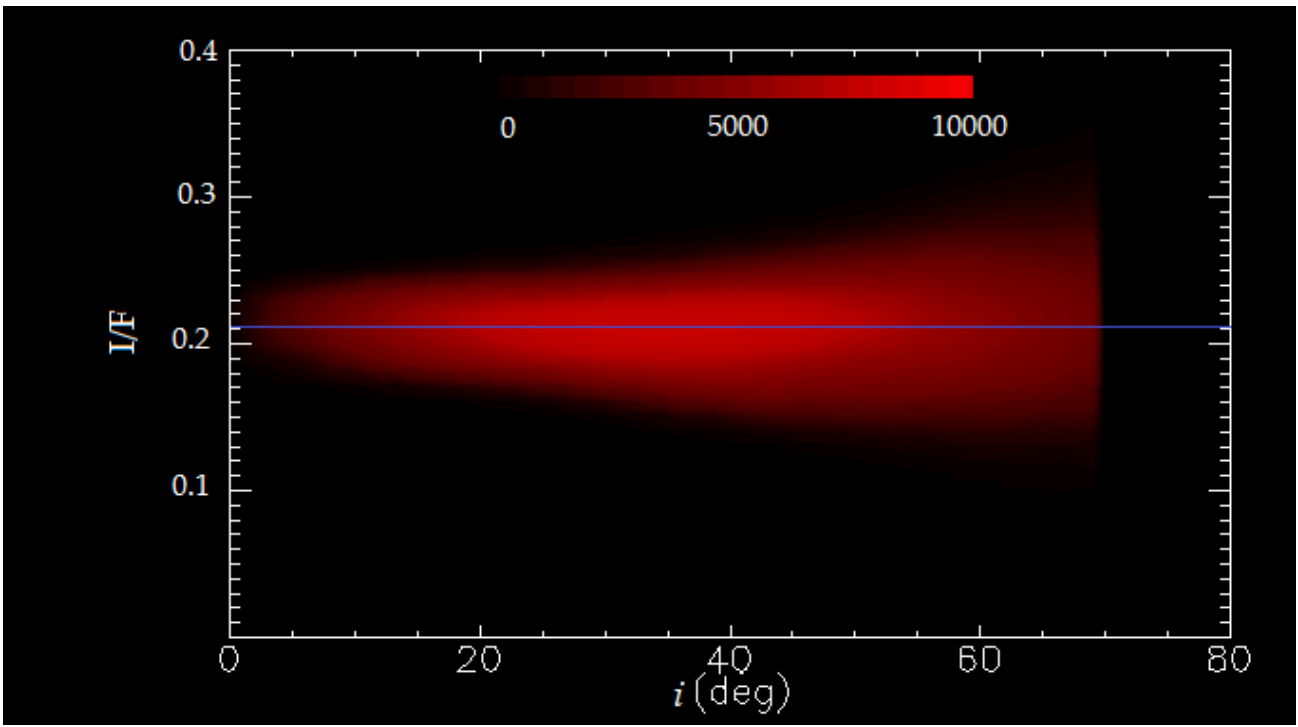

**Fig. 43 – Density plot of reflectance vs incidence angle, Survey –** density plot of the corrected reflectance (y axis) over the incidence angle (x axis) at 0.796 μm. A linear fit of the distribution is shown (I/F=0.20+i*5.29*E-4). It indicates that the distribution is flat with respect to the x axis.

As for the Survey, the effect of the correction on the Hamo data-set is shown in the density-plot of **Fig. 44**; the color bar represents a non-linear stretching, adopted in order to magnify differences that otherwise would be difficult to see. Comparing the linear fits before (**Fig. 29**) and after correction, we see that:

- corrected data-set → (I/F=0.20-$\alpha$*5.70*E-5),

- uncorrected data-set → (I/F=0.28+$\alpha$*3.79*E-3).

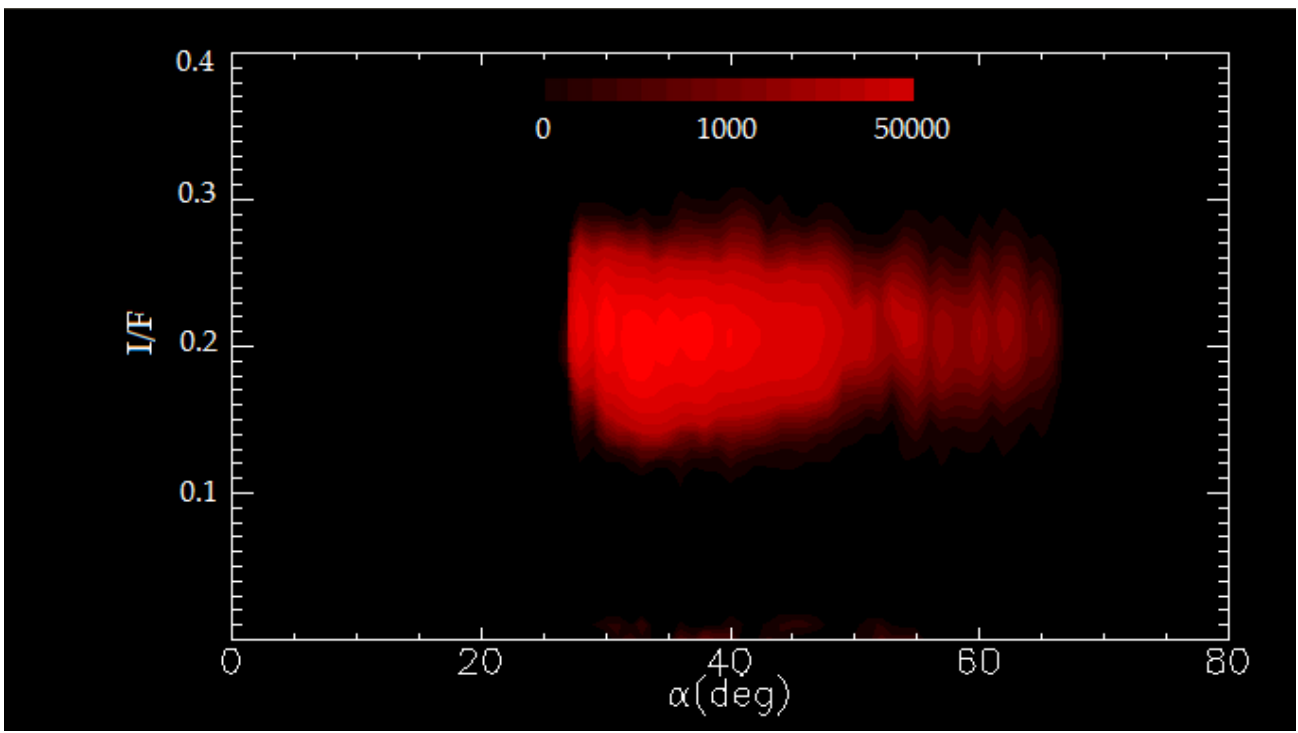

**Fig. 44 - correct I/F for Hamo data –** density plot for I/F at 0.796 after the correction, for the Hamo data-set. Brighter regions are more dense than the darker ones. Note that color scale is non-linear.

The two density plots of reflectance over emission and incidence angles are plotted in **Fig 45** and **46**. In the captions are reported the results of linear fits obtained, corresponding to the overplotted blue line.

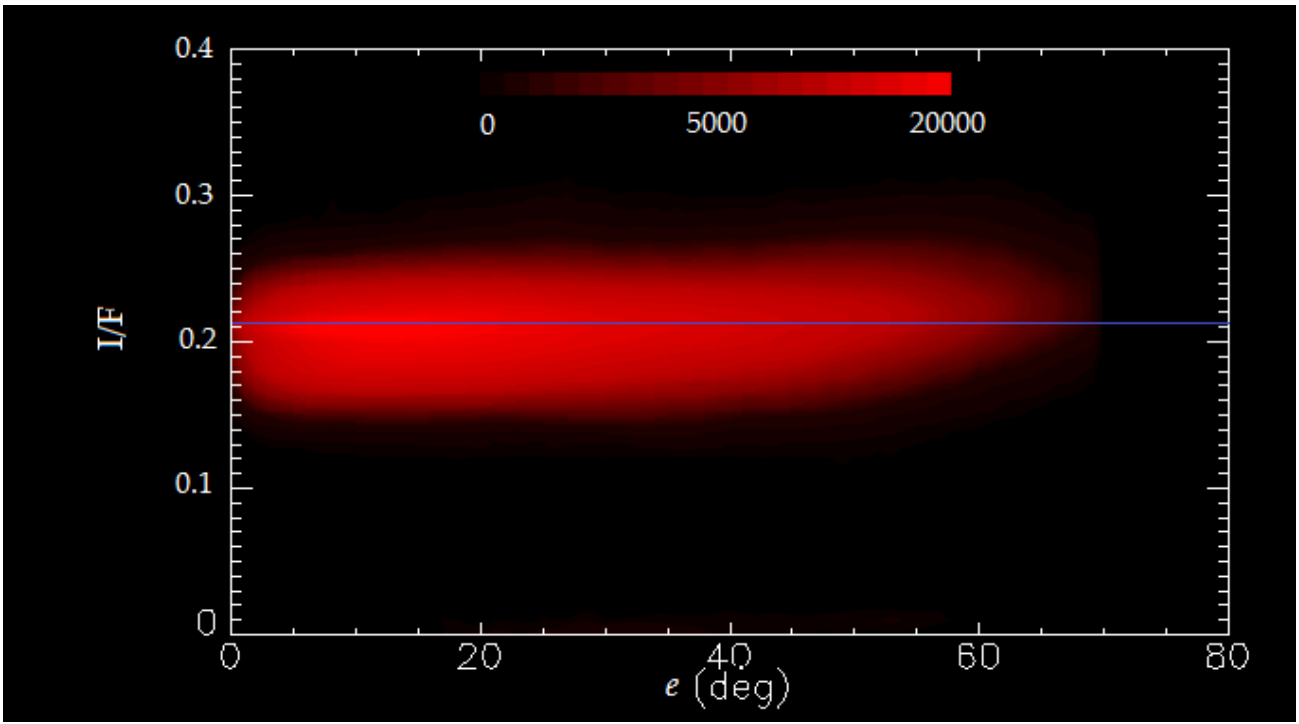

**Fig. 45 – Density plot of reflectance vs emission angle, Hamo –** density plot of the corrected reflectance (y axis) over the emission angle (x axis) at 0.796 μm. A linear fit of the distribution is shown (I/F=0.20+e*3.77*E-5). It indicates that the distribution is flat with respect to the x axis.

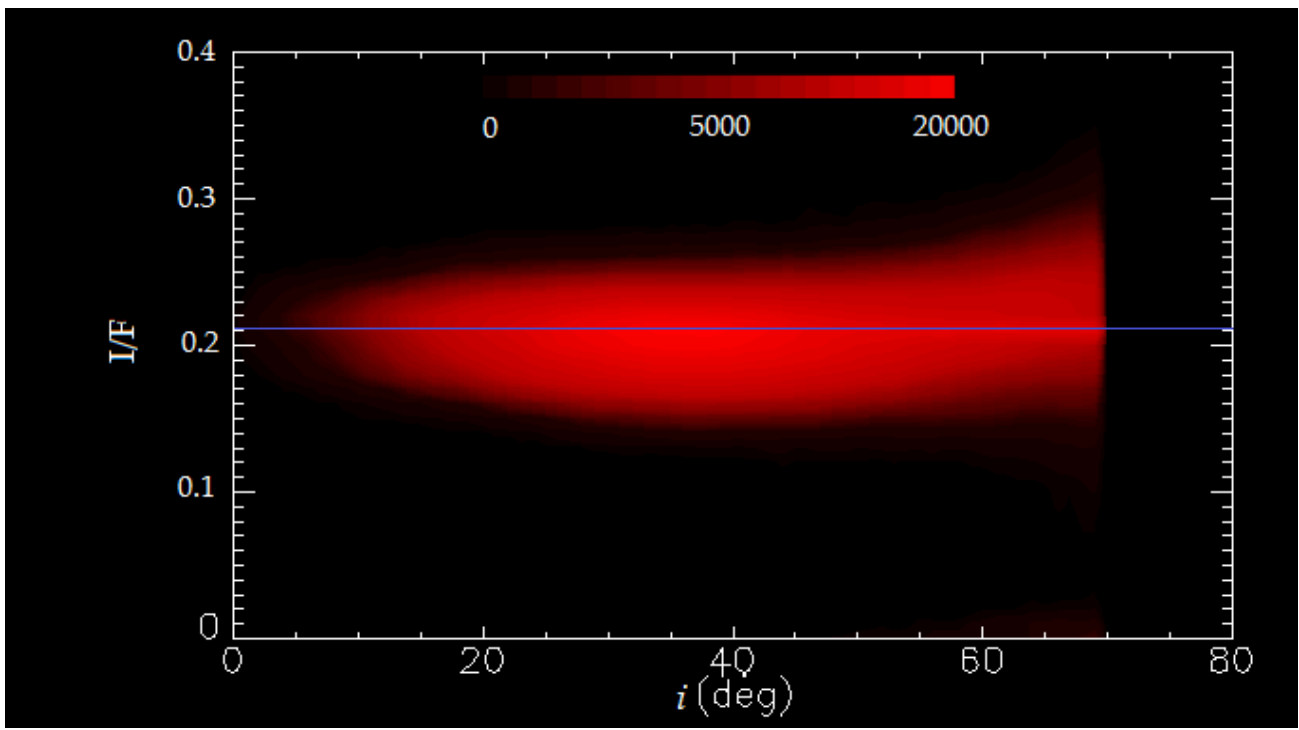

**Fig. 46 - Density plot of reflectance vs incidence angle, Hamo –** Density plot of the corrected reflectance (y axis) over the incidence angle (x axis) at 0.796 μm. A linear fit of the distribution is shown (I/F=0.21+i*2.33*E-5). It indicates that the distribution is flat with respect to the x axis.

## 4.6 Errors discussion

Our analysis is affected by errors related to measurements and modeling. Here, they are briefly discussed.

*4.6a Measurements errors*

A first source of uncertainty is the one related to the measurements. Another type of error is related to the calibration of the instruments in use, even if instrumental effects were removed on VIR data by means of spectral and geometrical calibrations (*Filacchione, Ammannito, 2014*).

The second discretization that we operate when we average over a pixel greater than the one observed, as in the case of maps like those of **Fig. 20** and **25** and most of the results of the last chapter, reduces the final spatial resolution, but improves the reflectance accuracy. Maps of redundancy (**Fig. 21** and **26**) provide the local statistic and the coverage we have. Averaging over more values, allow us to reduce all the non-systematic error source. The accuracy of the average reflectance for each of these pixels, is higher where there is a better coverage, being our uncertainty modulated by a factor $\frac{1}{\sqrt{(N)}}$. Being the Survey dataset smaller than the Hamo one, the typical error in the produce maps is in general greater in the former case with respect to the latter.

4.6b *Modeling errors*

The error discussed in this section is related to the model adopted.

- In order to evaluate the accuracy of the derived Hapke's model and its ability to describe the observed reflectance, a plot of the measured I/F (observed data) over the modeled I/F (simulated data at the same observation geometry by means of the Hapke model) is shown in **Fig. 47** and **Fig. 48** respectively for the Survey and the Hamo cases. An ideal model of the reflectance would have shown null residuals in these distributions and a trend following that of the bisect. In both the cases a deviation from a bisect-like trend at higher reflectances is observed. However, it is important to note that the most populated, central region of the two distributions, are well fitted by a bisect and that both the deviation and the residuals are on the tail of the gaussian distribution.

The RMS

$$\sqrt{\frac{\sum \left[(I/F)_{obs}^2 - (I/F)_{meas}^2\right]}{\sqrt{N}}} \, , \tag{64}$$

the root mean squared difference between model and data, is 2.3% for the Hamo and 6.8% for the Survey, probably because of the better statistic that we have on the Hamo with respect to the Survey. The RMS is the combination of the model error and the intrinsic variability of the surface; we would have a null value of the RMS only in the ideal case of a perfectly uniform surface and a perfectly working model.

- The error in the final values of each of the five parameters of the Hapke model could be deduced by a propagation of the errors through the Hapke formula; this way is impractical because would require a precise information about the uncertainty of the measurements as well as of the observation geometry. However, we can give an estimate of the errors (see **Table 13**), by comparing the parameters obtained for the Survey (**Table 9**) and the Hamo (**Table 10**) cases, as:

$$\frac{par_{HAMO} - par_{SURVEY}}{\frac{(par_{HAMO} + par_{SURVEY})}{2}} . \qquad (65)$$

- The typical relative error in the determination of SSA is about 2.9%, with a lower uncertainty inside the two absorption bands at 0.9 and 1.9 µm;

- h is affected by fluctuations of the same order of the parameter or even larger;

- b and θ error is about 10.0% and 21.8%, with no evident differences between IR and VIS.

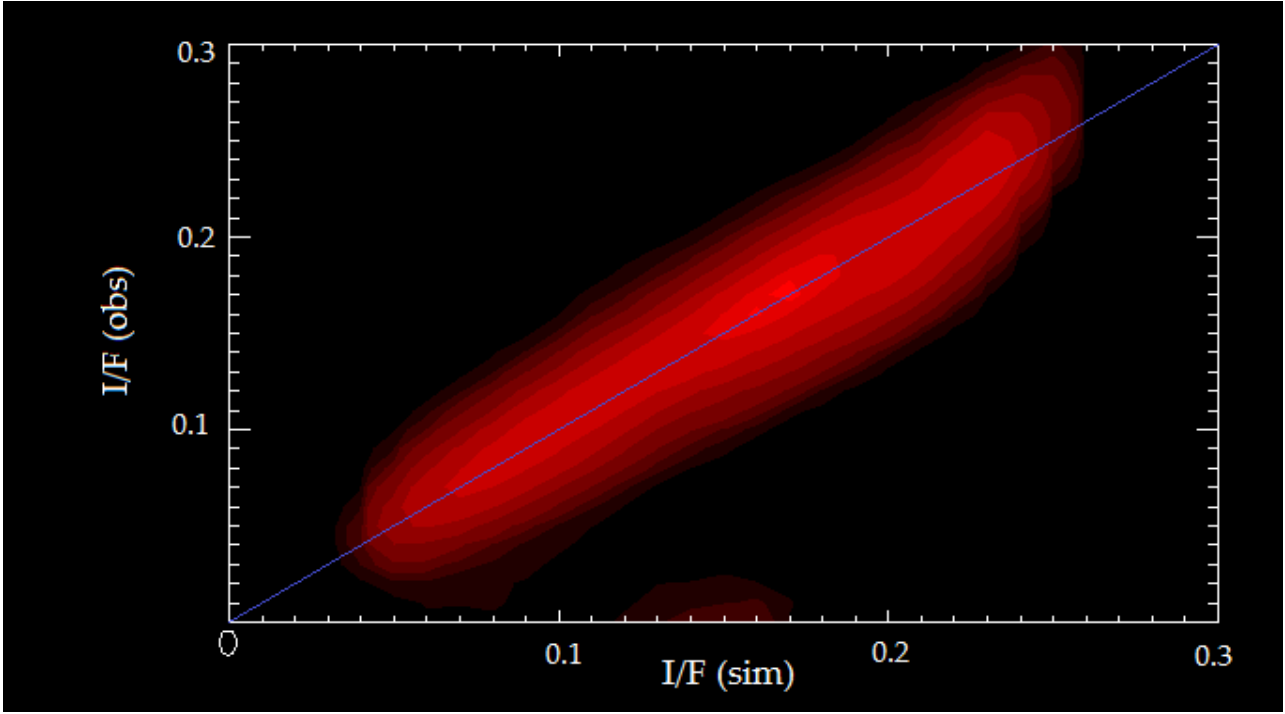

**Fig. 47 - Observed reflectance vs simulated, Survey -** density plot of the observed reflectance (I/F(obs)) against simulated reflectance (I/F(sim)) for the Survey at 0.796 µm. A discrepancy respect to the bisect (the blue line) can be observed at the higher values. The linear fit gives I/F(obs)=1.63*10$^{-3}$+I/F(sim)*0.99.

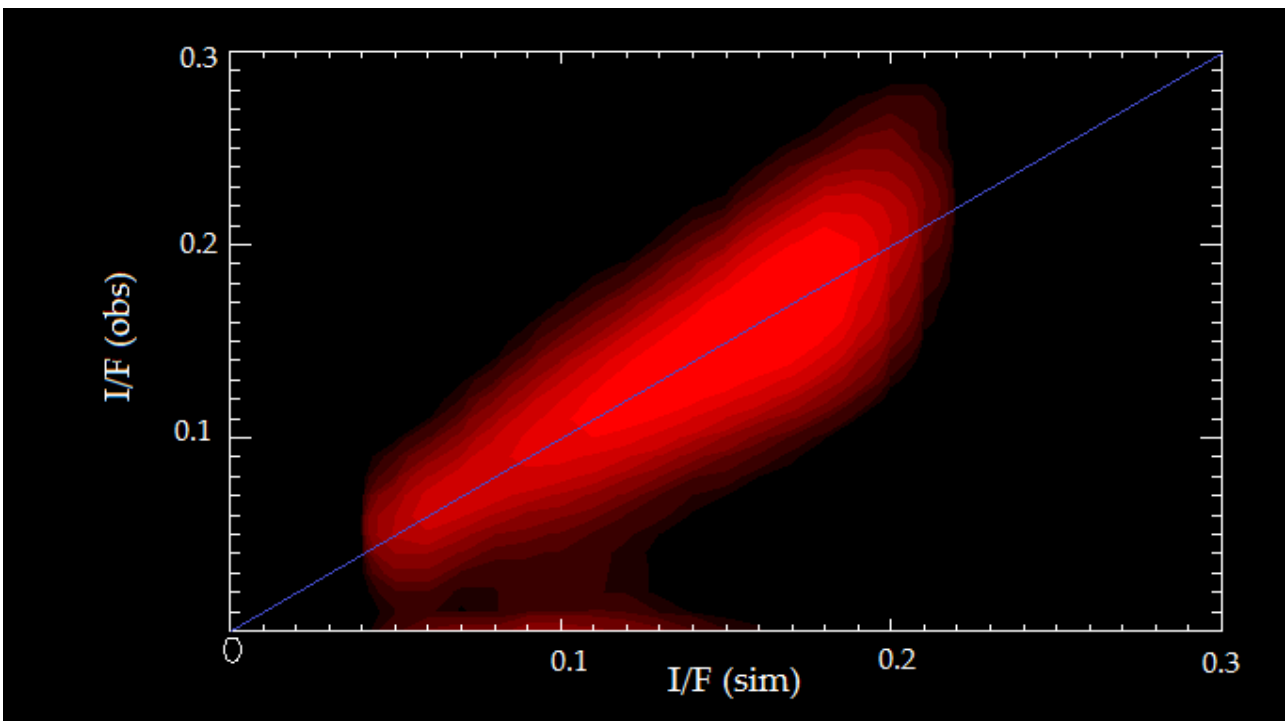

**Fig. 48 - Observed reflectance vs simulated, Hamo -** density plot of the observed reflectance (I/F(obs)) against simulated reflectance (I/F(sim)) for the Hamo at 0.796 µm. A discrepancy respect to the bisect (the blue line) can be observed at the higher values. The linear fit gives I/F(obs)=-1.54*10$^{-3}$+I/F(sim)*1.01)

| Table 13 – relative errors of Hapke's model parameters – h free ||||
| λ (μm) | w | h | b | θ |
| --- | --- | --- | --- | --- |
| 0.441 | 8.7% | 37.5% | 10.8% | 9.1% |
| 0.531 | 8.4% | 9.0% | 13.8% | 19.6% |
| 0.550 | 7.3% | 4.0% | 12.4% | 17.4% |
| 0.575 | 7.6% | 3.9% | 13.2% | 20.3% |
| 0.700 | 7.1% | 0.0% | 13.1% | 19.3% |
| 0.751 | 6.4% | 2.7% | 11.3% | 19.7% |
| 0.796 | 4.4% | 22.5% | 7.4% | 18.9% |
| 0.819 | 3.4% | 36.0% | 5.2% | 19.7% |
| 0.900 | 1.5% | 69.8% | 3.2% | 17.7% |
| 0.980 | 2.2% | 50.0% | 4.3% | 15.8% |
| 1.049 | 5.1% | 24.0% | 9.3% | 4.3% |
| 1.096 | 5.1% | 75.4% | 7.6% | 18.3% |
| 1.200 | 6.5% | 68.9% | 11.4% | 18.7% |
| 1.238 | 6.5% | 59.6% | 13.0% | 14.2% |
| 1.295 | 5.7% | 59.3% | 12.2% | 11.7% |
| 1.399 | 4.2% | 91.7% | 6.3% | 31.2% |
| 1.569 | 3.0% | 63.4% | 4.9% | 37.7% |
| 1.702 | 1.7% | 198.1% | 13.3% | 37.1% |
| 1.796 | 1.7% | 198.2% | 11.5% | 35.5% |
| 1.900 | 2.2% | 198.6% | 10.2% | 9.6% |
| 1.948 | 0.6% | 198.3% | 10.1% | 34.3% |
| 1.995 | 1.7% | 198.7% | 12.4% | 0.5% |
| 2.099 | 0.0% | 198.2% | 9.1% | 28.6% |
| 2.203 | 1.4% | 198.3% | 11.8% | 25.2% |
| 2.298 | 2.1% | 198.3% | 13.5% | 0.7% |
| 2.402 | 1.5% | 194.1% | 13.4% | 4.6% |
| 2.648 | 1.3% | 58.3% | 4.3% | 35.2% |
| 2.799 | 1.9% | 41.7% | 7.4% | 48.7% |
| 2.903 | 3.7% | 81.1% | 14.2% | 5.3% |
| 2.950 | 1.4% | 22.5% | 2.5% | 23.7% |
| 3.102 | 3.5% | 35.6% | 5.3% | 29.4% |
| 3.196 | 2.0% | 53.4% | 1.8% | 40.0% |
| 3.509 | 1.5% | 199.5% | 16.7% | 15.5% |

A similar estimation of the errors (see **Table 14**), is obtained for the sets derived with a fixed value of the h parameter (see **Table 11** and **Table 12**).

- The typical relative error in the determination of SSA is 3.6%, greater than that obtained for the h-free set.

- b is affected by fluctuations about 16.2%.

- θ error is about 20.9%, similar to that obtained in the case of a free-h parameter.

| Table 14 – relative errors of Hapke's model parameters – h fixed | | | | |
|---|---|---|---|---|
| λ (μm) | w | h | b | θ |
| 0.441 | 8.7% | | 23.9% | 10.9% |
| 0.531 | 8.4% | | 22.2% | 20.5% |
| 0.550 | 7.3% | | 19.4% | 18.2% |
| 0.575 | 7.6% | | 20.1% | 20.8% |
| 0.700 | 7.1% | | 19.7% | 19.7% |
| 0.751 | 6.4% | | 17.2% | 20.1% |
| 0.796 | 4.4% | | 9.0% | 19.0% |
| 0.819 | 3.4% | | 4.7% | 19.6% |
| 0.900 | 1.5% | | 2.1% | 17.1% |
| 0.980 | 0.2% | | 8.6% | 15.8% |
| 1.049 | 5.1% | | 22.2% | 16.9% |
| 1.096 | 5.1% | | 30.9% | 19.9% |
| 1.200 | 6.5% | | 34.1% | 20.2% |
| 1.238 | 6.5% | | 34.8% | 20.6% |
| 1.295 | 5.7% | | 34.0% | 24.0% |
| 1.399 | 4.2% | | 29.8% | 29.4% |
| 1.569 | 3.0% | | 22.4% | 38.2% |
| 1.702 | 1.7% | | 16.5% | 26.3% |
| 1.796 | 1.7% | | 13.8% | 26.3% |
| 1.900 | 2.2% | | 11.8% | 22.9% |

| | | | | | |
|---|---|---|---|---|---|
| 1.948 | 0.6% | | 9.5% | 23.4% |
| 1.995 | 1.7% | | 10.0% | 23.4% |
| 2.099 | 0.0% | | 8.6% | 23.0% |
| 2.203 | 1.4% | | 12.7% | 20.7% |
| 2.298 | 2.2% | | 14.6% | 19.4% |
| 2.402 | 1.5% | | 16.9% | 19.1% |
| 2.648 | 1.3% | | 18.2% | / |
| 2.799 | 1.9% | | 19.2% | / |
| 2.903 | 3.7% | | 19.0% | 23.4% |
| 2.950 | 1.4% | | 2.0% | 23.8% |
| 3.102 | 3.5% | | 2.6% | 24.2% |
| 3.196 | 2.0% | | 3.0% | 15.5% |
| 3.509 | 1.5% | | 1.2% | 5.0% |

Finally, a comparison with the study by *Li et al., (2013)*, is given in **Table 15**:

| Table 15 – Confrontation with Li et al., 2013 – The parameters labeled with Li13, refers to *Li et al., 2013* | | | | | | | |
|---|---|---|---|---|---|---|---|
| λ (Li13) | w (Li13) | b (Li13) | θ (Li13) | w | b | θ | λ |
| 438 | 0.408 | -0.233 | 18.0° | 0.300 (+-0.026) | -0.218 (+-0.052) | 22.8° (+-2.5°) | 441 |
| 554 | 0.500 | -0.229 | 17.7° | 0.494 (+-0.036) | -0.233 (+-0.045) | 22.7° (+-4.1°) | 550 |
| 652 | 0.547 | -0.229 | 18.0° | / | / | / | / |
| 749 | 0.534 | -0.222 | 17.6° | 0.546 (+-0.035) | -0.233 (+-0.040) | 23.2° (+-4.7°) | 751 |
| 828 | 0.465 | -0.232 | 17.8° | 0.501 (+-0.017) | -0.236 (+-0.012) | 23.0° (+-4.5%) | 819 |
| 916 | 0.377 | -0.231 | 18.7° | 0.402 (+-0.006) | -0.241 (+-0.005) | 23.6° (+-4.0°) | 900 |
| 961 | 0.391 | -0.230 | 17.8° | 0.449 (+-0.001) | -0.244 (+-0.021) | 22.5° (+-3.5°) | 980 |

According with the relative errors reported in **Table 15**, we note that the b parameter derived in this work is consistent with that obtained by *Li et al., (2013)*. On the contrary, θ and w by *Li et al., (2013)* are not within the error of our parameters. For θ, we can argue that difference in spatial resolution between VIR and the Framing Camera, produces a different characterization of the surface roughness.

In general, it must be noted that the error that we derived is just a rough estimation obtained comparing two modeled data-sets. Moreover, the little differences in wavelength between our set of parameters and that derived by *Li et al., (2013)*, can produce important changes in the modeled parameters in those spectral regions where reflectance changes rapidly with respect to the wavelength.

# Chapter 5

# Vesta albedo and spectral parameters maps

The photometric correction described in **Chapter 4**, allow us to exploit the intrinsic variability of Vesta. Hereafter, we will use the set of photometric parameters obtained for $B_0$=1.7 and h=0.07, for a direct comparison with (*Li et al., 2013)*. In this chapter will be presented the albedo maps obtained in the VIS and IR continuum (**section 5.1**), along with their interpretation. The relative difference of the obtained maps is computed in **section 5.2** to characterize the surface spectral variability. In **section 5.3**, RGB maps are shown both for the VIS and the IR, along with a comparison with previous works.

**5.1 Surface albedo maps**

Surface albedo maps in the continuum of Vesta's spectrum are computed in the spectral range away from the two pyroxene absorption bands at 0.9 and 1.9 μm (see **Table 8**). In order to make a comparison with *Li et al., (2013)* in the VIS range and *Longobardo et al., (2014)* in the IR range, these albedo maps have been obtained at 0.751 μm and 1.2 μm. Photometrically corrected data for both HAMO and Survey have been projected into a latitude-longitude grid (spatial resolution of 0.5° x 0.5°). The values assigned to each of the pixels of the maps, represents the average of all the photometric corrected reflectance values pertaining to the specific portion of the surface.

*5.1a Albedo maps at 0.751 μm*

In **Fig. 49** the albedo map at 0.751 μm for the survey case is shown. By comparing it with **Fig. 20,** it emerges that the equatorial gradient visible in the north-south direction as well as the difference between overlapping areas of different acquisitions, have been eliminated by means of the photometric correction.

The photometrically corrected maps displays a strong regional reflectance variability; in particular, we note a low reflectance area between 60°/210° in longitudes (typical albedo 0.18-0.20), while the surrounding regions are brighter (typical albedo 0.23-0.26). This is a confirmation of the east-west dichotomy suggested by HST data (*Binzel et al., 1997*), with the eastern hemisphere having a higher albedo than the western hemisphere. Structures like craters (es. the Marcia-Calpurnia system at 180°/210° long, 0°/30° lat) and the Saturnalia and Divalia Fossae, are clearly visible. Marcia-Calpurnia craters are embedded in a dark region and display a prominent dark ejecta field ($w$=0.16), while the craters walls and rims are brighter ($w$=0.24). These albedo differences have been associated to different composition of the surface and grain sizes (*De Sanctis et al., 2014*). In the South – with respect to Marcia and Calpurnia – are visible several different bright-spots associated to craters like Tuccia ($w$=0.3) and Vibidia ($w$=0.28), while in the Eastern part appear other small bright structures like Justina ($w$=0.28) and Canuleia($w$=0.29).

The dark, elongated arch-shaped structures at 60°/90° lon., -40°/-50° lat and 90°/130° lon, -40°/-50° lat., are due to corrupted acquisitions that the we have not been able to filter out at this stage.

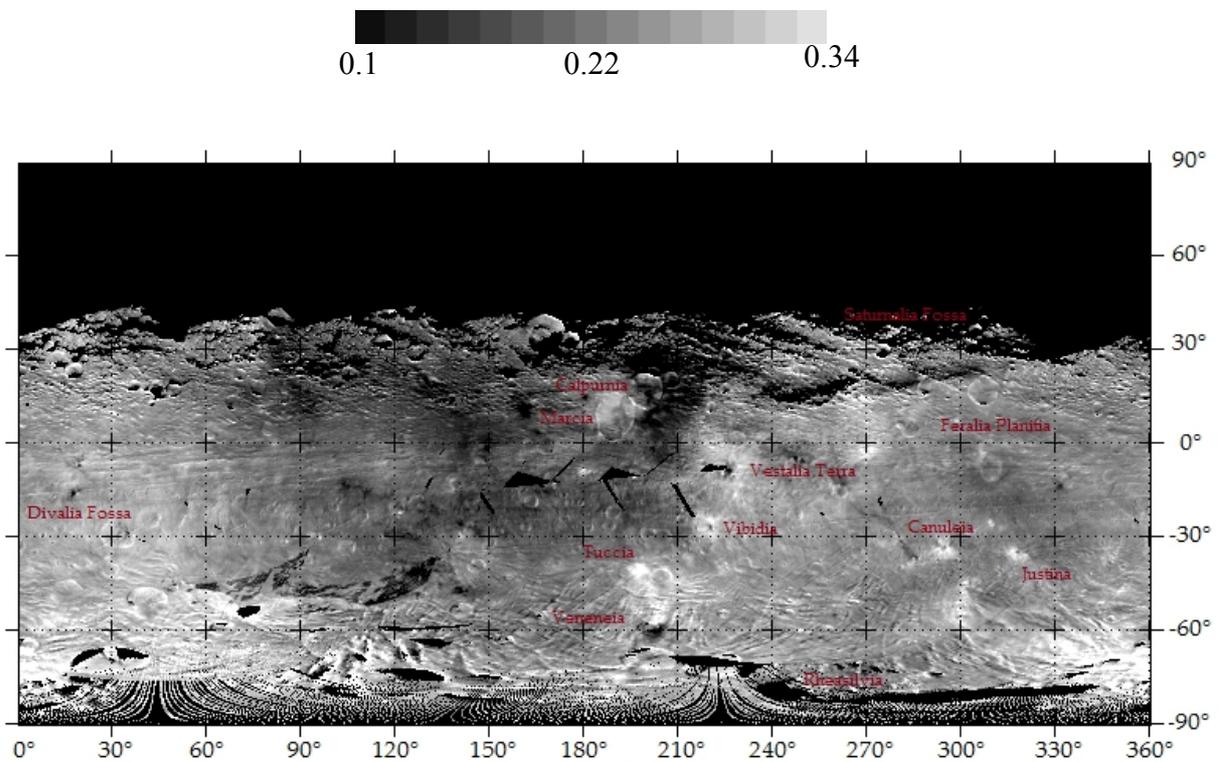

**Fig. 49 – Surface albedo map after correction at 0.751 μm, Survey -** longitudes and latitudes are indicated on the x and y axis, respectively. The adopted spatial resolution is 0.5° x 0.5°. Names of some of the most prominent features are reported.

The VIS HAMO photometrically corrected map is shown in **Fig. 50** (compare to the uncorrected HAMO map in **Fig. 25**). This map confirms the strong variability observed form low resolution data, with black small spots (es. Aricia Tholus, albedo ≈ 0.1-0.13) over the dark, central region.

Some of the most prominent structures in the Survey are visible (es. Marcia), and other appears in the northern hemisphere, like Bellicia, Caparronia and Domitia. It must be noted that the northern hemisphere was in more favorable lighting conditions during HAMOs with respect to the Survey conditions. No evident artifacts affect it.

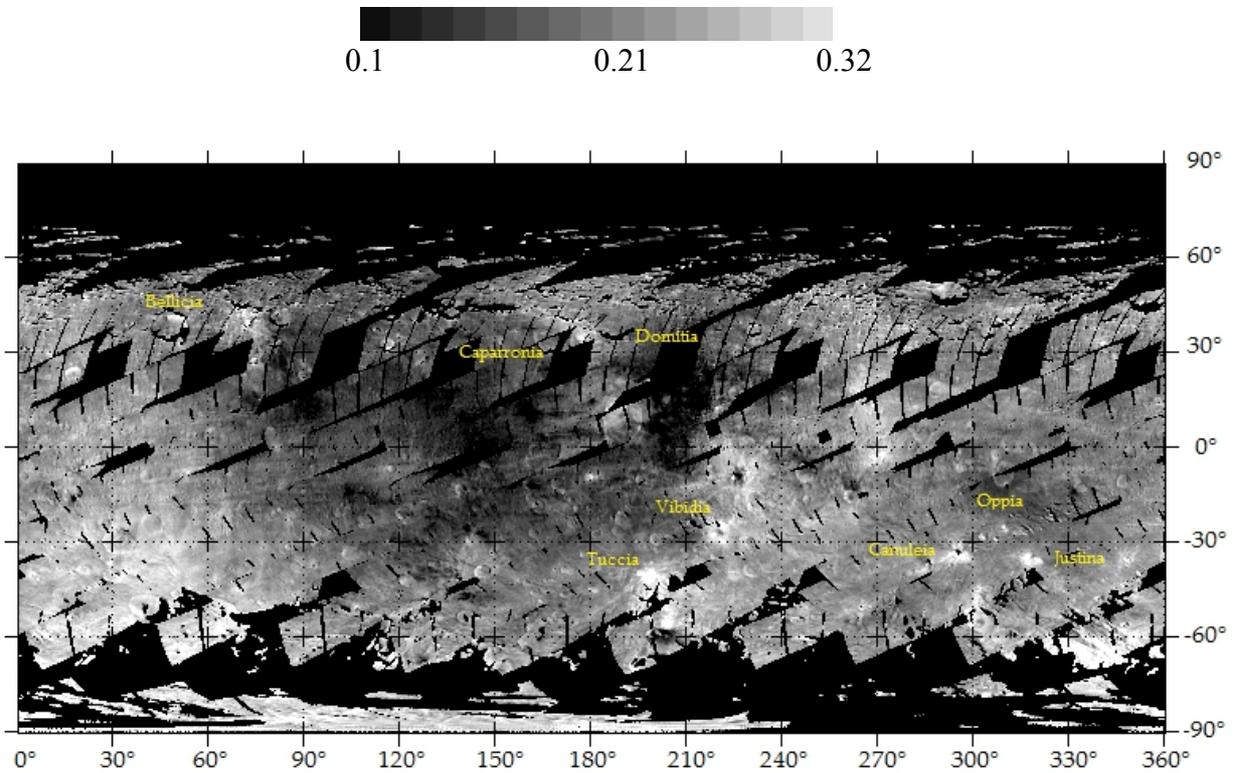

**Fig. 50 – Surface albedo map after correction at 0.751 µm, HAMO -** longitudes and latitudes are indicated on the x and y axis, respectively. The adopted spatial resolution adopted is 0.5° x 0.5°. Names of some of the most prominent features are reported.

*5.1b Albedo maps at 1.2 µm*

Albedo maps for the IR continuum are obtained at 1.2 µm (**Fig. 51** and **Fig. 52**). These maps are generally brighter than the 0.751 µm case, as expected because of the higher reflectances.

In general, the photometrically corrected HAMO map (**Fig. 52**) is less affected by artifacts with respect to the Survey case (**Fig. 52**). The bright linear features – both in the HAMO and in the Survey – and the two square-structures just above the equator of the Survey map are related to spurious pixels that the selection was unable to remove. Data related to the C7 cycle (see **Section 4.1**) have been removed from the Survey map, because they should have carried even more artifacts: the lack of observations in the central latitudes are due to this depletion.

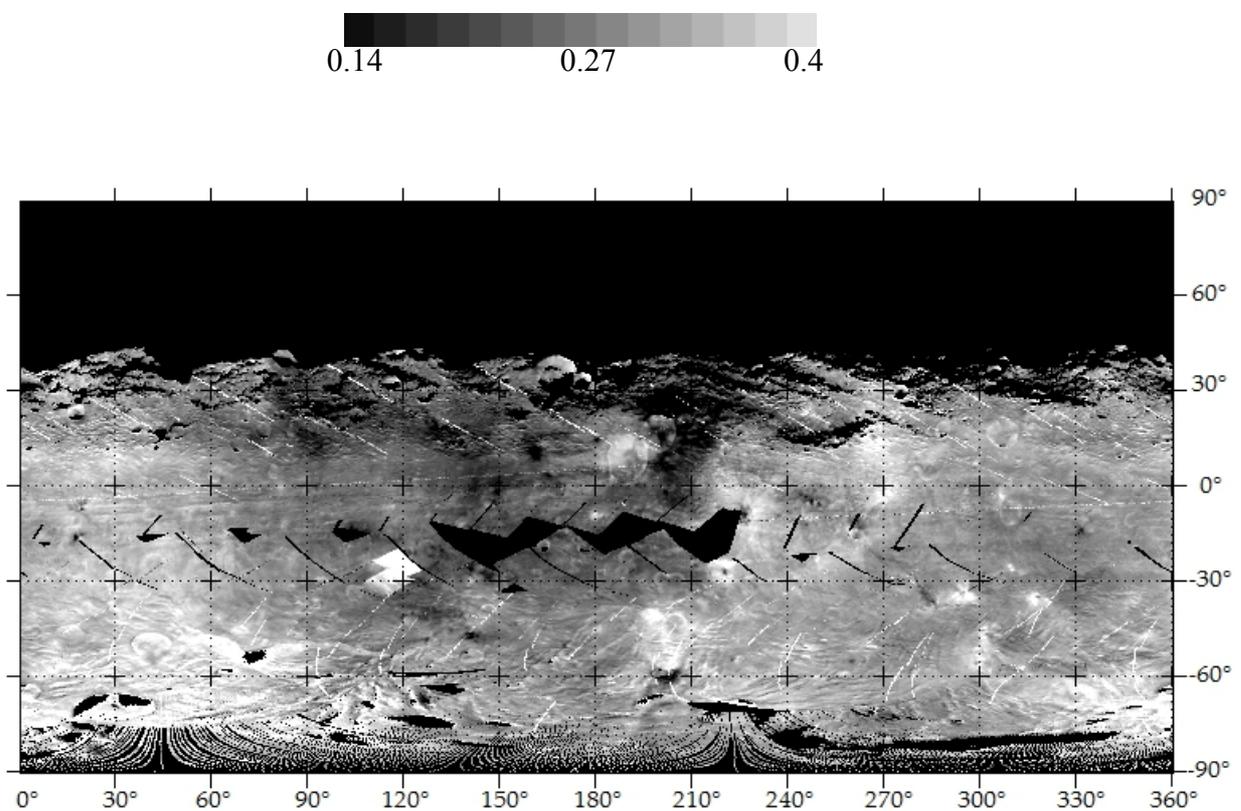

**Fig. 51 – Surface albedo map after correction at 1.2 μm, Survey -** longitudes and latitudes are indicated on the x and y axis, respectively. The adopted spatial resolution is 0.5° x 0.5°. The lack of equatorial data is explained in the text above.

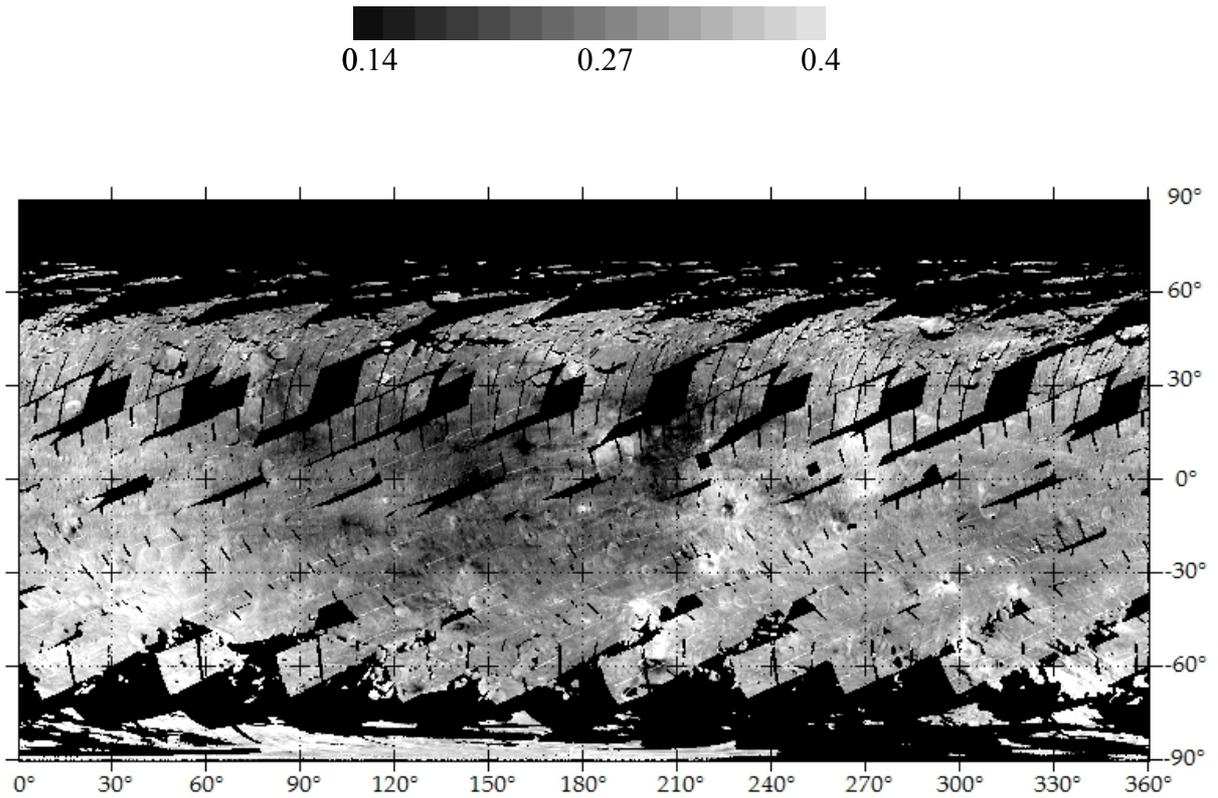

**Fig. 52 - Surface albedo map after correction at 1.2 μm, HAMO -** longitudes and latitudes are indicated on the x and y axis, respectively. The adopted spatial resolution is 0.5° x 0.5°.

**5.2 Spectral variability through albedo maps**

The two wavelengths chosen for the analysis of **section 5.1**, can be compared to show how Vesta's surface albedo changes between the IR and VIS spectral continuum. In order to enhance spectral differences between bands, reduce the effects of topography and isolate regions with maximum spectral variability, we report in **Fig. 53** (Survey) and **Fig. 54** (HAMO) the relative differences between 1.2 and 0.75 μm:

$$\Delta = 2 \frac{I/F_{std}(1.2\,\mu m) - I/F_{std}(0.75\,\mu m)}{I/F_{std}(1.2\,\mu m) + I/F_{std}(0.75\,\mu m)} \qquad (64)$$

I/F$_{std}$ are the I/F ratios at 1.2 and 0.75 µm, reported at standard geometries.

In general, both the ratio maps in **Fig. 53** and **54** are smoother than that in **Figs. 49**, **50**, **51** and **52**, showing only some variations, as:

- a positive Δ-signal in the albedo difference at 300°/330° long and -30°/0° lat, where the Oppia crater is located;

- a negative Δ-signal at 200° long and -40° lat., corresponding to Tuccia crater.

The ratio maps can be used to study the effects of space weathering on Vesta. In general, space weathering effects on spectra are to reduce the albedo, to weak absorption bands and to exhibit a red-sloped continuum (reflectance increasing towards longer wavelength): mature (old) terrains should show these characteristics. Thus, the ratio between the VIS and IR give us information on the spectral slope, while original maps, give us information of the darkening of the spectrum. Putting together the different maps can help in understanding the Vestan space weathering.

In particular, it can be seen that the central, dark region between 60° and 210° in longitudes, disappears, suggesting that the overall surface behaves in a similar way in the IR and VIS. This means that Vestan surface does not show a clear spectral reddening, as also noted by *Pieters et al. (2012)*, based on a small scale analysis of the surface. This has been interpret as due to material mixing. In fact, spectroscopic data reveal that on Vesta a locally homogenized upper regolith is generated with time through small-scale mixing of diverse surface components (*Pieters et al., 2012*).

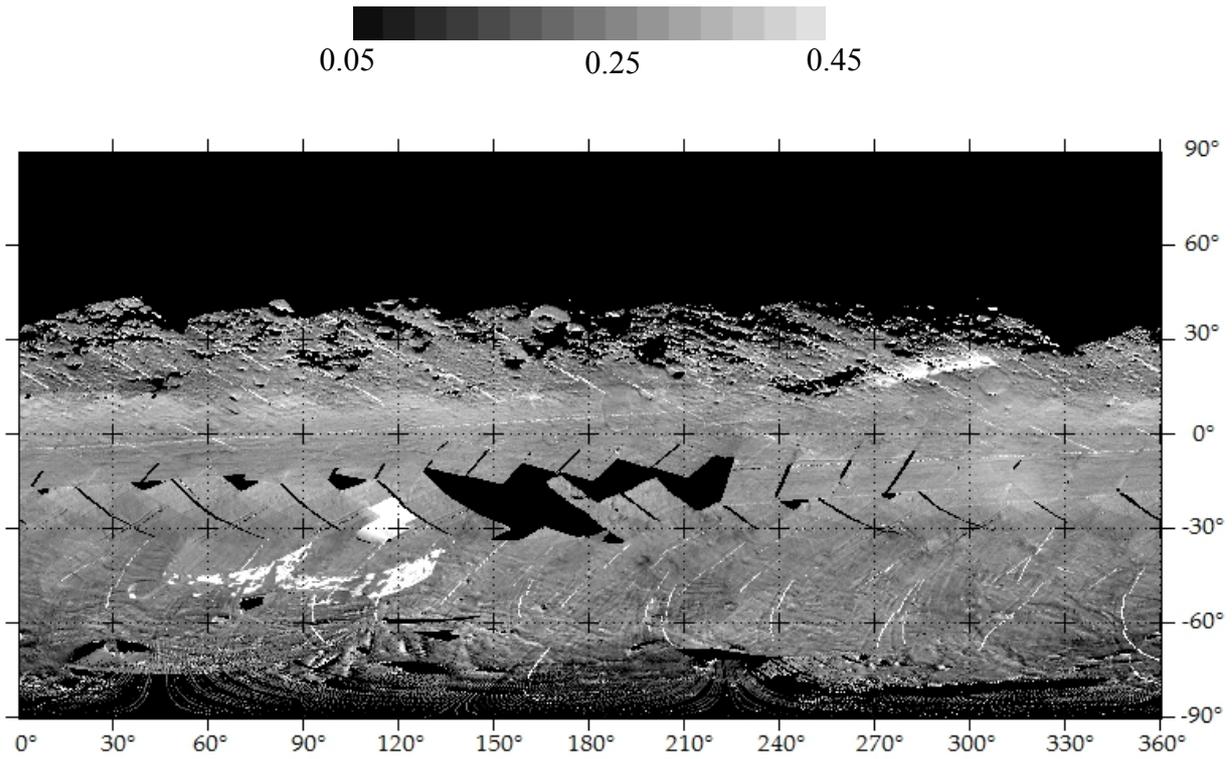

**Fig. 53 – Map of the albedo difference, Survey -** Map of the albedo difference at 1.2 μm and 0.751 μm for the Survey. Longitudes and latitudes are indicated on the x and y axis.

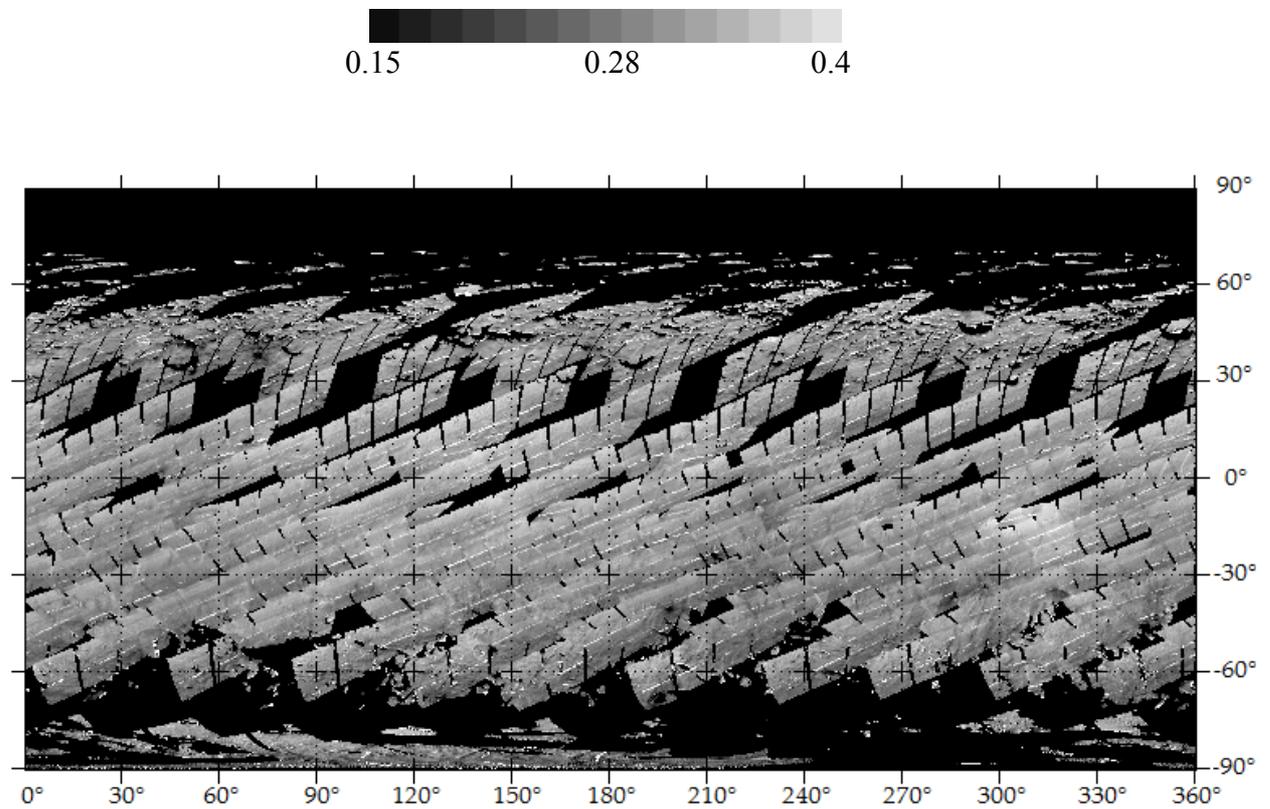

**Fig. 54 – Map of the albedo difference, HAMO -** Map of the albedo difference at 1.2 μm and 0.751 μm for the Hamo. Longitudes and latitudes are indicated on the x and y axis.

## 5.3 Overall and enhanced color maps

In the previous sections we have produced separately Survey and HAMO maps after the photometric correction and the maps of the relative differences between 0.751 and 1.200 µm.

Below (**Fig. 55-56**), it is shown an overall map of Vesta, comprehensive of both HAMO and Survey information, after correction. In particular, the albedo map from Hamo (**Fig.50**) has been filled with Survey values (**Fig. 49**) in regions with missing observations.

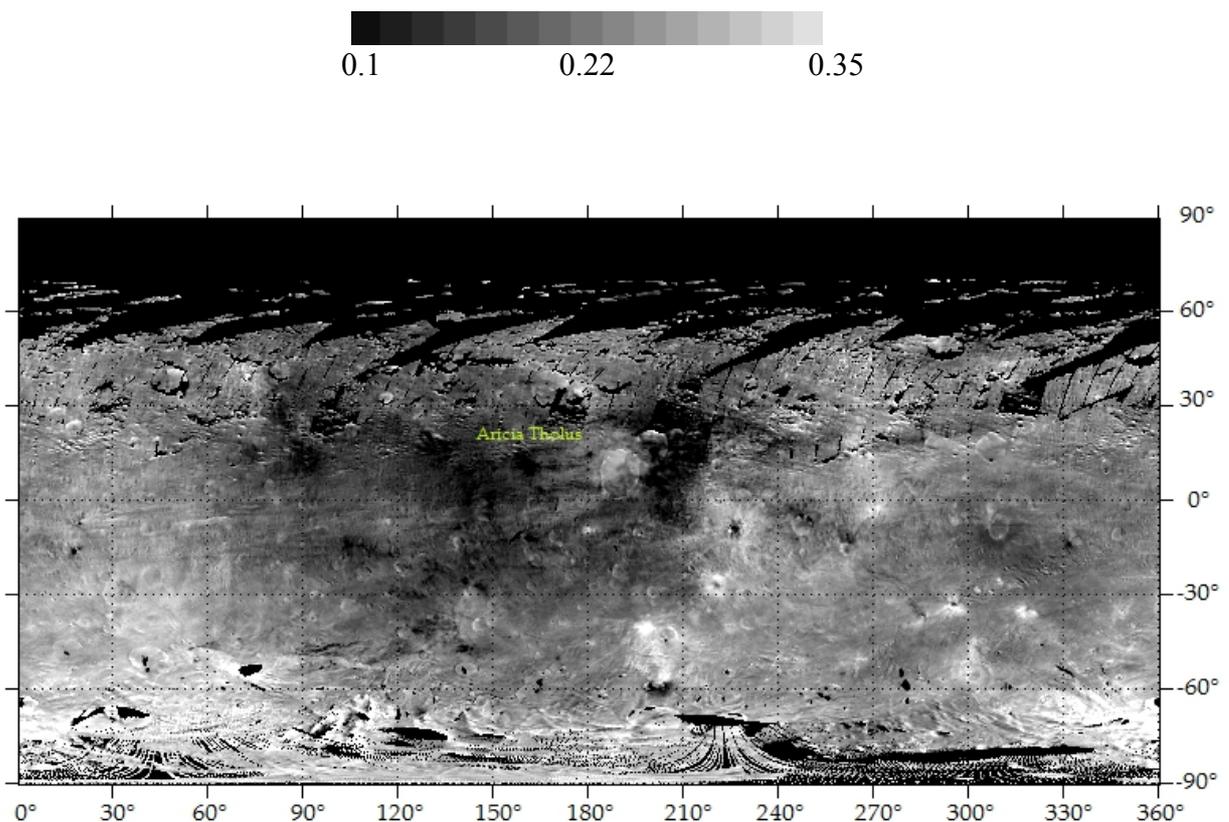

**Fig. 55 – Surface albedo map of Vesta-** corrected albedo map of Vesta at 0.751 µm, obtained from all the Survey and HAMO data-sets. Longitudes and latitudes are indicated on the x and y axis.

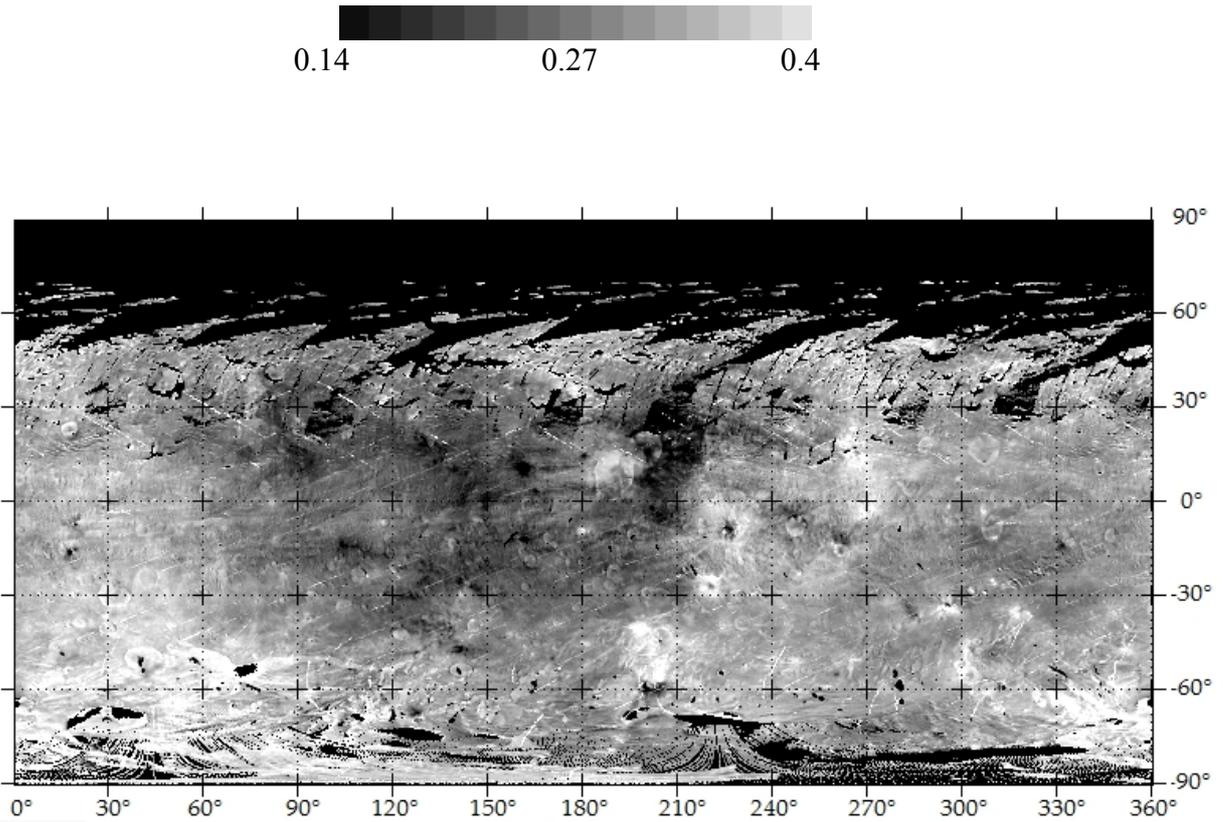

**Fig. 56 – Surface albedo map of Vesta at 1.2μm -** corrected albedo map of Vesta at 1.2μm, obtained from all the Survey and HAMO data-sets. Longitudes and latitudes are indicated on the x and y axis.

The map in the VIS has been compared with the map obtained by Li et al., (2013) (see **Fig. 57**) from the Framing Camera data. The predominant albedo variations seen across the map by *Li et al., (2013)* are consistent with our observation: in both **Fig. 55** and **56** we see the dark region between 60° and 190° in longitudes, and the bright areas outside of that region, in particular the Rheasilvia basin (the South corners). Similarly, Aricia Tholus, the darkest area on Vesta (*Li et al., 2013*) located at 10° in latitudes and 160° in longitudes, and the brightest area, corresponding to the wall of an unnamed crater at -65° in latitudes and 360° in longitudes (*Li et al., 2013*) are visible.

The data used by *Li et al., (2013)*, are obtained at a higher spatial resolution. The photometric correction was performed with the same values of $B_0$ ($B_0$=1.7) and h (h=0.07) that we used to produce all the corrected VIR maps. The two panels are obtained at 0.749 µm and refer to different ranges of phase angles (Upper panel=5°-15°, Bottom panel=30°-40°): for this reason, terrains in the upper panel map appear smoother than that in the bottom panel, since are less affected by shadows.

As suggested in *Li et al., (2013)*, dark areas observed on Vesta might be, in general, exogenous carbonaceous material delivered through low-speed impacts (*McCord et al., 2012, Reddy et al., 2012, De Sanctis et al., 2012*), while the bright areas could be related to the exposures of materials originally formed on Vesta (*Li et al., 2012, De Sanctis et al., 2014*).

In particular, in *McCord et al., (2012)*, it is stated that the broad dark region that we note on Vesta is associated with localized dark-material, which tend to clusters in this area, thus suggesting a causal relationship. Moreover, in their work, they haven't noted any correlation between dark and bright materials, suggesting a different origin for them. The bright material was interpreted as the uncontaminated indigenous Vesta basaltic soil, while dark material may come from low-albedo impactors, and diffused over time through the Vestan regolith by impact mixing, creating the broad dark region that we observe.

In **Fig. 58** it is shown the map of Vesta produced by *McCord et al., (2012)* at 1.7µm, using the Akimov photometric correction, while in **Fig. 59**, is reported our map at the same wavelength (1.7 µm). It can be noted how – thanks to the photometric correction that we applied – our map is improved and show the intrinsic surface variability.

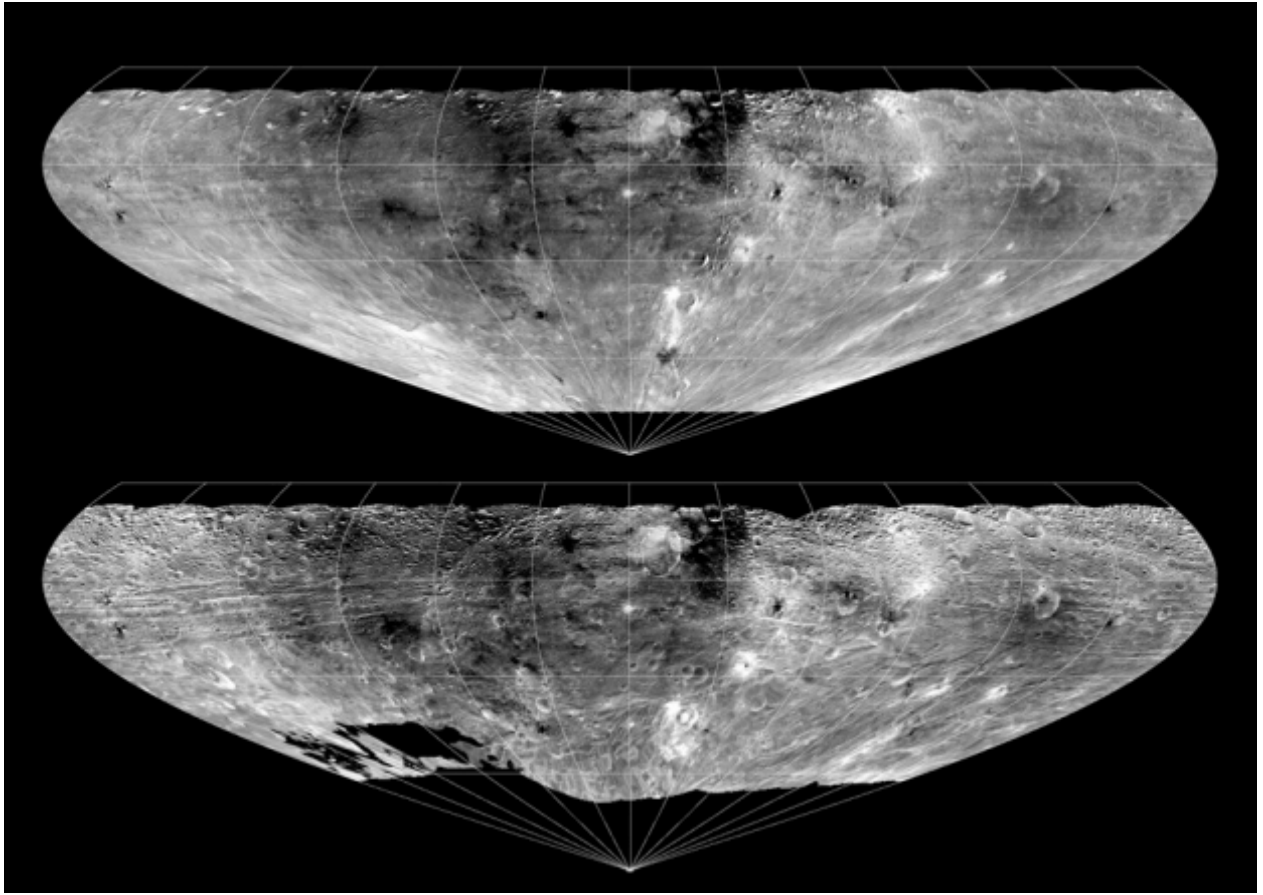

**Fig. 57 – FC Photometrically corrected map at 0.749μm,** *Li et al., 2013* **-** stereographic projection of the corrected map of Vesta at 0.749 μm. (*Li et al., 2013*)

In **Fig. 60** the map obtained by *Longobardo et al., (2014)*, is shown. In this work, they have applied an empirical approach, using the Akimov parameter-less disk function to remove the incidence and emission influence. The two maps are very similar and are as well suited to correct data at intermediate latitude. However, at the South Pole, the map in **Fig. 60**, shows regions where the Survey and the HAMO do not overlap well, maybe because of the presence of high phase angles, that the Akhimov model is unable to characterize well.

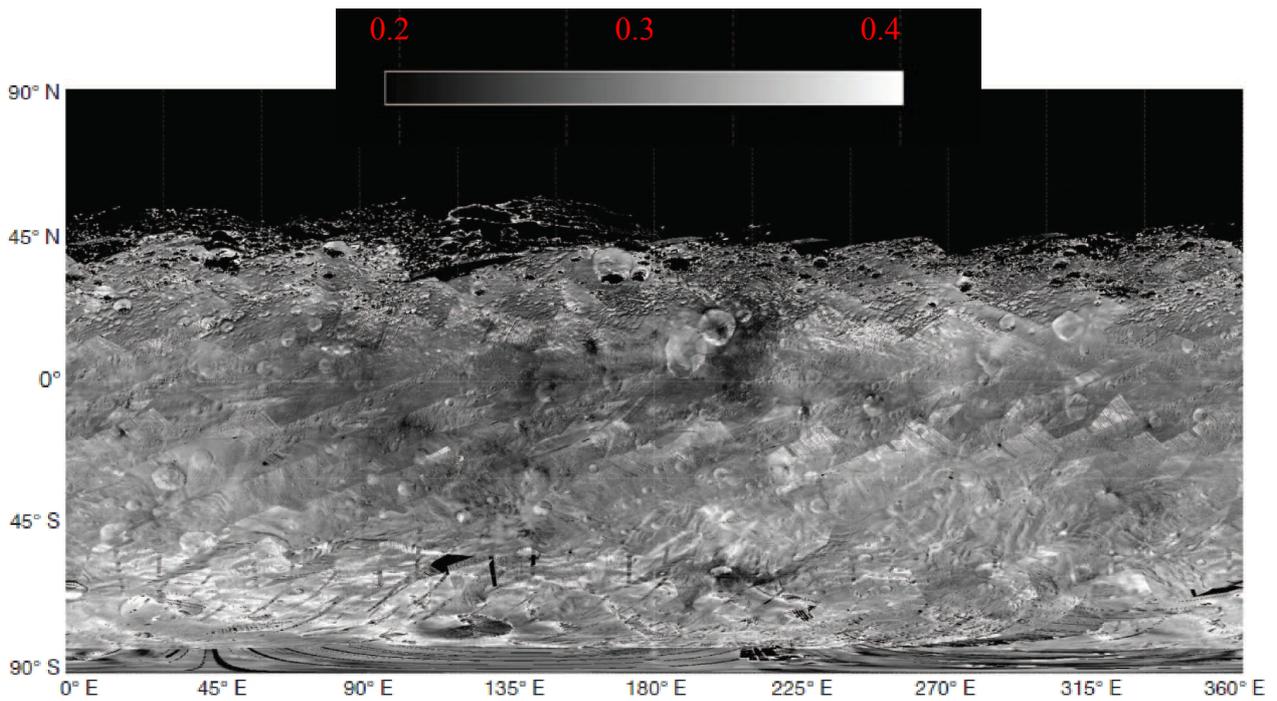

**Fig. 58 - McCord's map at 1.7 μm -** McCord's overall map of Vesta at 1.7 μm from the whole VIR infrared data set at 1.7 μm, as obtained from late Approach, Survey and HAMO. (Adapted from *McCord et al., 2012*)

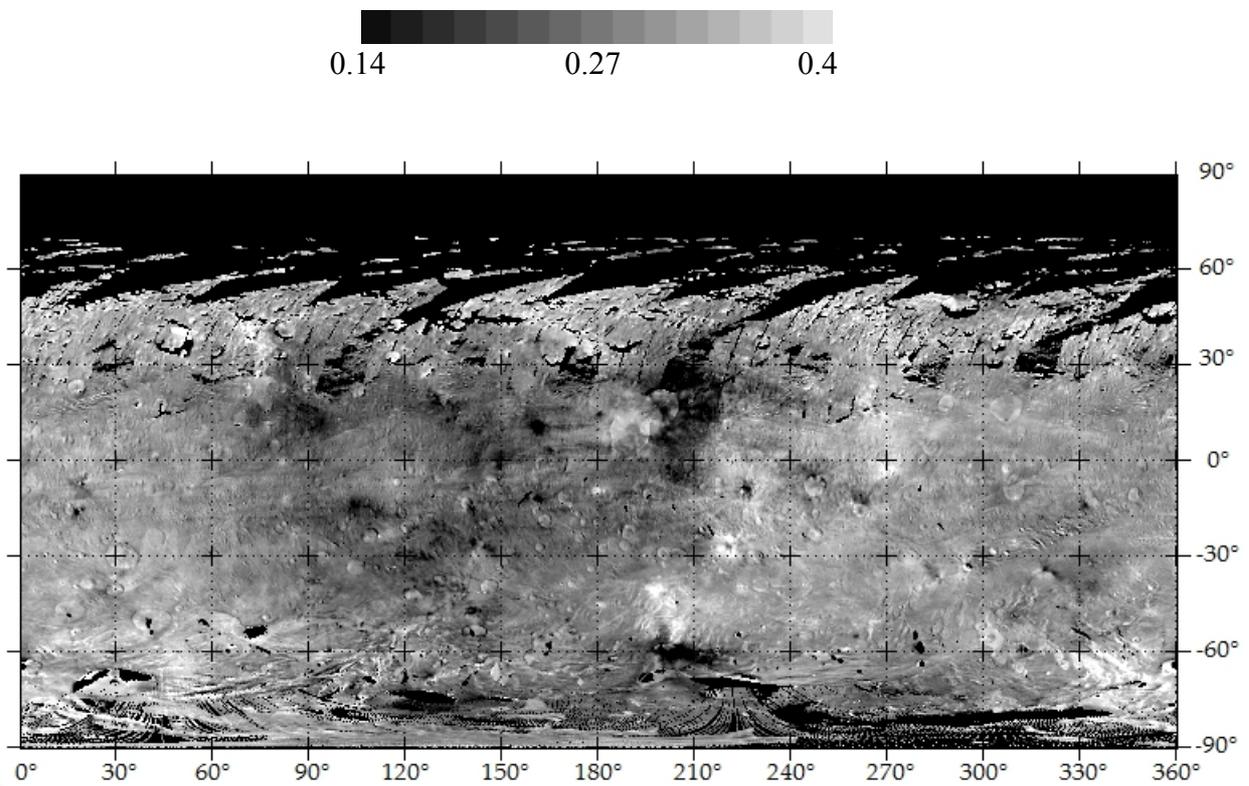

**Fig. 59 – Overall surface albedo map of Vesta at 1.7μm -** corrected albedo map of Vesta at 1.7μm, obtained from all the corrected Survey and HAMO data-sets. Longitudes and latitudes are indicated on the x and y axis. The artifacts of Fig. 56, are not present.

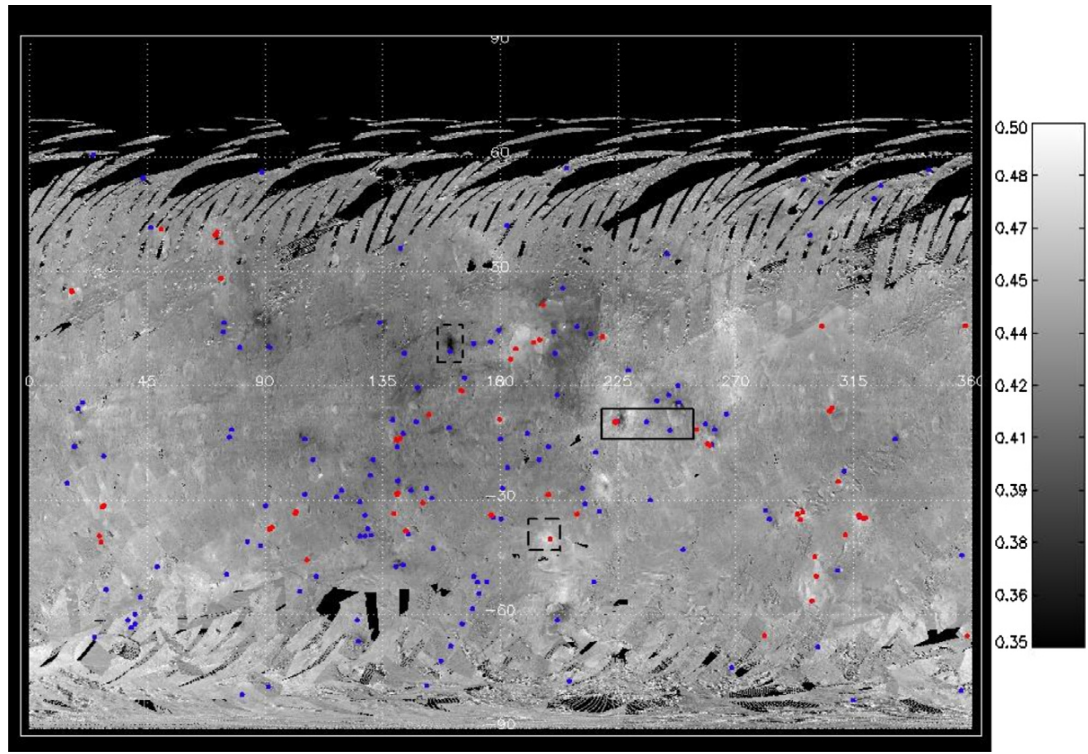

**Fig. 60 - Vesta albedo map at 1.2 μm, Longobardo et al., 2014 –** corrected map of Vesta at 1.2 μm. *(Longobardo et al., 2014)*

In order to perform a better characterization of Vesta's surface, enhanced color maps (RGB) in both VIS (**Fig. 61**) and IR (**Fig. 62**) are produced from the merging of HAMO and Survey data. The photometrically corrected maps for three different wavelengths of interest are associated to different colors (Red, Green and Blue) and their contribution are weighted with respect to each other; this gives the RGB maps of **Fig. 61** and **62**. The selected wavelengths adopted to produce the maps are chosen to give a proper description of the spectral variability along the VIS and the IR, avoiding the two absorption bands.

We choose the following wavelengths:

- 440nm, 550nm and 751 nm in the VIS;

- 1200nm, 1702nm and 2402 nm in the IR.

In the VIS map of **Fig. 61**, the most notable feature is the Oppia crater at 300°/330° long., -10° lat., which appears to show a red region, extended on the southern part of it. Also the equatorial region, just in the west of Aricia Tholus, seems redder than the surroundings. The blueish, southwest-northeast lines along all the map, are produced by instrumental artifacts; it may be also the case of the whole Rheasilvia region general blueish-pattern.

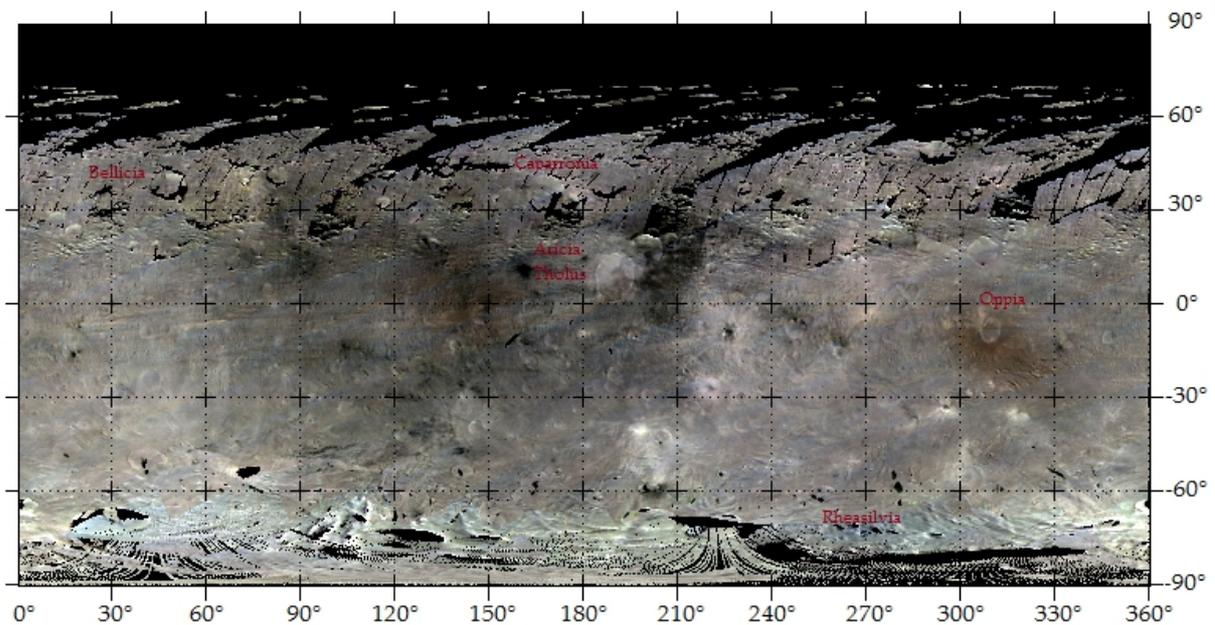

**Fig. 61 – RGB map in the visible -** Red (0.751 μm), Green (0.550 μm), Blue (0.440 μm) map in the visible. Longitudes and latitudes are indicated on the x and y axis.

In the IR map of **Fig. 62**, the surface appears as an overall green, with the exception of the South and the Western region, where there is a combination of blue and red. An elongated, particularly strong blue feature, is visible in association with Antonia, a crater on the South of Tuccia (200° longitudes, -60° latitudes).

This map is very well correlated with the results from *Ammannito et al., (2013)*, where lithological maps of Vesta's surface were presented (**Fig. 63**). In *Ammannito et al., (2013)* it was assumed that the background surface of Vesta is composed by howardite or cumulate eucrite (not distinguishable), while the southern region related to Rheasilvia and the western part of the asteroid, which goes from Rheasilvia to the equator, is mainly composed by howardite, and partially diogenites (in particular related to the blueish Antonia crater). Of course, different lithologies could be really distinguished only through a proper compositional analysis – which is not the focus of this master degree thesis work – and in particular studying the centers' position of the two pyroxene bands at 0.9 and 1.9 µm.

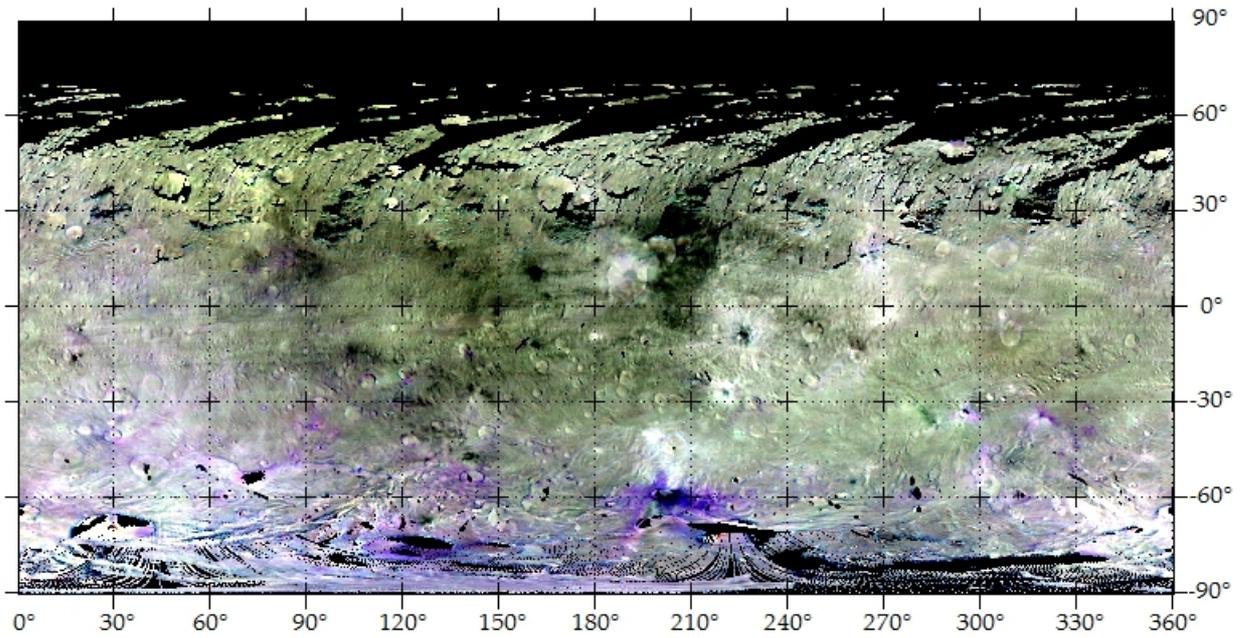

**Fig. 62 - RGB map in the infrared -** Red (2.402 μm), Green (1.702 μm), Blue (1.200 μm) map in the IR. Longitudes and latitudes are indicated on the x and y axis.

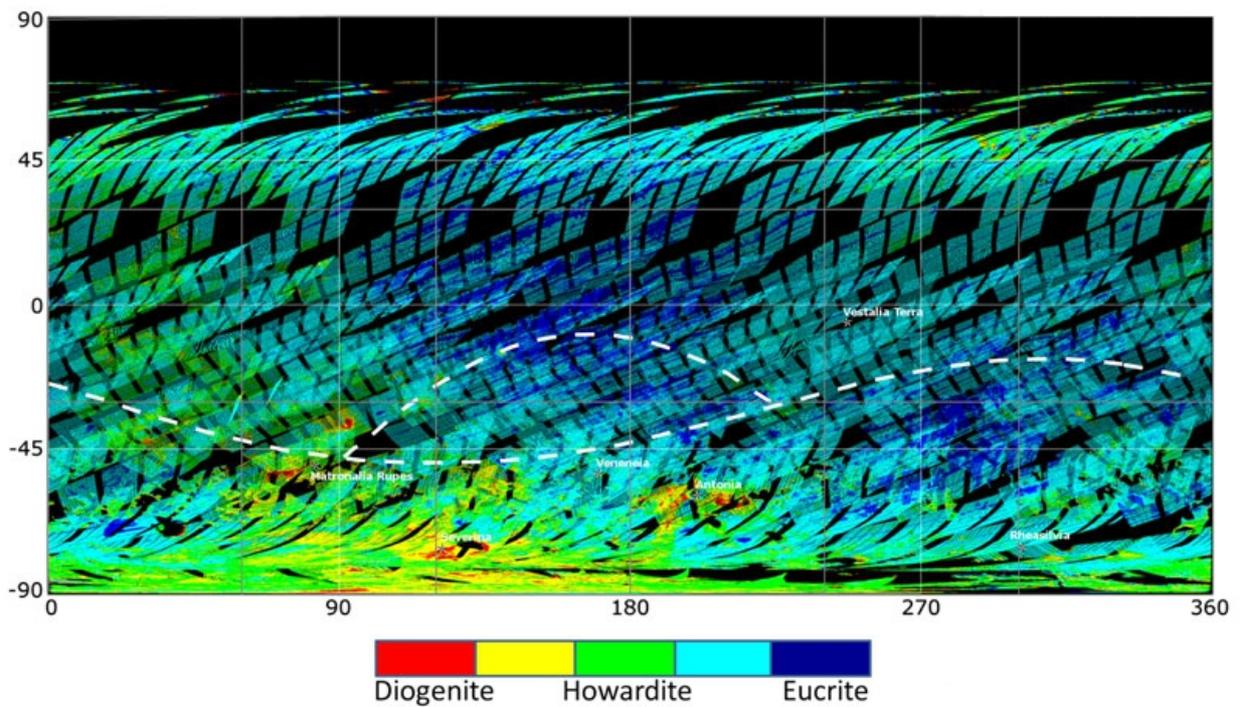

**Fig. 63 - Lithological map of Vesta -** Longitudes and latitudes are indicated on the x and y axis. (Adapted from *Ammannito et al., 2013*)

# SUMMARY AND CONCLUSIONS

We have performed the spectrophotometric correction of the asteroid Vesta data obtained by the VIR instrument.

The wide spectral range covered by VIR (0.25-1.05 μm and 1.0-5.0 μm) permits a proper characterization of the spectral properties of Vesta, both in the VIS and in the IR ranges, including the two pyroxene absorption bands at 0.9 μm and 1.9 μm. Even with the exception of the thermal emission range above ≈3.5 μm and the spectral bands related to the filters on the detectors, we can still produce our photometric correction over several hundreds of different wavelengths. Among these, the limited set of wavelengths that we have selected, allowed us to perform the photometric correction in a short computational time, preserving the trend of Vesta's spectrum and the wide spectral range.

We used the data acquired during both the HAMO phases and the Survey phase. The data have been filtered out of pixels corresponding to extremely high and low values of reflectance (0.001<reflectance<1.0) and of those acquisition obtained at high incidence and emission angles (i>70° and e>70°). Each of the two filtered data-sets has been photometrically corrected over selected wavelengths.

The photometric correction was obtained through a five parameters Hapke's model (*Hapke, 2012*), which takes into account the contributions of single scattering albedo for a typical particle of the medium (*w*), that of the Opposition Effect ($B_0$, *h*), the one related to the phase function (*b*) and that of the large scale roughness of the surface (*θ*).

The lack of data points at small phase angles (<2°) prevented us to model the Coherent Backscattering Opposition Effect. Consequently, we have assumed $B_{CB}$=1, since its effects at the larger phase angles typical of our data can be neglected. Also a proper characterization of SHOE is not possible, because of the given phase angle coverage (≈28°/≈67° for the HAMO, ≈12°/≈78° for the Survey). However, SHOE can still, partially, affect the phase curve up to 20°/30°: for this reason, in our analysis we assumed the amplitude $B_0$=1.03, as derived by *Helfenstein* and *Veverka, 1989,* and we let the angular width *h,* free to vary. A set of best-fit parameters is derived (see **Table 9** and **10**) for each of the wavelengths and for each observational phase (both the HAMO and the Survey).

The *h* parameter appears to be poorly constrained; it shows some degree of dependence from the wavelength, mainly due to fit degeneration with other parameters (see **Fig. 32, 33** and **34**).

As in *Li et al., (2013)*, we decided to fix the *h* parameter to 0.07 and we derived new sets of best-fit parameters (see **Table 11** and **12**). *w* shows no significant discrepancies with respect to the previous analysis (**Fig. 39**), suggesting that the degeneration with Opposition Effect parameters is limited. *θ* values are a 5% (in the VIS range) and a 10% (in the IR range) higher than the previous analysis; however, this parameter is qualitatively similar to that obtained with *h* free to vary, and shows a decreasing trend with the wavelength in the IR region, suggesting that the large scale roughness on the surface is caused by topographic shadows at scales much larger than the wavelengths of the FC data (*Li et al., 2013*) or that a larger *w* allows a higher contribution of inter-facets scattering. The negative *b* parameters that we found (in general spanning from -0.18 to -0.34), indicates a back-scattering phase function; for a fixed *h, b* appears to be higher at increasing values of the albedo, which means a more forward-scattering phase function in this condition and particles more transparent to the light.

The sets of parameters obtained, have been used to correct the reflectance distribution, and to report the radiance factor I/F to the standard observational geometry ($i$=30°, $e$=0°, $\alpha$=30°). For a test wavelength (0.796 µm), the trend of reflectance distribution is now flattened with respect to the phase angle. In fact, comparing the angular coefficient of a linear fit of the distribution, before and after the photometric correction, it can be noted (**Fig. 40** and **Fig. 44**) that it changes from 2.81*E-3 to 8.88*E-5 for the Survey case, while it goes from 3.79*E-3 to 5.70*E-5 in the HAMO sequence. Also the trend of reflectance with respect to the incidence and emission angle is removed (**Fig. 42, 43, 45** and **46**), confirming that the photometric correction that we have performed reduces the geometrical dependence.

In order to further test the accuracy of our photometric model we simulated the observations taken during Survey and HAMO and compared them with real VIR measurements obtaining a 6% RMS (root mean squared difference between model and data) and a 2.3% RMS, respectively.

We considered the solution obtained with fixed values of $B_0$ and $h$ as the nominal one. The corresponding parameters have been used to correct the data-set and produce albedo maps in the VIS and IR continuum, at 0.751 µm and 1.2 µm.

From the comparison of **Fig. 20 and 25** with **Fig. 49 and 50,** it can be noted that artifacts present in the albedo maps before the photometric correction and due to observation geometry effects are removed in the photometrically corrected ones. From the corrected maps obtained by merging HAMO and Survey data (**Fig. 55** and **56**), it emerges a strong regional reflectance variability; in particular, we note the east-west dichotomy suggested by HST data (*Binzel et al., 1997*), with the eastern hemisphere having a higher albedo (0.23-0.26) than the western one (0.18-0.20). A comparison with *Li et al., (2013)* (**Table 15**), confirm that Vesta is a really bright asteroid (at 550 nm we found $w$=0.494 against $w$=0.500 by *Li et al., 2013*), more than three times bright than Ceres ($w$=0.14, *Ciarniello et al., 2016*), the second focus of the Dawn mission, and its $w$ is among the higher of the observed asteroid and cometary nuclei (see *Ciarniello et al., (2016)*, for a comparison with other minor bodies). At a smaller scale we confirm, through our photometric correction, that Marcia-Calpurnia craters are embedded in a dark region and display a prominent dark ejecta field, while the craters walls and rims are brighter (a difference of about 75% in the IR and 50% in the VIS). These albedo differences have been associated to different composition of the surface and grain sizes (*De Sanctis et al., 2014*). Also other smaller craters emerge as bright (es. Justina, Canuleia, and Tuccia) or dark areas (es. Aricia Tholus).

A first analysis of the dark (*Palomba et al., 2014*) and bright regions (*Zambon et al., 2014*) have been done, but the precise photometric behavior of spectral parameters in these terrains (band depth, band area and band ratio) must be derived. The correction is needed to really distinguish regions in terms of compositions, multiple scattering efficiency and grain size.

We have shown useful applications of the photometric correction through relative difference of the maps obtained in the IR continuum and in the VIS continuum (**Fig. 53** and **54**), and enhanced color maps in the VIS (R=440nm, G=550nm and B=751 nm) and in the IR (R=2402nm, G=1702nm and B=1200nm) regions (**Fig. 61** and **62**).

These ratio maps, provides information on the spectral slope, that can be used to study the space weathering. According to the relative difference of the maps, it can be noted that Vestan surface does not show a clear spectral reddening. This has been interpreted as due to material mixing (*Pieters et al., 2012*).

Structures like the Oppia crater clearly emerges in the RGB map produced with the VIS wavelengths and appear to be related to terrain with a different composition with respect to the average surrounding surface (*Le Corre et al, 2013*). The RGB map obtained at IR wavelengths is very well correlated with the lithological maps presented in *Ammannito et al., (2013)*, and allow us to confirm a difference among dark terrains (more eucritic) and brighter ones (less eucritic).

Comparisons with respect to previous works (*Li et al., 2013, Longobardo et al., 2014, McCord et al., 2012*), show different maps obtained for Vestan surface. In particular, the photometrically corrected map of *Li et al., (2013)* at 0.749 µm, derived using the Framing Camera data-set (which is characterized by higher spatial resolution, lower spectral resolution and limited spectral coverage with respect to VIR), show a strong correlation with our corrected map at 0.751 µm. In addition, the map obtained after the photometric correction by *Longobardo et al., (2014)* at 1.2 µm has been compared to our map. These maps show small differences due to the different approaches (in *Longobardo et al., (2014)* they used an empirical approach based on Akimov photometric model) and the solutions adopted during the data analysis.

The photometrically corrected maps described in this work represent few examples of the future applications and different studies that can be performed, thanks to the photometric correction that we have produced.

As a further step of this work it could be useful to extend the study to a larger set of wavelengths, producing an even finer characterization of the spectral properties of Vesta. However new analysis strategies have to be adopted, in order to reduce the computational time of the model parameters derivation. The possibility to map different spectral parameters as band depths, slopes and band centers can provide valuable information to infer compositional and physical properties of Vestan surface, and their variability on local and regional scale.

# SOMMARIO E CONCLUSIONI

In questo lavoro di tesi abbiamo effettuato la correzione spettrofotometrica dei dati dell'asteroide Vesta, ottenuti per mezzo dello spettrometro ad immagine VIR, a bordo della missione Dawn.

L'ampio range di lunghezze d'onda a nostra disposizione (0.25-1.05 μm e 1.0-5.0 μm), rende possibile un'adeguata caratterizzazione delle proprietà spettrali di Vesta, sia nel continuo che nelle bande di assorbimento tipiche dei pirosseni a 0.9 μm e 1.9 μm. Anche dopo l'esclusione della regione di emissione termica oltre 3.5 μm e delle posizioni dei filtri di banda, siamo comunque in grado di produrre una valida correzione fotometrica su diverse centinaia di lunghezze d'onda. Dati i lunghi tempi computazionali, si è scelto di operare su un set più limitato di lunghezze d'onda, in grado comunque di caratterizzare il trend dello spettro osservato di Vesta nell'ampio range spettrale offerto dallo strumento VIR.

Entrambe le sequenze HAMO e Survey, sono state adoperate in questo lavoro. In primo luogo si è proceduto ad un'adeguata selezione dei dati, acquisiti in corrispondenza di geometrie particolarmente sfavorevoli (i>70°, e >70°) o caratterizzati da valori spuri della riflettanza misurata (0.001<riflettanza<1.0). I due data-set sono stati poi separatamente corretti su ciascuna delle lunghezze d'onda in esame.

La correzione fotometrica è stata ottenuta per mezzo di un modello di Hapke a 5 parametri (*Hapke, 2012*), che tiene in conto i contributi del singolo scattering su una particella tipica del mezzo ($w$), l'effetto di opposizione ($B_0$, $h$), quello relativo alla funzione di fase ($b$) e la rugosità superficiale a larga scala ($\theta$).

La mancanza di dati ad angoli di fase molto piccoli *(<2°)*, rende impossibile la caratterizzazione dell'Effetto di Opposizione dovuto al Backscattering Coerente. Dal momento che tale effetto ad angoli di fase più alti può essere trascurato, si è deciso di eliminarne la dipendenza all'interno del nostro modello (assumendo $B_{CB}$=1). Anche un'adeguata modellizzazione dell'Effetto di Opposizione di Shadow Hiding può essere difficoltosa, a causa di una copertura non favorevole ad angoli bassi (l'angolo di fase coperto dalle due sequenze è ≈28°/≈67° per la HAMO e ≈12°/≈78° per la Survey). Tuttavia tale effetto può ancora incidere fino a 20°/30° in angolo di fase; in una prima analisi è stato perciò necessario includerlo, assumendo da letteratura (*Helfenstein* e *Veverka, 1989*) $B_0$=1.03 e lasciando libera di variare l'ampiezza angolare *h*. Un set di parametri è ottenuto (**Tab. 9** e **10**) per ciascuna delle lunghezze d'onda in esame e per entrambe le sequenze osservative (HAMO e Survey).

Il parametro *h* sembra essere difficile da caratterizzare per mezzo di questo approccio, poiché mostra una forte variabilità con la lunghezza d'onda, principalmente a causa di un fattore degenerativo tra parametri (cfr. **Fig. 32, 33 e 34**).

Decidiamo di seguire lo stesso approccio di *Li et al., (2013)*, fissando *h* a 0.07 e derivando i nuovi set per i parametri liberi del modello (**Tab. 11** e **12**). *w* non mostra rilevanti differenze rispetto all'analisi condotta precedentemente (**Fig. 39**) e ciò ci permette di dedurre una bassa degenerazione di questo parametro con i due che descrivono l'Effetto di Opposizione. *θ* è qualitativamente simile nei due casi (*h* libero e *h* fissato a 0.07), ma in entrambe le sequenze osservative mostra un incremento, rispettivamente del 5% nel range del Visibile e del 10% nel range dell'Infrarosso; l'anticorrelazione notata con w, suggerisce un effetto di scattering tra le diverse faccette delle particelle del mezzo, ma può essere anche dovuta alla topografia superficiale a scale tipiche delle lunghezze d'onda dell'IR (*Li et al., 2013*). Il parametro *b* è negativo per tutti i set derivati (varia tra -0.18 e -0.34), e ciò si lega a una funzione di fase back-scattering, che diventa più forward-scattering in corrispondenza di valori più alti dell'albedo.

I set di parametri ottenuti, sono stati poi usati per correggere i valori di riflettanza osservata e riportare il fattore di radianza I/F alle condizioni di geometria standard ($i=30°$, $e=0°$, $\alpha=30°$). Per una data lunghezza d'onda (0.796 µm), è stato dimostrato come la riflettanza sia corretta rispetto alle geometrie di osservazione e in particolare l'angolo di fase (**Fig. 40** e **44**). Infatti, effettuando un fit lineare della distribuzione, si osserva come il coefficiente angolare cambi da 2.81*E-3 a 8.88*E-5 nella Survey e da 3.79*E-3 a 5.70*E-5 nel caso HAMO. Analogamente è eliminata anche la dipendenza della riflettanza dall'angolo di incidenza e di emissione (**Fig. 42, 43 45** e **46**).

Per testare l'accuratezza del nostro modello fotometrico, abbiamo simulato le osservazioni acquisite durante la Survey e la HAMO e le abbiamo confrontate con quelle realmente ottenute da VIR, ottenendo un RMS (root mean squared difference) pari al 6% per la Survey e al 2% per la HAMO.

La soluzione in cui l'Effetto di Opposizione è fissato in entrambi i suoi parametri, è adottata per correggere il dataset e produrre le mappe di albedo sia nel Visibile che nell'Infrarosso, a due lunghezze d'onda di riferimento (0.751 µm e 1.2 µm). Confrontando **Fig. 20** e **Fig. 25** con **Fig. 49** e **Fig. 50**, si può vedere che gli artefatti dovuto alle geometrie sono rimossi per mezzo della correzione fotometrica apportata. La grande variabilità intrinseca dell'intera superficie di Vesta può essere allora osservata (**Fig. 55** e **56**). La dicotomia emisferica inferita dai dati dell'HST (*Binzel et al., 1997*) tra la regione a Est (w=0.23-0.26) e quella a Ovest (0.18-0.20) di Vesta, è confermata. Un confronto con *Li et al., (2013)* (**Tabella 15**), conferma che Vesta è un asteroide brillante (a 550 nm *w*=0.494 contro il *w*=0.500 trovato da *Li et al., 2013*), tre volte più brillante di Cerere (*w*=0.14, *Ciarniello et al., 2016*), il secondo obiettivo di Dawn, e tra i più chiari tra gli asteroidi e i nuclei cometari osservati (cfr *Ciarniello et al., (2016),* per un confronto con gli altri corpi minori).

A scale più piccole, si osservano regioni a basso (es. Aricia Tholus) e alto albedo (es. Justina, Canuleia, and Tuccia,). Grazie alla nostra correzione fotometrica si può notare che i crateri Marcia e Calpurnia mostrano un forte contrasto tra le pareti chiare e il campo dei detriti (una variazione di albedo dell'ordine del 75% nell'IR e del 50% nel VIS). Come mostrato in *De Sanctis et al., (2014),* questa differenza di albedo è associata a diverse composizioni e dimensioni dei grani.

Una prima analisi delle regioni chiare di Vesta (*Zambon et al., 2014*) e di quelle scure (*Palomba et al., 2014*) è stato effettuato, ma un set fotometricamente corretto può permettere uno studio più accurato dei parametri spettrali (profondità di banda, area di banda, rapporto di banda) e quindi dei processi fisici e della composizione delle diverse regioni superficiali.

Utili applicazioni di questo lavoro di correzione fotometrica sono le mappe di differenza spettrale tra il continuo dell'IR e del VIS (**Fig. 53** e **54**), o le mappe in falsi colori nel VIS (R=440nm, G=550nm and B=751 nm) e nell'IR (R=2402nm, G=1702nm and B=1200nm) (**Fig. 61** e **62**).

Le mappe di differenza spettrale dà informazioni sulla spectral slope, che permette di studiare lo space weathering. Nelle mappe di differenza spettrale non si può osservare un chiaro arrossamento spettrale. Ciò è stato interpretato come effetto di un "material mixing". (*Pieters et al., 2012*).

Strutture peculiari, come per esempio il cratere Oppia, si notano nella mappa RGB nelle lunghezze d'onda del Visibile; questa struttura sembra essere associata a una differente composizione rispetto alla superficie circostante (*Le Corre et al, 2013*). La mappa RGB dell'IR è estremamente correlata con le mappe litologiche presentate in *Ammannito et al., (2013)*, e permette ti confermare la differenza tra terreni più scuri (più eucritici) e quelli più chiari (meno eucritici).

Confronti con altri lavori (*Li et al., 2013, Longobardo et al., 2014, McCord et al., 2012*) mostrano diverse mappe ottenute per la superficie di Vesta. In particolare, la mappa fotometricamente corretta di *Li et al., (2013)* a 0.749 μm, ottenuta usando i dati della Framing Camera (che ha una migliore risoluzione spaziale, ma una peggiore risoluzione e copertura spettrale più limitata rispetto a VIR), mostra una forte correlazione con la nostra mappa a 0.751 μm. Inoltre, la mappa ottenuta dopo la correzione fotometrica da *Longobardo et al., (2014)* a 1.2 μm mostra alcune differenze dovute ai diversi approcci (in *Longobardo et al., (2014)* è stato usato un metodo empirico) e alle soluzioni adoperate nel corso dell'analisi.

Le mappe fotometricamente corrette che abbiamo descritto in questo lavoro, rappresentano solo pochi esempi di future applicazioni e dei diversi studi che possono essere condotti a partire dai parametri del modello di Hapke che abbiamo ottenuto in questo lavoro.

Potrebbe essere utile estendere il lavoro a un più largo set di lunghezze d'onda, per caratterizzare in modo ancor più fine le proprietà spettrali di Vesta. Per ridurre il tempo computazionale, sarà necessario adottare nuove strategie di analisi. La possibilità di creare mappe di differenti parametri spettrali, come profondità di banda, pendenze spettrali e centri di banda, può fornire gli strumenti per studiare le proprietà fisiche e la composizione della superficie di Vesta, e la loro variabilità su scala regionale e locale.

# Bibliography

Akimov, L. A., 1988. Light reflection by the moon. Kinematika i Fizika Nebesnykh Tel (ISSN 0233-7665), vol. 4, Jan.-Feb. 1988, p. 3-10. In Russian.

Aikio M., 2001. Hyperspectral prism-grating-prism imaging spectrograph.

Ammannito, E., de Sanctis, M. C., Palomba, E., Longobardo, A., Mittlefehldt, D. W., McSween, H. Y., Marchi, S., Capria, M. T., Capaccioni, F., Frigeri, A., Pieters, C. M., Ruesch, O., Tosi, F., Zambon, F., Carraro, F., Fonte, S., Hiesinger, H., Magni, G., McFadden, L. A., Raymond, C. A., Russell, C. T., Sunshine, J. M., 2013. Olivine in an unexpected location on Vesta's surface. Nature, Volume 504, Issue 7478, pp. 122-125 (2013).

Baer, J., Chesley, S. R., 2008. Astrometric masses of 21 asteroids, and an integrated asteroid ephemeris. Celestial Mechanics and Dynamical Astronomy, Volume 100, Issue 1, pp.27-42.

Binzel, R. P., Xu, S., 1993. Chips off of asteroid 4 Vesta - Evidence for the parent body of basaltic achondrite meteorites. Science (ISSN 0036-8075), vol. 260, no. 5105, p. 186-191.

Binzel, R. P., Gaffey, M. J., Thomas, P. C., Zellner, B. H., Storrs, A. D., Wells, E. N., 1997. Geologic Mapping of Vesta from 1994 Hubble Space Telescope Images. Icarus, Volume 128, Issue 1, pp. 95-103.

Bowell, E., Hapke, B., Domingue, D., Lumme, K., Peltoniemi, J., Harris, A. W., 1989. Application of photometric models to asteroids. Asteroids II, Proceedings of the Conference, Tucson, AZ, Mar. 8-11, 1988 (A90-27001 10-91). Tucson, AZ, University of Arizona Press, 1989, p. 524-556.

Brown, R. H., Baines, K. H., Bellucci, G., Bibring, J.-P., Buratti, B. J., Capaccioni, F.,

Cerroni, P., Clark, R. N., Coradini, A., Cruikshank, D. P., Drossart, P., Formisano, V., Jaumann, R., Langevin, Y., Matson, D. L., McCord, T. B., Mennella, V., Miller, E., Nelson, R. M., Nicholson, P. D., Sicardy, B., Sotin, C., 2004. Mapping Spectrometer (Vims) Investigation. Space Science Reviews, Volume 115, Issue 1-4, pp. 111-168.

Capaccioni, F., Coradini, A., Filacchione, G., Erard, S., Arnold, G., Drossart, P., De Sanctis, M. C., Bockelee-Morvan, D., Capria, M. T., Tosi, F., Leyrat, C., Schmitt, B., Quirico, E., Cerroni, P., Mennella, V., Raponi, A., Ciarniello, M., McCord, T., Moroz, L., Palomba, E., Ammannito, E., Barucci, M. A., Bellucci, G., Benkhoff, J., Bibring, J. P., Blanco, A., Blecka, M., Carlson, R., Carsenty, U., Colangeli, L., Combes, M., Combi, M., Crovisier, J., Encrenaz, T., Federico, C., Fink, U., Fonti, S., Ip, W. H., Irwin, P., Jaumann, R., Kuehrt, E., Langevin, Y., Magni, G., Mottola, S., Orofino, V., Palumbo, P., Piccioni, G., Schade, U., Taylor, F., Tiphene, D., Tozzi, G. P., Beck, P., Biver, N., Bonal, L., Combe, J.-Ph., Despan, D., Flamini, E., Fornasier, S., Frigeri, A., Grassi, D., Gudipati, M., Longobardo, A., Markus, K., Merlin, F., Orosei, R., Rinaldi, G., Stephan, K., Cartacci, M., Cicchetti, A., Giuppi, S., Hello, Y., Henry, F., Jacquinod, S., Noschese, R., Peter, G., Politi, R., Reess, J. M., Semery, A, 2015. The organic-rich surface of comet 67P/Churyumov-Gerasimenko as seen by VIRTIS/Rosetta. Science, Volume 347, Issue 6220, article id. aaa0628.

Capria M.T., 2014. VIR standard data products and archive volume software interface specification.

Chamberlin, 2007. http://ssd.jpl.nasa.gov/images/ast_histo.ps.

Ciarniello, M., De Sanctis, M. C., Ammannito, E., Raponi, A., Longobardo, A., Palomba, E., Carrozzo, F. G., Tosi, F., Li, J.-Y., Schröder, S., Zambon, F., Frigeri, A., Fonte, S., Giardino, M., Pieters, C. M., Raymond, C. A., Russell, C. T., 2016. Spectrophotometric properties of dwarf planet Ceres from VIR onboard Dawn mission. eprint ArXiv:1608.04643.

Clark, B.E., 1993. Spectral reflectance studies and optical surface alteration in the search for links between meteorites and asteroids.

Clark, R. N. 1999. Spectroscopy of Rocks and Minerals, and Principles of Spectroscopy. Derived from Chapter 1 in: Manual of Remote Sensing.

Consolmagno, G. J., Drake, M. J., 1977. Composition and evolution of the eucrite parent body - Evidence from rare earth elements. Geochimica et Cosmochimica Acta, vol. 41, Sept. 1977, p. 1271-1282.

Coradini, A., Capaccioni, F., Drossart, P., Semery, A., Arnold, G., Schade, U., Angrilli, F., Barucci, M. A., Bellucci, G., Bianchini, G., Bibring, J. P., Blanco, A., Blecka, M., Bockelee-Morvan, D., Bonsignori, R., Bouye, M., Bussoletti, E., Capria, M. T., Carlson, R., Carsenty, U., Cerroni, P., Colangeli, L., Combes, M., Combi, M., Crovisier, J., Dami, M., DeSanctis, M. C., DiLellis, A. M., Dotto, E., Encrenaz, T., Epifani, E., Erard, S., Espinasse, S., Fave, A., Federico, C., Fink, U., Fonti, S., Formisano, V., Hello, Y., Hirsch, H., Huntzinger, G., Knoll, R., Kouach, D., Ip, W. H., Irwin, P., Kachlicki, J., Langevin, Y., Magni, G., McCord, T., Mennella, V., Michaelis, H., Mondello, G., Mottola, S., Neukum, G., Orofino, V., Orosei, R., Palumbo, P., Peter, G., Pforte, B., Piccioni, G., Reess, J. M., Ress, E., Saggin, B., Schmitt, B., Stern, A., Taylor, F., Tiphene, D., Tozzi, G., 1998. Virtis : an imaging spectrometer for the rosetta mission. Planetary and Space Science, Volume 46, Issue 9-10, p. 1291-1304.

Cruikshank, D. P., Tholen, D. J., Bell, J. F., Hartmann, W. K., Brown, R. H., 1991. Three basaltic earth-approaching asteroids and the source of the basaltic meteorites. Icarus (ISSN 0019-1035), vol. 89, Jan. 1991, p. 1-13.

De Sanctis, M. C., Coradini, A., Ammannito, E., Filacchione, G., Capria, M. T., Fonte, S., Magni, G., Barbis, A., Bini, A., Dami, M., Ficai-Veltroni, I., Preti, G., 2011. The VIR Spectrometer. Space Science Reviews, Volume 163, Issue 1-4, pp. 329-369.

De Sanctis, M. C., Ammannito, E., Capria, M. T., Tosi, F., Capaccioni, F., Zambon, F.,

Carraro, F., Fonte, S., Frigeri, A., Jaumann, R., Magni, G., Marchi, S., McCord, T. B., McFadden, L. A., McSween, H. Y., Mittlefehldt, D. W., Nathues, A., Palomba, E., Pieters, C. M., Raymond, C. A., Russell, C. T., Toplis, M. J., Turrini, D., 2012. Spectroscopic Characterization of Mineralogy and Its Diversity Across Vesta. Science, Volume 336, Issue 6082, pp. 697- (2012).

De Sanctis, M. Cristina, Ammannito, Eleonora, Capria, M. Teresa, Capaccioni, Fabrizio, Combe, Jean-Philippe, Frigeri, Alessandro, Longobardo, Andrea, Magni, Gianfranco, Marchi, Simone, McCord, Tom B., Palomba, Ernesto, Tosi, Federico, Zambon, Francesca, Carraro, Francesco, Fonte, Sergio, Li, Y. J., McFadden, Lucy A., Mittlefehldt, David W., Pieters, Carle M., Jaumann, Ralf, Stephan, Katrin, Raymond, Carol A., Russell, Christopher T., 2013. Vesta's mineralogical composition as revealed by the visible and infrared spectrometer on Dawn. Meteoritics & Planetary Science, Volume 48, Issue 11, pp. 2166-2184.

De Sanctis, M., Ammannito, E., Frigeri, A., Tosi, F., Zambon, F., Raymond, C., Russell, C., 2014. Mineralogy of the Marcia crater on Vesta. Asteroids, Comets, Meteors 2014. Proceedings of the conference held 30 June - 4 July, 2014 in Helsinki, Finland. Edited by K. Muinonen et al.

Ermakov, Anton I., Zuber, Maria T., Smith, David E., Raymond, Carol A., Balmino, Georges, Fu, Roger R., Ivanov, Boris A., 2014. Constraints on Vesta's interior structure using gravity and shape models from the Dawn mission. Icarus, Volume 240, p. 146-160.

Gaffey, M.J., 1997. Surface lithologic heterogeneity of Asteroid 4 Vesta. Icarus, 127 (1997), pp. 130–157.

Ghosh, A., McSween, H. Y., 1998. A Thermal Model for the Differentiation of Asteroid 4 Vesta, Based on Radiogenic Heating. Icarus, Volume 134, Issue 2, pp. 187-206.

Fienga, A., Laskar, J., Morley, T., Manche, H., Kuchynka, P., Le Poncin-Lafitte, C.,

Budnik, F., Gastineau, M., Somenzi, L., 2009. INPOP08, a 4-D planetary ephemeris: from asteroid and time-scale computations to ESA Mars Express and Venus Express contributions. Astronomy and Astrophysics, Volume 507, Issue 3, 2009, pp.1675-1686.

Filacchione G., 2006. Calibrazioni a terra e prestazioni in volo di spettrometri ad immagine nel visibile e nel vicino Infrarosso per l'esplorazione planetaria.

Filacchione G., Ammannito E., 2014. Dawn Vir calibration document.

Formisano, M., Federico, C., Turrini, D., Capaccioni, F., 2012. Modelling the thermal history of Vesta: time scale of accretion and differentiation. EGU General Assembly 2012, held 22-27 April, 2012 in Vienna, Austria., p.8555.

Gehrels, T., Coffeen, T., Owings, D., 1965. Wavelength dependence of polarization. III. The lunar surface. Astronomical Journal, Vol. 70, p. 447.

Hapke, B., van Horn, H., 1963. Photometric studies of complex surfaces, with applications to the Moon. Journal of Geophysical Research, vol. 68, p. 4545-4570 (1963).

Hapke, B., 1981. Bidirectional reflectance spectroscopy. I - Theory.

Hapke, B., 1984. Bidirectional reflectance spectroscopy. III - Correction for macroscopic roughness.

Hapke, B., 1986. Bidirectional reflectance spectroscopy. IV - The extinction coefficient and the opposition effect

Hapke, B., 1993. Theory of reflectance and emission spectroscopy.

Hapke, B., 2002. Bidirectional Reflectance Spectroscopy. 5. The Coherent Backscatter Opposition Effect and Anisotropic Scattering. Icarus, Volume 157, Issue 2, p. 523-534.

Hapke, B., 2008. Bidirectional reflectance spectroscopy. 6. Effects of porosity. Icarus, Volume 195, Issue 2, p. 918-926.

Hapke, B., 2012. Bidirectional reflectance spectroscopy 7. The single particle phase function hockey stick relation. Icarus, Volume 221, Issue 2, p. 1079-1083.

Helfenstein, P., Veverka, J., 1989. Physical characterization of asteroid surfaces from photometric analysis. Asteroids II, Proceedings of the Conference, Tucson, AZ, Mar. 8-11, 1988 (A90-27001 10-91). Tucson, AZ, University of Arizona Press, 1989, p. 557-593.

Helfenstein, P., Veverka,J., Hillier,J., 1997. The Lunar Opposition Effect: A Test of Alternative Models. Icarus, Volume 128, Issue 1, pp. 2-14.

Henyey, L. G., Greenstein, J. L., 1941. Diffuse radiation in the Galaxy. Astrophysical Journal, vol. 93, p. 70-83 (1941).

Ikeda, Y., Takeda, H., 1985. A model for the origin of basaltic achondrites based on the Yamato 7308 howardite. Journal of Geophysical Research, Vol. 90, Suppl., p. C649 – C663.

Le Corre,L., Reddy, V., Schmedemann,N., Becker, K. J., O'Brien, D. P., Yamashita, Naoyuki, Peplowski, P. N., Prettyman, T. H., Li, J.-Y., Cloutis, E. A., Denevi, B. W., Kneissl, T., Palmer, Eric, Gaskell, R. W., Nathues, A., Gaffey, M. J., Mittlefehldt, D. W., Garry, W. B., Sierks, H., Russell, C. T., Raymond, C. A., De Sanctis, M. C., Ammanito, E., 2013. Olivine or impact melt: Nature of the "Orange" material on Vesta from Dawn. Icarus, Volume 226, Issue 2, p. 1568-1594.

Li, J.-Y., Le Corre, L., Schröder, S. E., Reddy, V., Denevi, B. W., Buratti, B. J., Mottola, S., Hoffmann, M., Gutierrez-Marques, P., Nathues, A., Russell, C. T., Raymond, C. A., 2013. Global photometric properties of Asteroid (4) Vesta observed with Dawn Framing Camera . Icarus, Volume 226, Issue 2, p. 1252-1274.

Li, J.-Y., Buratti, B. J., De Sanctis, C. M., Denevi, B. W., Hoffmann, M., Longobardo, A., Mottola, S., Nathues, A., Reddy, V., Russell, C. T., Schröder, S. E., 2014. The Photometric Properties of Vesta and the Implications. 45th Lunar and Planetary

Science Conference, held 17-21 March, 2014 at The Woodlands, Texas. LPI Contribution No. 1777, p.1583.

Li J.-Y., Helfenstein P., Buratti B.J., Takir D., Clark B.E., 2015. Asteroid photometry.

Longobardo, A., Palomba, E., Capaccioni, F., De Sanctis, M. C., Tosi, F., Ammannito, E., Schröder, S. E., Zambon, F., Raymond, C. A., Russell, C. T., 2014. Photometric behavior of spectral parameters in Vesta dark and bright regions as inferred by the Dawn VIR spectrometer. Icarus, Volume 240, p. 20-35.

Konopliv, A. S., Yoder, C. F., Standish, E. M., Yuan, Dah-Ning, Sjogren, W. L., 2006. A global solution for the Mars static and seasonal gravity, Mars orientation, Phobos and Deimos masses, and Mars ephemeris. Icarus, Volume 182, Issue 1, p. 23-50.

Konopliv, A. S., Asmar, S. W., Bills, B. G., Mastrodemos, N., Park, R. S., Raymond, C. A., Smith, D. E., Zuber, M. T., 2011. The Dawn Gravity Investigation at Vesta and Ceres. Space Science Reviews, Volume 163, Issue 1-4, pp. 461-486.

Kuzmanoski, M., Apostolovska, G., Novaković, B., 2010. The Mass of (4) Vesta Derived from its Largest Gravitational Effects. The Astronomical Journal, Volume 140, Issue 3, pp. 880-886 (2010).

Lambert, J.H., 1759. L perspective affranchie de l'embaras du Plan geometral. Zurich : Heidegguer, 1759, VIII, 192 p. : 6 tavv. f. t. , in 8., DCC.16.29.

Longhi, J., Pan, V., 2008. Phase equilibrium constraints on the howardite-eucrite-diogenite association. Lunar and Planetary Science Conference, 18th, Houston, TX, Mar. 16-20, 1987, Proceedings (A89-10851 01-91). Cambridge and New York/Houston, TX, Cambridge University Press/Lunar and Planetary Institute, 1988, p. 459-470.

Mandler, B. E., Elkins-Tanton, L. T., 2013. The origin of eucrites, diogenites, and olivine diogenites: Magma ocean crystallization and shallow magma chamber processes on Vesta. Meteoritics & Planetary Science, Volume 48, Issue 11, pp. 2333-2349.

Markwardt, C.B., 2008. Non-linear least squares fitting in IDL with MPFIT. Astronomical Data Analysis Software and Systems XVII.

McSween, H. Y., Binzel, R. P., de Sanctis, M. C., Ammannito, E., Prettyman, T. H., Beck, A. W., Reddy, V., Corre, L., Gaffey, M. J., McCord, T. B., Raymond, C. A., Russell, C. T., 2013. Dawn, the Vesta-HED connection, and the geologic context for eucrites, diogenites, and howardites. Meteoritics & Planetary Science, Volume 48, Issue 11, pp. 2090-2104.

Michalak, G., 2000. Determination of asteroid masses --- I. (1) Ceres, (2) Pallas and (4) Vesta. Astronomy and Astrophysics, Vol. 360, p.363-374.

Mie, G., 1908. Beiträge zur Optik trüber Medien, speziell kolloidaler Metallösungen. Annalen der Physik, vol. 330, Issue 3, pp.377-445.

McCord, T. B., Adams, J. B., Johnson, T. V., 1970. Asteroid Vesta: Spectral Reflectivity and Compositional Implications. Science, Volume 168, Issue 3938, pp. 1445-1447.

McCord, T. B., Li, J.-Y., Combe, J.-P., McSween, H. Y., Jaumann, R., Reddy, V., Tosi, F., Williams, D. A., Blewett, D. T., Turrini, D., Palomba, E., Pieters, C. M., de Sanctis, M. C., Ammannito, E., Capria, M. T., Le Corre, L., Longobardo, A., Nathues, A., Mittlefehldt, D. W., Schröder, S. E., Hiesinger, H., Beck, A. W., Capaccioni, F., Carsenty, U., Keller, H. U., Denevi, B. W., Sunshine, J. M., Raymond, C. A., Russell, C. T., 2012. Dark material on Vesta from the infall of carbonaceous volatile-rich material. Nature, Volume 491, Issue 7422, pp. 83-86 (2012).

McSween, H. Y., Grove, T. L., Wyatt, M. B., 2003. Constraints on the composition and petrogenesis of the Martian crust. Journal of Geophysical Research, Volume 108, Issue E12, pp. 9-1, CiteID 5135, DOI 10.1029/2003JE002175.

Minnaert, M., 1941. The reciprocity principle in lunar photometry. Astrophysical Journal, vol. 93, p. 403-410 (1941).

Neumann, W., Breuer, D., Spohn, T., 2013. Early magma ocean and core formation on Vesta. EGU General Assembly 2013, held 7-12 April, 2013 in Vienna, Austria, id. EGU2013-13487.

Nicodemus, F.E., 1970. Reflectance Nomenclature and Directional Reflectance and Emissivity. Applied Optics, vol. 9, issue 6, p. 1474.

Nicodemus, F. E., Richmond, J. C., Hsia, J. J., Ginsberg, I. W., Limperis, T., 1977. Geometrical considerations and nomenclature for reflectance. Final Report National Bureau of Standards, Washington, DC. Inst. for Basic Standards.

Palomba, E., Longobardo, A., De Sanctis, M. C., Zambon, Francesca, Tosi, F., Ammannito, E., Capaccioni, F., Frigeri, A., Capria, M. T., Cloutis, E. A., Jaumann, R., Combe, J.-P., Raymond, C.l A., Russell, C. T., 2014. Composition and mineralogy of dark material units on Vesta. Icarus, Volume 240, p. 58-72.

Pieters, C. M., Ammannito, E., Blewett, D. T., Denevi, B. W., de Sanctis, M. C., Gaffey, M. J., Le Corre, L., Li, J.-Y., Marchi, S., McCord, T. B., McFadden, L. A., Mittlefehldt, D. W., Nathues, A., Palmer, E., Reddy, V., Raymond, C. A., Russell, C. T., 2012. Distinctive space weathering on Vesta from regolith mixing processes. Nature, Volume 491, Issue 7422, pp. 79-82 (2012).

Piccioni, G., Drossart, P., Coradini, A., Arnold, G., 2002. Virtis Experiment For Venus Express. EGS XXVII General Assembly, Nice, 21-26 April 2002, abstract #5923.

Pitjeva, E. V., 2005. High-Precision Ephemerides of Planets—EPM and Determination of Some Astronomical Constants. Solar System Research, Volume 39, Issue 3, pp.176-186.

Prettyman, T. H., Feldman, W. C., McSween, H. Y., Dingler, R. D., Enemark, D. C., Patrick, D. E., Storms, S. A., Hendricks, J. S., Morgenthaler, J. P., Pitman, K. M., Reedy, R. C., 2011. Dawn's Gamma Ray and Neutron Detector. Space Science Reviews, Volume 163, Issue 1-4, pp. 371-459.

Rayman, M. D., Fraschetti, T. C., Raymond, C. A., Russell, C. T., 2006. Dawn: A mission in development for exploration of main belt asteroids Vesta and Ceres. Acta Astronautica, Volume 58, Issue 11, p. 605-616.

Raymond, C. A., Jaumann, R., Nathues, A., Sierks, H., Roatsch, T., Preusker, F., Scholten, F., Gaskell, R. W., Jorda, L., Keller, H.-U., Zuber, M. T., Smith, D. E., Mastrodemos, N., Mottola, S., 2011. The Dawn Topography Investigation. Space Science Reviews, Volume 163, Issue 1-4, pp. 487-510.

Reddy, V., Nathues, A., Le Corre, L., Sierks, H., Li, J.-Y., Gaskell, R., McCoy, T., Beck, A. W., Schröder, S. E., Pieters, C. M., Becker, K. J., Buratti, B. J., Denevi, B., Blewett, D. T., Christensen, U., Gaffey, M. J., Gutierrez-Marques, P., Hicks, M., Keller, H. U., Maue, T., Mottola, S., McFadden, L. A., McSween, H. Y., Mittlefehldt, D., O'Brien, D. P., Raymond, C., Russell, C., 2012. Color and Albedo Heterogeneity of Vesta from Dawn. Science, Volume 336, Issue 6082, pp. 700- (2012).

Russell, C. T., Raymond, C. A., Coradini, A., McSween, H. Y., Zuber, M. T., Nathues, A., De Sanctis, M. C., Jaumann, R., Konopliv, A. S., Preusker, F., Asmar, S. W., Park, R. S., Gaskell, R., Keller, H. U., Mottola, S., Roatsch, T., Scully, J. E. C., Smith, D. E., Tricarico, P., Toplis, M. J., Christensen, U. R., Feldman, W. C., Lawrence, D. J., McCoy, T. J., Prettyman, T. H., Reedy, R. C., Sykes, M. E., Titus, T. N., 2012. Dawn at Vesta: Testing the Protoplanetary Paradigm. Science, Volume 336, Issue 6082, pp. 684- (2012).

Ruzicka, A., Snyder, G. A., Taylor, L. A., 1997. Vesta as the HED Parent Body: Implications for the Size of a Core and for Large-Scale Differentiation. Meteoritics & Planetary Science, vol. 32, no. 6, pages 825-840.

Schenk, P., O'Brien, D. P., Marchi, S., Gaskell, R., Preusker, F., Roatsch, T., Jaumann, R., Buczkowski, D., McCord, T., McSween, H. Y., Williams, D., Yingst, A., Raymond, C., Russell, C., 2012. The Geologically Recent Giant Impact Basins at Vesta's South Pole. Science, Volume 336, Issue 6082, pp. 694- (2012).

Schroder S.E., Mottola S., Keller H.U., Raymond C.A., Russell C.T., 2013. Resolved photometry of Vesta reveals physical properties of crater regolith.

Seeliger, H., 1887. Revue des publications astronomiques. Seeliger (H.). Ueber den Einfluss dioptrischer Felder des Auges auf das Resultat astronomischer Messungen (extrait des Mémoires de l'Académie des Sciences de Munich, t. XV). Munich, 1886; in-4°. Bulletin Astronomique, Serie I, vol. 4, pp.107-108.

Seeliger H., 1895. Über das Newton'sche Gravitationsgesetz. Astronomische Nachrichten, volume 137, Issue 9, p.129.

Shafer, D., 1982. Simple infrared telescope with stray-light rejection.

Shkuratov, Y., Starukhina, L., Hoffmann, H., Arnold, G., 1999. A Model of Spectral Albedo of Particulate Surfaces: Implications for Optical Properties of the Moon. Icarus, Volume 137, Issue 2, pp. 235-246.

Sierks, H., Keller, H. U., Jaumann, R., Michalik, H., Behnke, T., Bubenhagen, F., Büttner, I., Carsenty, U., Christensen, U., Enge, R., Fiethe, B., Gutiérrez Marqués, P., Hartwig, H., Krüger, H., Kühne, W., Maue, T., Mottola, S., Nathues, A., Reiche, K.-U., Richards, M. L., Roatsch, T., Schröder, S. E., Szemerey, I., Tschentscher, M., 2011. The Dawn Framing Camera. Space Science Reviews, Volume 163, Issue 1-4, pp. 263-327.

Stolper, E., 1977. Experimental petrology of eucritic meteorites. Geochimica et Cosmochimica Acta, vol. 41, May 1977, p. 587-611.

Takeda H., 1997. Mineralogical records of early planetary processes on the HED parent body with reference to Vesta. Meteoritics & Planetary Science, vol. 32, no. 6, pages 841-853.

Thomas, P. C., Binzel, R. P., Gaffey, M. J., Zellner, B. H., Storrs, A. D., Wells, E., 1997. Vesta: Spin Pole, Size, and Shape from HST Images. Icarus, Volume 128, Issue 1, pp. 88-94.

Thomas, P. C., Binzel, R. P., Gaffey, M. J., Storrs, A. D., Wells, E. N., Zellner, B. H., 1997.

Impact excavation on asteroid 4 Vesta: Hubble Space Telescope results. Science, vol. 277, p. 1492-1495.

Thomas, V. C., Makowski, J. M., Brown, G. M., McCarthy, J. F., Bruno, D., Cardoso, J. C., Chiville, W. M., Meyer, T. F., Nelson, K. E., Pavri, B. E., Termohlen, D. A., Violet, M. D., Williams, J. B., 2011. The Dawn Spacecraft. Space Science Reviews, Volume 163, Issue 1-4, pp. 175-249.

Toplis, M. J., Mizzon, H., Monnereau, M., Forni, O., McSween, H. Y., Mittlefehldt, D. W., McCoy, T. J., Prettyman, T. H., De Sanctis, M. C., Raymond, C. A., Russell, C. T., 2013. Chondritic models of 4 Vesta: Implications for geochemical and geophysical properties. Meteoritics & Planetary Science, Volume 48, Issue 11, pp. 2300-2315.

Warren, P. H., 1985. The magma ocean concept and lunar evolution. Annual review of earth and planetary sciences. Volume 13. Palo Alto, CA, Annual Reviews, Inc., 1985, p. 201-240.

Willoughby, C. T., Folkman, M. A., Figueroa, M. A., 1996. Application of hyperspectral imaging spectrometer systems to industrial inspection. Proceedings of the SPIE, Volume 2599, p. 264-272 (1996).

Yingst, R. A., Mest, S. C., Berman, D. C., Garry, W. B., Williams, D. A., Buczkowski, D., Jaumann, R., Pieters, C. M., De Sanctis, M. C., Frigeri, A., Le Corre, L., Preusker, F., Raymond, C. A., Reddy, V., Russell, C. T., Roatsch, T., Schenk, P. M., 2014. Geologic mapping of Vesta. Planetary and Space Science, Volume 103, p. 2-23.

Zambon, F., De Sanctis, M. C., Schröder, S., Tosi, F., Longobardo, A., Ammannito, E., Blewett, D. T., Mittlefehldt, D. W., Li, J.-Y., Palomba, E., Capaccioni, F., Frigeri, A., Capria, M. T., Fonte, S., Nathues, A., , Russell, C. T., Raymond, C. A., 2014. Spectral analysis of the bright materials on the asteroid Vesta. Icarus, Volume 240, p. 73-85.

Zuber, M. T., McSween, H. Y., Binzel, R. P., Elkins-Tanton, L. T., Konopliv, A. S., Pieters, C. M., Smith, D. E., 2011. Origin, Internal Structure and Evolution of 4 Vesta.